\documentclass{aa}

\usepackage{eurosym,rotating, booktabs}
\usepackage{xcolor}
\usepackage{graphicx}
\usepackage{amsmath}
\usepackage{txfonts}
\usepackage{csquotes}
\usepackage{xurl}
\usepackage{amsmath}
\usepackage{amssymb}
\usepackage{longtable}
\usepackage{threeparttable}
\usepackage{comment}
\usepackage{supertabular}
\usepackage{lscape}
\usepackage{multicol}
\usepackage{multirow}
\usepackage{pifont}

\usepackage{hyperref}

\begin{document}

\def\nodata{...}

\definecolor{mygreen}{RGB}{60,179,113} 
\definecolor{myblue}{RGB}{0, 123, 255} 
\definecolor{myorange}{RGB}{255, 159, 0} 
\definecolor{myred}{RGB}{191, 10, 48}
\definecolor{myyellow}{RGB}{255, 221, 51} 
\definecolor{mypurple}{RGB}{153, 102, 255} 

\newcommand{\green}[1]{\textcolor{mygreen}{#1}}
\newcommand{\red}[1]{\textcolor{myred}{#1}}
\newcommand{\yellow}[1]{\textcolor{myyellow}{#1}}
\newcommand{\purple}[1]{\textcolor{mypurple}{#1}}

\def\yes{\green{\ding{51}}}
\def\no{\red{\ding{55}}}
\def\maybe{\yellow{\ding{51}$^{?}$}}

\def\ngc6357{NGC\,6357}
\def\mys{MYStIX}

\def\Hii{H\,{\sc ii}}

\def\micron{$\mu$m}
\def\kms{km\,s$^{-1}$}
\def\cmss{cm\,s$^{-2}$}
\def\lsol{L$_{\odot}$}
\def\msun{M$_{\odot}$}
\def\msol{M$_{\odot}$}
\def\rsol{R$_{\odot}$}
\def\rsun{R$_{\odot}$}
\def\Rsun{R$_{\odot}$}
\def\s{$\sigma$}
\def\w{$\omega$}
\def\vsini{$v \sin i$}
\def\sigrms{$\sigma_\mathrm{rms}$}
\def\srv{$\sigma_\mathrm{RV}$}

\def\Msol{M$_\odot$}
\def\Msun{M$_\odot$}
\def\Lsol{L$_\odot$}
\def\Lsun{L$_\odot$}
\def\s{$\sigma$}
\def\feros{{\sc feros}}
\def\lco{{\sc lco}}
\def\uves{{\sc uves}}
\def\iacob{{\sc iacob}}

\def\l{$\lambda$}
\def\ll{$\lambda\lambda$}
\def\palp{Pa~$\alpha$}
\def\palph{Pa~$\alpha$}
\def\palpha{Pa~$\alpha$}
\def\pbet{Pa~$\beta$}
\def\pbeta{Pa~$\beta$}
\def\pdelt{Pa~$\delta$}
\def\pgam{Pa~$\gamma$}
\def\peps{Pa~$\epsilon$}
\def\halp{H~$\alpha$}
\def\halph{H~$\alpha$}
\def\halpha{H~$\alpha$}
\def\hbet{H~$\beta$}
\def\hdelt{H~$\delta$}
\def\hgam{H~$\gamma$}
\def\ha{H\,{\sc i}}
\def\hb{H\,{\sc ii}}
\def\hea{He\,{\sc i}}
\def\heb{He\,{\sc ii}}
\def\nc{N\,{\sc iii}}
\def\fea{Fe\,{\sc i}}
\def\nd{N\,{\sc iv}}
\def\ne{N\,{\sc v}}
\def\mgb{Mg\,{\sc ii}}
\def\ob{O\,{\sc ii}}
\def\sic{Si\,{\sc iii}}
\def\sid{Si\,{\sc iv}}
\def\H2O{H$_{2}$O}
\def\C2H2{C$_{2}$H$_{2}$}
\def\CO2{$^{12}$CO$_{2}$}
\def\13CO{$^{13}$CO$_{2}$}
\def\13CO2{$^{13}$CO$_{2}$}
\def\CH3+{CH$_{3}^{+}$}

\title{XUE: JWST spectroscopy of externally irradiated disks\\ around young intermediate-mass stars\thanks{The data described here may be obtained from
\href{https://doi.org/10.17909/tkjy-f210}{doi:10.17909/tkjy-f210}.}$^,$\thanks{Table~\ref{tab:full_properties}, containing all the properties of the XUE sources is available in online form.}} \titlerunning{MIRI spectroscopy of externally irradiated IMTT disks} \authorrunning{Ram\'irez-Tannus, Bik et al.}

\author{Mar\'ia Claudia Ram\'irez-Tannus\inst{1}
\and 
Arjan Bik\inst{2}
\and
Konstantin V. Getman\inst{3}
\and
Rens Waters\inst{4}
\and
Bayron Portilla-Revelo\inst{3, 5}
\and 
Christiane G{\"o}ppl\inst{6}
\and
Andrew J. Winter\inst{1}
\and
Jenny Frediani\inst{2}
\and
Germán Chaparro\inst{7}
\and
Eric D. Feigelson\inst{3, 5}
\and
Thomas J. Haworth\inst{8}
\and 
Thomas Henning\inst{1}
\and
Sebastián Hernández\inst{7}
\and
M. Alejandra Lemus-Nemocón\inst{9}
\and
Michael Kuhn\inst{10}
\and
Thomas Preibisch\inst{6}
\and 
Veronica Roccatagliata\inst{11,12}
\and 
Elena Sabbi\inst{13}
\and
Roy van Boekel\inst{1}
\and 
Peter Zeidler\inst{14}
}

\institute{Max-Planck Institut f{\"u}r Astronomie (MPIA), K{\"o}nigstuhl 17, 69117 Heidelberg, Germany
\and
Department of Astronomy, Stockholm University, AlbaNova University Center, 10691 Stockholm, Sweden
\and 
Department of Astronomy \& Astrophysics, Pennsylvania State University, 525 Davey Laboratory, University Park, PA 16802, USA
\and
Department of Astrophysics/IMAPP, Radboud University, PO Box 9010, 6500 GL Nijmegen, The Netherlands
\and
Center for Exoplanets and Habitable Worlds, Pennsylvania State University, 525 Davey Laboratory, University Park, PA 16802, USA
\and
Universit{\"a}ts-Sternwarte M{\"u}nchen,
Ludwig-Maximilians-Universit{\"a}t,
Scheinerstr.~1, 81679  M{\"u}nchen, Germany
\and
FACom, Instituto de Física – FCEN, Universidad de Antioquia, Calle 70 No. 52-21, Medellín, Colombia
\and
Astronomy Unit, School of Physics and Astronomy, Queen Mary University of London, London E1 4NS, UK
\and
Observatorio Astron\'omico Nacional, Universidad Nacional de Colombia, Bogot\'a, Colombia
\and
Centre for Astrophysics Research, University of Hertfordshire, Hatfield, AL10 9AB, UK
\and
Alma Mater Studiorum, Università di Bologna, Dipartimento di Fisica e Astronomia (DIFA), Via Gobetti 93/2, 40129 Bologna,Italy
\and
INAF-Osservatorio Astrofisico di Arcetri, Largo E. Fermi 5, 50125 Firenze, Italy
\and
Gemini Observatory/NSFs NOIRLab, 950 N. Cherry Ave., Tucson, AZ 85719, USA
\and
AURA for the European Space Agency (ESA), ESA Office, Space Telescope Science Institute, 3700 San Martin Drive, Baltimore, MD
21218, USA
}

\abstract
{Our knowledge of the initial conditions of terrestrial planet formation is mainly based on the study of  protoplanetary disks around nearby isolated low-mass stars. However, most young stars and therefore planetary systems form in high-mass star forming regions and are exposed to ultraviolet radiation, affecting the protoplanetary disk. These regions are located at large distances and only now with JWST become accessible to study the inner disks surrounding young stars.}
{
We present the eXtreme UV Environments (XUE) program, which provides the first detailed characterization of the physical and chemical properties of the inner disks  around young intermediate-mass ($1$–$4$~\msun)  stars exposed to external irradiation from nearby massive stars. We present high signal to noise MIRI-MRS spectroscopy of 12 disks located in three sub-clusters of the high-mass star-forming region NGC~6357 ($d\sim1690$~pc). }
{Based on their mid-infrared spectral energy distribution, we classify the XUE sources into Group~I and II based on the Meeus scheme. We analyze their molecular emission features, and compare their spectral indices and 10~\micron\ silicate emission profiles to those of nearby Herbig and intermediate T~Tauri (IMTT) disks.}
{The XUE program provides the first detailed characterization of the rich molecular inventory in IMTT disks, including water, CO, CO$_2$, HCN, and \C2H2. In the XUE sample, the detected emission likely originates from within 10~au, although this inner disk origin may not be typical for all externally irradiated disks.
Despite being more massive, the XUE stars host disks with molecular richness comparable to isolated T~Tauri systems. The spectral indices are also consistent with similar-mass stars in nearby regions. 
The 10~\micron\ silicate features in the XUE sample exhibit lower F$_{11.3}$/F$_{9.8}$ ratios at a given F$_{\mathrm{peak}}$, suggesting that the disk surfaces may be dominated by smaller grains compared to nearby disks. However, uncertainties in extinction prevent us from drawing firm conclusions about their inner disk properties.
The majority of disks display water emission from the inner disk, suggesting that even in these extreme environments rocky planets can form in the presence of water. Only one object shows PAH emission, contrasting with the higher PAH detection rates in IMTT surveys from lower-UV environments.
}
{The absence of strong line fluxes and other irradiation signatures suggests that the XUE disks have been truncated by external UV photons. However, this truncation does not appear to significantly impact the chemical richness of their inner regions. These findings indicate that even in extreme environments, IMTT disks can retain the ingredients necessary for rocky planet formation, comparable to those of lower mass T~Tauri disks in low-mass star-forming regions.}


\maketitle

\section{Introduction} \label{sec:intro}

Planets form in circumstellar disks around pre-main-sequence (PMS) stars, which are broadly classified into T~Tauri stars \citep[typically of spectral type F or later;][]{1945ApJ...102..168J, 1989A&ARv...1..291A}, and Herbig Ae/Be (HAeBe) stars \citep[spectral type A or B; ][]{1960ApJS....4..337H, 2010PASP..122..905J, 2023SSRv..219....7B, 2025A&A...693A..49S}. 
Intermediate-mass T Tauri stars (IMTTs) are considered the evolutionary predecessors of Herbig Ae/Be stars. With masses exceeding $1.5$~\msun, IMTTs are at an earlier evolutionary stage and have lower effective temperatures than HAeBe stars, typically exhibiting spectral types from early K to late F \citep{2004AJ....128.1294C, 2021A&A...652A.133V}.

\textit{Spitzer} mid-infrared (MIR) and ground based near-infrared (NIR) spectroscopic observations of T~Tauri disks located in near-by regions, such as Lupus an Taurus have revealed a rich molecular inventory in their inner 10~AU \citep[e.g.][]{2008Sci...319.1504C, 2008ApJ...676L..49S, 2009ApJ...696..143P, 2013ApJ...779..178P, 2010ApJ...720..887P, 2012ApJ...747...92M}, detecting molecules such as CO, \H2O, OH, \C2H2, HCN and \CO2. 
The improved sensitivity of the \textit{James Webb} Space Telescope \citep[JWST;][]{2023PASP..135d8001R} has enabled the detection of more and more complex molecules. Observations with the Medium Resolution Spectrometer \citep[MRS;][]{2015PASP..127..646W} of the Mid-InfraRed Instrument \citep[MIRI;][]{2015PASP..127..595W, 2015PASP..127..584R, 2023PASP..135d8003W} reveal molecules such as water, CO, \CO2\ isotopologues and several hydrocarbons \citep[e.g.][]{2023ApJ...945L...7K, 2023ApJ...947L...6G, 2023A&A...679A.117G, 2024ApJ...964...36R, 2025A&A...693A.278V, 2025AJ....169..184S, 2025arXiv250713921F}. Interestingly, disks surrounding very low-mass stars ($< 0.3$~\msun) appear to be carbon-rich and show a deficiency in O-rich species \citep[e.g.][]{2023NatAs...7..805T, 2024Sci...384.1086A}.
In contrast, \textit{Spitzer} observations of Herbig disks in nearby star-forming regions, suggest that they are relatively line-poor at IR wavelengths \citep[e.g.][]{2010ApJ...720..887P, 2011ApJ...731..130S, 2023SSRv..219....7B}, which \citet{2016A&A...585A..61A} attribute to the high continuum flux levels. MIRI observations of nearby IMTT disks are very challenging because of their brightness. Therefore, little is know about the inner disk molecular inventory of these objects.

With the advent of JWST, it is now possible to target protoplanetary disks (PPDs) in massive star forming regions, typically located at large distances \citep[$d > 1.5$~kpc;][]{2013ApJS..209...26F}. JWSTs improved sensitivity and spectral resolution allows us to detect spectral lines on top of a bright continuum. The large distances of these regions enable us to observe PPDs around intermediate-mass young stars, which would otherwise be too bright and saturate in nearby star-forming regions when observed with JWST. 

More than half of all stars, and thus planets, form under high external irradiation \citep[$\approx10^{3-6}$\,G$_0$, where G$_0$ is the Habing field;][]{2022EPJP..137.1132W}. 
These FUV fields are several orders of magnitude higher than those of nearby, low-mass star-forming regions such as the well-studied Taurus or Ophiuchus clouds. 
Therefore, characterizing PPDs in these extreme radiation environments is crucial to understanding the diversity of the detected exoplanet population \citep[][]{2020MNRAS.491..903W, 2023MNRAS.522.1939Q, 2025OJAp....8E..54A}. 
ALMA studies show that the cold dust reservoir in the outer disks near massive stars in Orion is depleted by photo-evaporation \citep[][]{2017AJ....153..240A, 2019A&A...628A..85V, 2020A&A...640A..27V}. In contrast, inner disk studies of a large sample of young massive star clusters, combining X-ray ($Chandra$) and near-infrared (\textit{Spitzer}) data, have 
obtained contradictory conclusions about the impact of cluster environment on the dissipation timescale of the warm, inner disk \citep[][]{2011ApJ...733..113R, 2012A&A...539A.119F, 2015ApJ...811...10R,2018MNRAS.477.5191R, 2023ApJS..269...13G}. 
Theoretical studies demonstrate the impact of cluster membership and UV irradiation on disk properties \citep{2008ApJ...675.1361F, 2020MNRAS.491..903W}. 
For instance, dynamical processes and photo-evaporative loss may result in disk truncation, leading to different planetary system architectures lacking gas giants in wide orbits. 
The inner disk chemistry in truncated disks may be considerably different due to a rapid inward drift of large dust grains, and a simultaneous depletion of small dust grains by entrainment in the photo evaporative wind \citep[e.g.][]{2016MNRAS.457.3593F, 2020MNRAS.492.1279S, 2023MNRAS.522.1939Q, 2025MNRAS.539.1414P}. This would result into a lower influx of dust ices transported to the inner disk. Since the main oxygen carriers are transported with the ices, a more carbon-rich inner disk chemistry may be then be prevalent.
Additionally, external FUV flux can heat surface layers of the disk up to several tens of AU, increasing the emitting area and therefore boosting the observed line fluxes for species like, OH, \C2H2\ and water \citep[][Hernandez Arboleda, \textit{in prep}]{2015A&A...582A.105A}.

JWST observations of externally irradiated disks around low-mass stars in the Orion Nebular Cluster (ONC) show the effect of the far-ultraviolet radiation (FUV) on the disk chemistry. For example, \citet{2023Natur.621...56B} present the first detection \CH3+ in the photo-dissociation region of the ionized proto-planetary disk (proplyd) d203-506, showing that photochemistry may allow for different pathways to complex organic chemistry. OH observations of the same source show evidence of photo-dissociation of \H2O\ in the presence of FUV radiation \citep[][]{2023A&A...671A..41Z}. Observations of CH$^+$ and \CH3+ emission \citep[][]{2025A&A...696A..99Z} as well as the detection of C$_{\rm I}$ fluorescence tracing the irradiated outer disk and photo-evaporative wind \citep[][]{2024A&A...692A.137A}, provide additional evidence for FUV-pumping \citep[][]{2024A&A...689L...4G}. 

In the ONC, where a single O star ($\theta^1$ Ori C) dominates the UV flux, irradiated disks are expected to migrate in and out of the proplyd regime ($\sim 5 \times 10^4$ G$_0$) on short timescales ($\sim 0.1$ Myr). In contrast, high-mass star-forming regions are usually powered by numerous O-type stars \citep[e.g.][]{2013ApJS..209...26F}, exposing PPDs to consistently higher radiation fields ($10^3$–$10^6$~G$_0$) throughout their lifetimes. To investigate the impact of such extreme environments, we proposed the eXtreme UV Environments (XUE) program, which aims to characterize the physical and chemical properties of highly irradiated PPDs in NGC 6357.   

\begin{figure*}
    \centering
    \includegraphics[width=\textwidth]{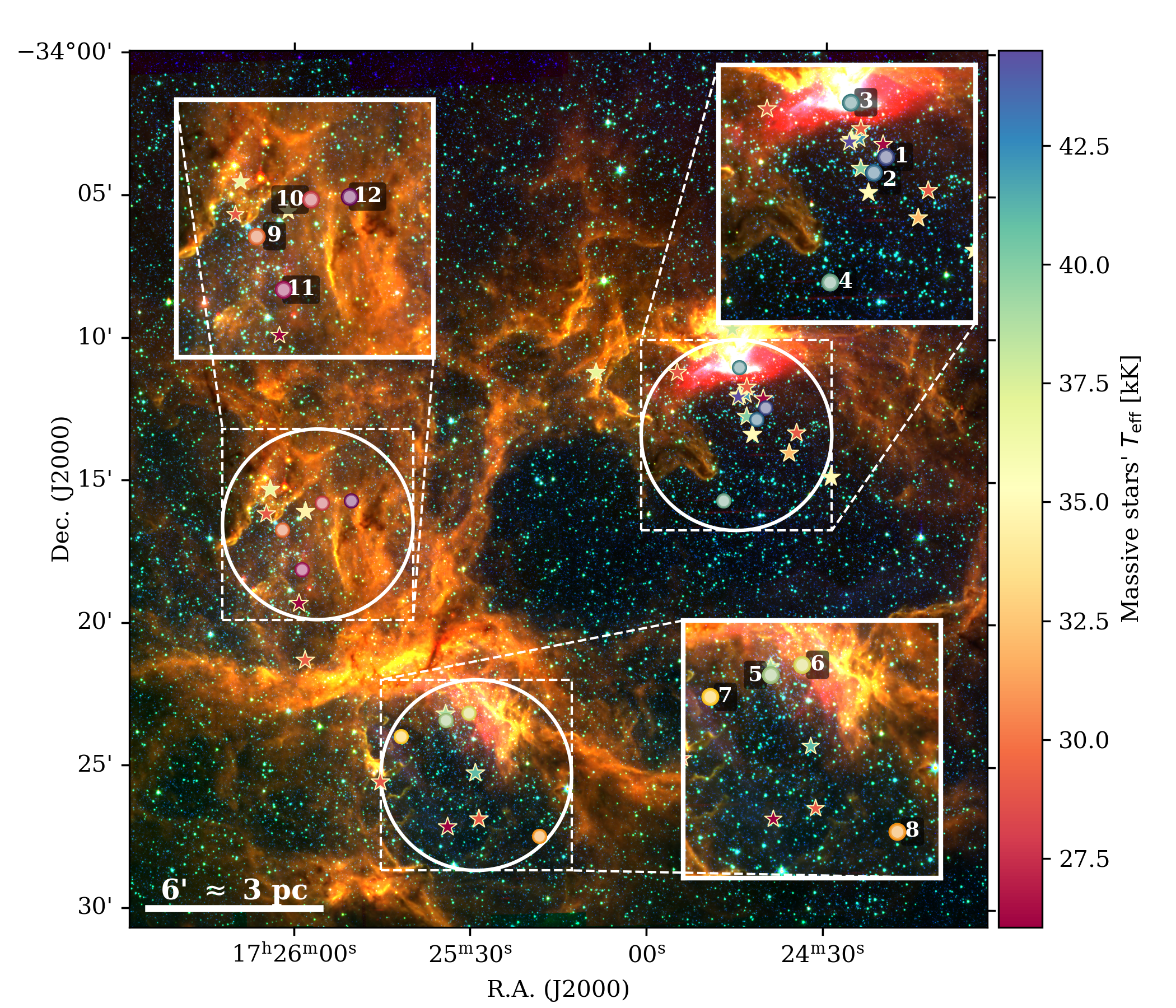}
    \caption{Color composite image of the massive star forming region NGC~6357 combining UKIRT Ks-band (blue) with \textit{Spitzer}-IRAC 4.5~\micron\ (green) and \textit{Spitzer}-IRAC 8.0~\micron\ (red) data. The small colored points show the location of the XUE sources with respect to the massive stars ($T_{\rm eff} > 25000$~K) in the region which are shown with the stars. The color bar indicates the temperature of the massive stars.
    The white circles show the location of the three sub-clusters: Pis~24 containing XUE~1, 2, 3 and 4, G353.1+0.6 containing XUE~5, 6, 7 and 8, and G353.1+0.7 containing XUE~9, 10, 11 and 12.}
    \label{fig:RGB}
\end{figure*}

With an estimated age of 1–1.6 Myr \citep{2014ApJ...787..108G} and located at a distance of 1.69~kpc \citep[][]{2023ApJ...958L..30R}, NGC~6357 is among the youngest and nearest massive star-forming regions. It hosts one of the most massive stars in the Galaxy \citep[Pis24-1, O4III(f+)+O3.5If*;][]{2003IAUS..212...13W} along with more than 20 additional O-type stars scattered across the field (Figure~\ref{fig:RGB}). 
NGC~6357's three sub-clusters - Pismis~24, G353.1+0.6, and G353.2+0.7 \citep{2007ApJ...670..428C, 2012A&A...539A.119F, 2014ApJ...787..108G} - share a similar age and distance but exhibit considerable variation in the far-ultraviolet (FUV) radiation fields that influence their PPDs \citep{2010A&A...515A..55R, 2020A&A...633A.155R}, thus offering us the opportunity to study in details the impact of diverse radiation environments on the early evolution of disks.

In this paper we present the XUE sample. Section~\ref{sec:obs} describes the sample selection, the ancillary data used to characterize the stellar properties and the JWST MIRI observations and data reduction. In sections~\ref{sec:SEDs} and \ref{sec:molecular_inventory} we describe the spectral energy distributions (SEDs) and the molecular content of our sources. Section~\ref{sec:comparisons} compares the XUE sources with nearby isolated T~Tauri, IMTT and Herbig disks as well as with a sample of PPDs in Orion. Finally, in sections~\ref{sec:discussion} and \ref{sec:conclusion} we discuss and conclude our work. 

\section{Observations}\label{sec:obs}

In this section we introduce the XUE sample in terms of selection criteria, observations, data reduction, and the auxiliary photometry used for our analysis. A detailed description of the Gaia and photometric properties of our sample is given in Appendix~\ref{sec:Gaia_mass_det}.

\subsection{Sample selection}\label{sec:selection}

In order to control for age and FUV flux history we selected disks in three sub-regions in NGC~6357 where the FUV flux regime differs, but formed in the same molecular cloud and environment (see Figure~\ref{fig:RGB}). We selected PPDs surrounding stars from the study of \citet{2020A&A...633A.155R} which selected cluster members from the \mys\ catalog \citep[][]{2018ApJS..235...43T, 2013ApJS..209...26F}.  
The selection included sources which had spectral type determinations based on NIR spectra obtained with the $K$-band Multi Object Spectrograph \citep[KMOS;][]{2013Msngr.151...21S} on the ESO Very Large Telescope (VLT) as well as signatures of mid-infrared excess based on their \textit{Spitzer} photometry. 
From all the objects that met the previous two criteria,
we selected sources with SEDs that are characteristic of Class II PMS stars, namely, disk-bearing stars.  
We aimed to cover a range of spectral types (and thus masses) from G to A evenly distributed between the sub-regions, as close as possible to the massive stars. 

\begin{figure}
    \centering
    \includegraphics[width=0.48\textwidth]{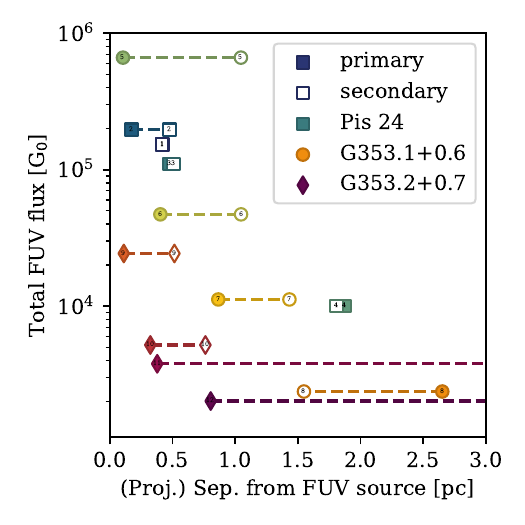}
    \caption{FUV flux experienced by our targets with respect to their projected separation to ionising sources in each NGC\,6357 sub-region. The full and empty symbols show the primary and secondary ionising stars in each sub-region. The
    names of the sources are displayed on the markers and are visible when zooming in.
    }
    \label{fig:FUV_flux_vs_distance}
\end{figure}

To calculate the external FUV flux towards the XUE sources, we integrated Phoenix atmosphere models \citep[][]{2013A&A...553A...6H} over the wavelength range $912-2000$~\AA\ to obtain the FUV luminosity from the massive stars. The FUV flux exposure to the XUE sources was computed based on the projected distance to the massive stars, assuming no extinction between the massive stars and our sources. 
For the input parameters to the Phoenix models, we selected the MIST \citep[MESA Isochrones and Stellar Tracks][]{2016ApJS..222....8D} isochrones that matched the effective temperature and luminosity of each OB star \citep[from][]{2020A&A...633A.155R, 2007ApJ...660.1480M}, assuming that the rotational velocity of the star is given by $v = 0.4 v_{\text{crit}}$, where $v_{\text{crit}}$ is the critical rotational velocity of the star.
The assumed properties for the massive stars are listed in Table~\ref{tab:OB_stars_props}. The FUV luminosities obtained using this method agree within 10\% with those computed using the scaling formula between stellar mass and
FUV luminosity from \citet{2003ApJ...584..797P}. 
Figure~\ref{fig:FUV_flux_vs_distance} shows the external FUV flux for each of the XUE sources as a function of the projected distance to the two massive stars that contribute the most to their irradiation.
Due to their location near several massive stars in NGC~6357, we expect the XUE sources to have been constantly exposed to a high radiation field throughout their lifetime.

The final selection contained 15 disks, 5 per sub-region, in order to have a wide spread of stellar mass and external FUV irradiation. 
Out of those 15 disks, one observation failed and two turned out to be foreground stars, leaving us  with a final sample of 12 sources.
The final sample of 12 sources contains 4 disks per sub-regions and spans a range of external FUV fluxes from $\sim10^3$ to $10^6$\,G$_0$ (Figure~\ref{fig:FUV_flux_vs_distance}).  
Two of our sources are late-A or early-F type stars (XUE~3 and XUE~10), two are K-type stars (XUE~2 and XUE~5), and the remaining systems are G-type stars (Table~\ref{tab:full_properties}). In Section~\ref{sec:aux_data} and Appendix~\ref{sec:Gaia_mass_det} we present the mass estimates for our sources and show that, with the exception of the T~Tauri disk XUE~1, they are IMTT disks.

\subsection{JWST MIRI spectra}

The observations were taken as part of the XUE project in Cycle~1 \citep[GO-1759][]{2021jwst.prop.1759R} with the Mid-Infrared Instrument in the MIRI-MRS \citep{2015PASP..127..584R, 2015PASP..127..646W, 2015PASP..127..595W}. 
All three wavelength settings (SHORT, MEDIUM and LONG) were used for the observations.
A four-point dither optimized for a point source was performed in the negative direction. 
The observations were obtained in the FASTR1 readout mode with 40 groups per integration and two integration per dither position. 
No separate background observations were taken for these observations as the background in these regions is highly spatially variable and therefore would not be representative of the background around our sources.

The data were reduced using the JWST pipeline version:  1.14.0 \citep{2024zndo..10870758B}. 
The \texttt{Detector1} part of the pipeline was run without modifications. In \texttt{Spec2} we implemented a custom background subtraction; after the WCS assignment, we performed a background subtraction by removing the two nodding positions from each other. This minimizes the effect of the variable background in comparison to taking a dedicated background observation. 
We then run the standard steps in \texttt{Spec2}, including the residual fringe correction. 
The \texttt{Spec3} run without any modifications. To extract the spectra, we measured the position of each source in the datacube and ran the \texttt{extract1d} step with the position of the sources as input.

We then stitched the 1D spectra by keeping the short wavelength ends of each band and, if necessary, scaling the flux of each band to match that of the previous one. 
Given the very strong nebular and emission from  Polycyclic aromatic hydrocarbons (PAHs) from the surroundings, the resulting 1D spectra still have residuals from the background subtraction. To mitigate the effect of the over- or under-subtraction of nebular lines, we used a peak removal algorithm that finds all local maxima by a simple comparison of neighboring values. These maxima were then replaced with the mean value of the neighboring pixels.

\subsection{Auxiliary data}\label{sec:aux_data}

Additionally to the MIRI-MRS spectra we use photometric data, taken from the \mys\ catalogue \citep{2013ApJS..209...26F, 2013ApJS..209...28K, 2013ApJS..209...29K} which includes the $JHK$-band magnitudes from the {\it United Kingdom Infra-red Telescope} (UKIRT) wide field camera \citep{2007MNRAS.379.1599L, 2008MNRAS.391..136L}, plus the four {\it Spitzer} IRAC bands \citep[GLIMPSE;][]{2003PASP..115..953B}. In the specific case of XUE~10, we used VVV NIR photometry for the $J$, $H$ and $K$-bands.

In Appendix~\ref{sec:Gaia_mass_det} we derive the best mass estimates for the XUE sources. The inferred stellar masses for most XUE stars fall within $(2-4)$~M$_{\odot}$, classifying them as intermediate-mass young stars. This aligns with their initial $K_s$-band selection criteria ($M_K < 10$~mag) for JWST observations \citep[Figure~\ref{fig:CMD_CCD}d;][]{2020A&A...633A.155R}.  The exception is XUE~1, a binary system, whose primary component has a mass of $\sim 1$~M$_{\odot}$ \citep[][]{2023ApJ...958L..30R}.
Based on the observed photometry and MIRI spectroscopy of the XUE sources we construct the spectral energy distributions (SEDs) between 1 and 28~\micron\ which are shown in Appendix~\ref{sec:full_SEDs}. 

\section{MIRI-MRS Spectral Energy Distributions}\label{sec:SEDs}

\begin{figure*}
    \centering
    \includegraphics[width=\textwidth]{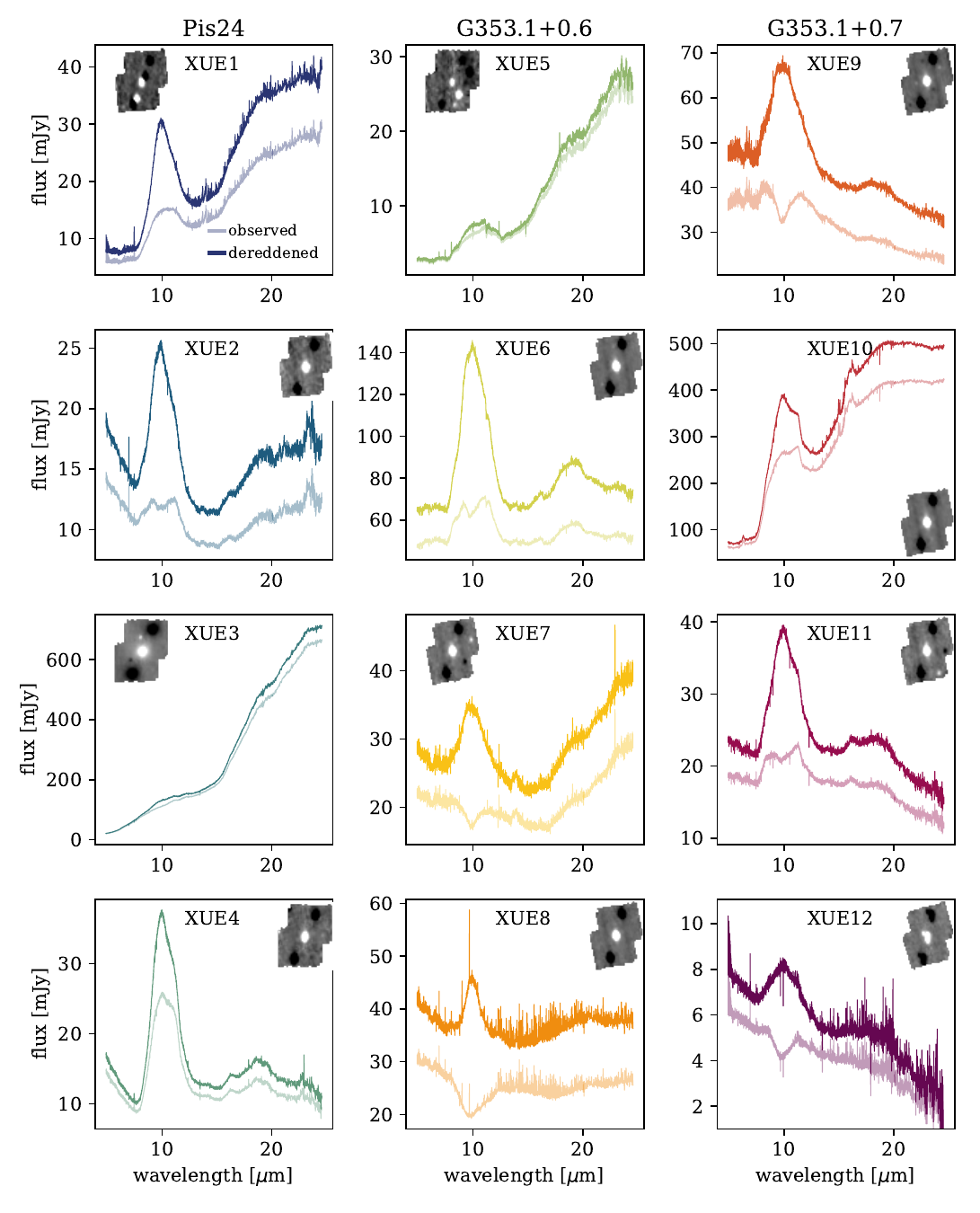}
    \caption{Spectral energy distributions of the XUE sources. Each column shows the sources in one region: Pis~24, G353.1+0.6 and G353.1+0.7 from left to right. The light colors show the observed data, the dark ones show the dereddened data. The panels in the top of each subfigure show the MRS data cube at 5.5~\micron.}
    \label{fig:SED_overview}
\end{figure*}

The observed MIR SEDs between 4.8 and 25~\micron\ are shown with the lighter colors in Figure~\ref{fig:SED_overview}. 
To deredden our data, we adopt the \citet{2023ApJ...950...86G} extinction law \citep[][]{2009ApJ...705.1320G, 2019ApJ...886..108F, 2021ApJ...916...33G, 2022ApJ...930...15D}, R$_V = 3.3$, and the A$_V$ values derived in Section~\ref{sec:stellar_props_section}. The dereddened SEDs are shown with the darker colors in Figure~\ref{fig:SED_overview}.
The panels at the top left of each SED show a slice of the MRS data cubes at 5.5~\micron. In the cases of XUE~1, 3, 5, 7 and 12 it is possible to see that there are extra sources in the detector with separations of at most 2\arcsec. The extra sources are probably contaminating the IRAC photometric points, therefore care must be taken when interpreting the existing MIR photometry for these targets (e.g. Figure~\ref{fig:SED_wStellarPhot_overview}). The spectra presented in this paper correspond to the the sources at the center of each cube. 
The additional sources detected in the field are either foreground stars or background dust; as such, their analysis falls outside the scope of this paper.

We classify our sources based on their MIRI-MRS spectral shape in Group~I and II following \citet{2001A&A...365..476M} classification. 
Group~I disks show a strong mid- and far-infrared excess in their SEDs, indicating that a significant portion of the disk surface is directly illuminated by the central star. These disks are thought to be geometrically flared, meaning that the outer disk regions receive and reprocess stellar radiation efficiently and therefore, their SEDs are brighter towards redder wavelengths. Imaging of Group~I disks showed that their SEDs can be understood as a result of large disk gaps, and/or the presence of an inner hole \citep[e.g.][]{2014A&A...563A..78M, 2017A&A...603A..21G}. 
Group~II disks exhibit a weaker mid- to far-IR excess compared to Group~I. The SED declines steeply beyond the NIR range, suggesting that the outer disk is receiving less direct stellar radiation. The disk structure is considered to be self-shadowed, meaning that a puffed-up inner disk rim blocks stellar light from reaching the outer disk or that the disk has a small outer dust radius. 
These disks are often more compact and show weaker mid- to far-IR emission \citep[][]{2001A&A...365..476M, 2021A&A...652A.133V}.

The SEDs of XUE~2, 4, 6, 7, 8, 9, 11 and 12 resemble those of Group~II disks. 
The SEDs of XUE~1, 3, 5, and 10 resemble Group~I disks.
XUE~1 and 10 are consistent with either flaring disks or disks with a gap \citep[][]{{2025ApJ...985...72P}}; however, given the available data, it is challenging to distinguish between these two scenarios. 
XUE~3 and 5 exhibit rising spectra, suggesting the presence of a substantial amount of cold dust. 
Specifically, the infrared excess of XUE~5 begins at approximately 8~\micron, which is indicative of an inner hole. A similar interpretation may apply to XUE~3; however, due to the limitations of the current data, this remains uncertain.
Following \citet{2001A&A...365..476M} all our disks belong either to Group~Ia or to Group~IIa given the ubiquitous presence of the 10-\micron\ silicate feature in the XUE sample. 

\section{Molecular inventory}\label{sec:molecular_inventory}

\begin{figure*}[ht!]
    \centering
    \includegraphics[width=\textwidth]{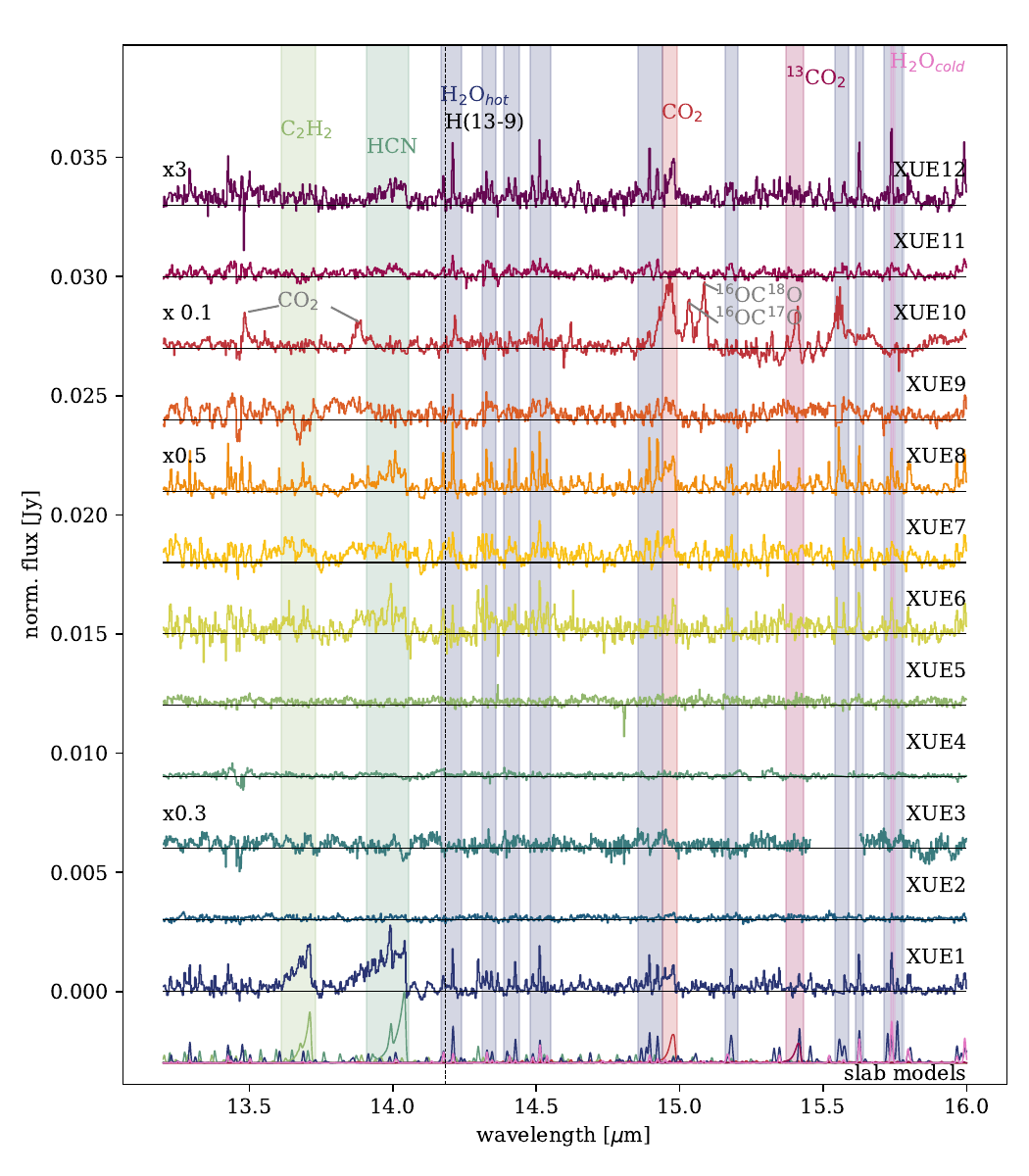}
    \caption{Overview of the spectral region between 13.2 and 16~\micron\ for all XUE sources. The lowermost spectrum shows the most prominent molecules in this region: HCN at 400~K and $10^{16}$~cm$^{-2}$ (dark green), H$_2$O at 400~K and $2.2\times10^{16}$~cm$^{-2}$ (magenta), C$_2$H$_2$ at 400~K and $2.2\times10^{16}$~cm$^{-2}$ (light green), CO$_2$ at 400~K and $10^{16}$~cm$^{-2}$ (red), H$_2$O at 850~K and $10^{18}$~cm$^{-2}$ (blue), $^{13}$CO$_2$ 400~K and $10^{16}$~cm$^{-2}$ (dark red).}
    \label{fig:13.5-15.5_region}
\end{figure*}

The XUE sample presents a great diversity in their molecular content. 
Figures~\ref{fig:13.5-15.5_region}, \ref{fig:4.9-5.3_region}, \ref{fig:6.5-6.9_region}, \ref{fig:15.8-17.0_region},  and \ref{fig:23.2-24.25_region} show an overview of continuum subtracted spectral regions containing key molecules. 
To estimate the continuum of our MIRI observations we used the \texttt{ctool} package \citep[][]{2024ApJ...963..158P}. 
Table~\ref{tab:molecules} summarizes the molecular content of all the sources in our sample, and a detailed description of each source is presented in Appendix~\ref{sec:individual_sources}.

\begin{table*}[ht]
    \centering
    \setlength{\tabcolsep}{1.5pt}
    \renewcommand{\arraystretch}{1.3}
    \caption{Presence of molecular emission in the spectra.}
    \begin{tabular}{c|c|c|c|c|cccccccccccc}
    \hline
    \hline
           Group &XUE& Sp. Type & M$\star$ & FUV & CO&  H$_2$O$_{h}$\tablefootmark{(a)} & H$_2$O$_{w}$\tablefootmark{(b)} &  HCN&  C$_2$H$_2$& OH & CO$_2$ &$^{13}$CO$_2$ & CH$_3^+$ & PAHs \\
            & & & [M$_\odot$] & [$\log{G_0}$] & & & & & & & & & & \\
           \hline

           I&3 & F3V & 2.7 & 5.04 & \no         &  \no         &  \no          &  \no&  \no&  \no&   \no&\no &\no &\no \\
        (inn. hole)&5 & K1V & 2.6 & 5.82 &  \no       &  \no         &  \no          &  \no&  \no&  \no&   \no&\no &\no &\no\\
           
           \hline

           I&1 & G9V & 1.2 & 5.19 &  \yes    & \yes     &  \yes     &  \yes&  \yes& \yes&   \yes & \no &\no &\no\\
           (flare/gap)&10 & F0V & 2.8 & 3.71 & \yes        & \maybe &  \no & \no& \no& \no &  \yes\yes& \yes\yes\tablefootmark{(*)} &\no &\yes\\
         \hline
          GI Fraction & & & & & 2/4 & [1-2]/4 & 1/4 & 1/4 & 1/4 & 1/4 & 2/4 & 1/4 & 0/4&1/4\\
           \hline
           &2 & K2V & 2.9 & 5.29 &  \yes       & \maybe &  \no          &  \maybe&  \no& \maybe&   \maybe & \no &\maybe &\no\\

           &4 & G4V & 3.6 & 4.00 & \no         &  \no         &  \no          &  \no&  \no&  \no&   \no&\no &\no &\no \\
           &6 & G6V & 3.5 & 4.67 & \maybe       &  \yes    &  \no          &  \yes&  \yes & \no &   \yes & \no &\no &\no \\
           II&7 & G8V & 3.4 & 4.05 & \yes    &  \yes    &  \no          &  \maybe &  \maybe & \maybe&   \yes&\no &\no &\no \\
           (self-&8 & G6V & 3.5 & 3.37 & \yes    &  \yes \yes    &  \yes \yes     &  \yes&  \maybe & \yes&   \yes& \maybe &\no &\no \\
           shadowed)&9 & G7V & 3.5 & 4.39 &  \no         &  \maybe         &  \no          &  \no&  \no& \no&   \no&\no &\no &\no \\           
           &11 & G9V & 3.0 & 3.57 & \no &  \yes    &  \no          & \yes & \no& \no&  \no&\no &\no &\no \\
           &12 & G8V & 3.2 & 3.30 & \yes     &  \yes    &  \yes      & \yes & \maybe & \yes&  \yes & \maybe &\no &\no \\
         \hline

         \hline
          GII Fraction & & & & & [4-5]/8 & [5-7]/8 & 2/8 & [4-6]/8 & [1-4]/8 & [2-4]/8 & [4-5]/8 & [0-2]/8 & [0-1]/8& 0/8\\
          \hline
          Total Fraction & & & & & [6-7]/12 & [6-9]/12 & 3/12 & [5-7]/12 & [2-5]/12 & [3-5]/12 & [6-7]/12 & [1-3]/12 & [0-1]/12 & 1/12\\          
        \hline
        \hline
    \end{tabular}
    \label{tab:molecules}
    \tablefoot{ The double check marks indicate a strong detection. The last row below each category shows the fraction of sources that present each molecule in their spectrum. In some cases, it is not possible to confirm or reject the presence of a certain molecule in the spectrum; in such cases, the fraction of sources is indicated in square brackets, with the numerator indicating a range of sources that might contain the molecule. \\
    \tablefootmark{(a)}{Water component at $\sim$900~K.} \\
    \tablefootmark{(b)}{Water component at $\sim$400~K.} \\
    \tablefootmark{(*)}{Also presents $^{16}$OC$^{17}$O and $^{16}$OC$^{18}$O.}    
    }
\end{table*}

We detect C- and O- bearing molecules such as CO, HCN, \C2H2, OH, and \CO2, which have also been detected in samples of nearby T~Tauri disks \citep[e.g.][]{2023ApJ...945L...7K, 2024PASP..136e4302H, 2023ApJ...957L..22B}. 
We also look for signatures of hot ($\sim900$~K), warm ($\sim400$~K) and cold ($\sim200$~K) water using the temperature-dependent flux ratio between the two low-energy lines near 23.85~\micron\ \citep[see][Figure~\ref{fig:23.2-24.25_region}]{2024A&A...689A.330T, 2025AJ....169..165B}. Most of our sources have a hot \H2O\ component, while only three show signatures of warm \H2O\ and none show a cold \H2O\ component. 
In one case (XUE~10) we detect multiple \CO2\ isotopologues as \13CO2, $^{16}$OC$^{18}$O and $^{16}$OC$^{17}$O \citep[][]{2025arXiv250713921F}, which have been detected in two nearby T~Tauri stars \citep[][]{2025A&A...693A.278V, 2025AJ....169..184S}. In the following we discuss the molecular content in the context of the classification presented  in Section~\ref{sec:SEDs}.

XUE~3 and 5, which are the sources classified as Group~I and with SEDs consistent with having an inner hole, do not show any molecular emission. 
The other Group~I sources are XUE~1 and 10. XUE~1 is a 1~\msun\ star and has a rich molecular inventory, similar to nearby T~Tauri disks as presented in \citet{2023ApJ...958L..30R}. In this work we were able to identify OH which was not reported in \citet{2023ApJ...958L..30R}, this is thanks to the improvement in the data reduction with respect to previous versions of the JWST pipeline.
A characterization of the thermochemical structure of XUE~1 is presented in \citet{2025ApJ...985...72P}. 
XUE~10 is a Herbig star with a very peculiar molecular content. It presents extremely abundant \CO2, \13CO2, $^{16}$OC$^{18}$O and $^{16}$OC$^{17}$O and it is very poor in water. A detailed investigation this source XUE~10 is presented in \citet{2025arXiv250713921F}. 

Among the eight Group~II sources, half (XUE~6, 7, 8 and 12) show a rich molecular inventory including CO, warm water, HCN, \C2H2 and \CO2. 
XUE~8 is remarkable due to its very abundant water content. In XUE~8 and 12 we also detect the presence of warm ($\sim400$~K) water lines between 23 and 24~\micron\ \citep[e.g.][]{2025A&A...694A.147G, 2025AJ....169..165B}.
XUE~4 does not exhibit any molecular emission in the MIRI wavelength range. Compared to other Group II sources, XUE~4 also lacks significant IR excess in the 2–7~\micron\ region (Figure~\ref{fig:SED_wStellarPhot_overview}). As with XUE~3 and XUE~5, the absence of molecular emission could point to a gas-depleted inner disk. However, the substantial IR excess at most MIRI wavelengths and the prominent 10~\micron\ silicate feature indicate that dust is still present, suggesting that XUE~4 may host a dust-rich but gas-poor inner disk.
XUE~9 does not show any molecular emission, except for a tentative detection of warm water, nevertheless this source does have IR excess starting at around 2~\micron. Therefore, the lack of molecules in XUE~9 cannot be attributed to a gas depleted inner disk and needs to be investigated further.  
XUE~2 and 11 are also poor in molecular content, where the spectrum of XUE~11 only contains warm water and HCN and XUE~2 only has a definite detection of CO, and tentative detections of warm water, HCN, OH and \CO2. The spectrum of XUE~2 has an emission feature at 7.15~\micron\ which could be consistent with the detection of the methyl cation (CH$_{3}^{+}$) as reported by \citet{2023Natur.621...56B} and later detected in TW Hya \citep[][]{2024PASP..136e4302H}. Nevertheless, given the possible presence of water emission at those same wavelengths and the S/N of our observations, it is not possible to claim a strong detection of CH$_{3}^{+}$ in XUE~2 (see Lemus-Nemocón, \textit{in prep.}).
XUE~2 and 11 show relatively low 7~\micron\ excess, which could explain the poor molecular inventory. 

Within the XUE sample only XUE~10 presents PAH emission at 6.2~\micron\ and 11.3~\micron. This is unexpected, as the central masses of our sources and the surrounding FUV flux suggest that PAHs inherited from the ISM should be excited in these disks (see Section~\ref{sec:truncation}).

\section{Comparison to nearby disks}\label{sec:comparisons}

In this section we present comparisons of the XUE sample with different samples of nearby PPDs. We first discuss the SED shapes of our sample in the context of other young intermediate mass stars (Section~\ref{sec:MIR_SED_shape}). Then we compare the 10~\micron\ silicate feature to samples of T~Tauri and Herbig disks (Section~\ref{sec:silicate_feature}) and discuss the spectral indices in comparison to a sample of disks in Orion (Section~\ref{sec:spectral_index}).  
Given the mass range of our sample, direct comparisons of the molecular content with similar observations of nearby disks are challenging. JWST observations of nearby Herbig and IMTT stars are difficult due to detector saturation from their intrinsic brightness. Meanwhile, MIR observations with \textit{Spitzer} lack the necessary signal-to-noise ratio and spectral resolution to detect molecular lines on top of the bright continuum of IMTT or Herbig disks. Consequently, we compare our sample’s molecular content with that of nearby T~Tauri stars, while recognizing that the central star's properties significantly influence inner disk properties, especially towards higher masses. In Section~\ref{sec:H2O_lum} we compare the water line luminosity at 17~\micron\ with a samples of nearby disks observed with \textit{Spitzer} and JWST.

\subsection{Spectral energy distribution}\label{sec:MIR_SED_shape}

Herbig Ae/Be stars and IMTTs typically exhibit bright disk emission at 7~\micron\ \citep[][]{2001A&A...365..476M, 2013A&A...559A...3S}.
A notable characteristic of the XUE sample is the relative weak disk emission compared to the stellar contribution at these wavelengths (see Figure~\ref{fig:SED_wStellarPhot_overview}).
This difference may be attributed to our sample selection, which focused on sources that could be classified based on their $K$-band spectra. As a result, the selected disks likely exhibit weak near-infrared emission, which could extend to the 7~\micron\ region as well. 

This trend in mid-infrared emission further aligns with the overall properties of the XUE sample, which also exhibits distinct PAH emission characteristics.
The XUE sample does not contain any PAH-dominated disks, which contrasts with observations in Herbig Ae/Be stars and IMTT disks where the detection rate is 70\% \citep[][]{2010ApJ...718..558A} and 44\% \citep[][]{2021A&A...652A.133V}, respectively. Notably, only one object in the XUE sample (XUE~10) exhibits clear PAH emission. This discrepancy cannot be solely attributed to the presence of cooler stars, as approximately half of the objects in the \citep[][]{2021A&A...652A.133V} sample show PAH emission despite having similar spectral types as the XUE sample. Instead, this suggests that the XUE sample predominantly behaves like Group II disks, which have a lower PAH detection rate \citep[][]{2021A&A...652A.133V}. The predominance of compact disks and/or disks without a large gap and an outer disk could be attributed to our sample selection as described above, or to prolonged exposure to high UV radiation, which may lead to outer disk evaporation.

\subsection{10~\micron\ silicate feature}\label{sec:silicate_feature}
  
We analyze the shape of the 10-\micron\ silicate feature in a similar way to \citet{2005A&A...437..189V} and \citet{2006ApJ...639..275K}. In order to minimize the uncertainty of the background subtraction, especially around the PAH band at 11.3~\micron\, we smoothed the MIRI spectrum using a median filter excluding the region between 10.6 and 11.68~\micron. For this we used \texttt{ctool}\footnote{https://github.com/pontoppi/ctool}, setting the threshold parameter to 0.5 such that the smoothed spectrum, $F_{\nu, smooth}$, passes through the median of our observations.
We normalized the the smoothed spectrum around the 10-\micron\ feature by fitting a linear function to the mean flux of two windows: [$7.5-7.7$]~\micron\ and [$13.2-13.6$]~\micron. The normalized spectrum, $S_{\nu}$ is given by:

\begin{equation}
    S_{\nu} = 1 + \frac{F_{\nu, smooth} - F_{\nu, cont}}{\langle F_{\nu, cont}\rangle},
\end{equation}

\noindent where $F_{\nu, smooth}$ is the smoothed observed flux, $F_{\nu, cont}$ is the continuum and $\langle F_{\nu, cont}\rangle$ is the mean value of the continuum between 7.6 and 13.4~\micron. The strength of the feature, $F_{peak}$ is then defined as the maximum value of $S_{\nu}$ between 6 and 14~\micron, and the shape of the feature, $F_{11.3}/F_{9.8}$ is given by the ratio of $S_{\nu}$ at 11.3 and 9.8. 

\begin{figure}[htbp]
    \centering
    \includegraphics[width=0.45\textwidth]{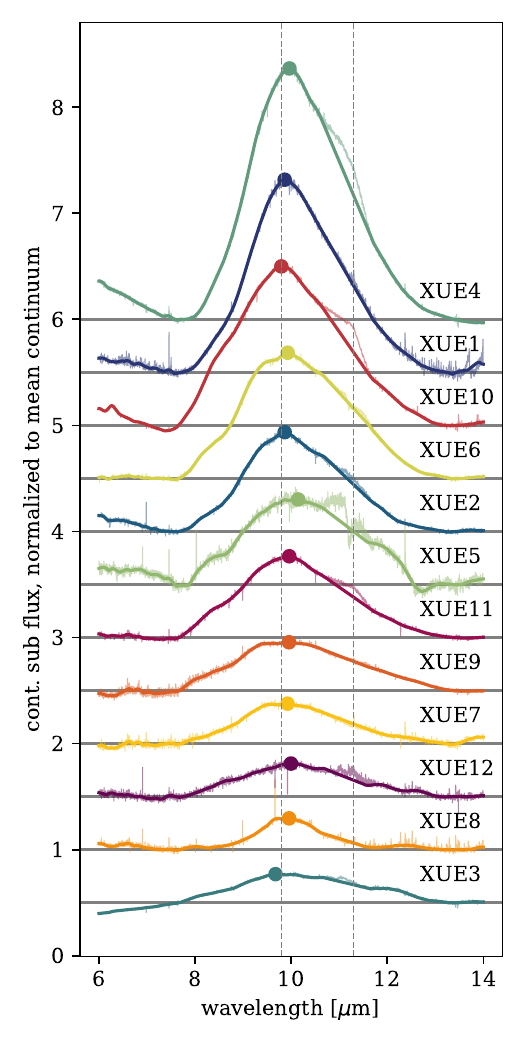}
    \caption{10~\micron\ silicate feature for the XUE sources. The features are continuum subtracted and normalized to the maximum value. The thick solid lines show the smoothed spectra and the thin lines the MIRI observations. The dashed lines mark 9.8 and 11.3~\micron\ and the dot on top of each feature shows the position of the maximum flux.}
    \label{fig:silicate_feature}
\end{figure}

Figure~\ref{fig:silicate_feature} shows the normalized silicate feature for all our sources in ascending order of silicate strength. The thick-solid lines show the smoothed spectra, $F_{\nu, smooth}$, used to calculate the silicate strength and shape, and the thin lines show the observed spectra. The vertical dashed lines mark 9.8 and 11.3~\micron\ and the dots show the strength, $F_{peak}$ of each feature. 

\begin{figure}[htbp]
    \centering
    \includegraphics[width=0.45\textwidth]{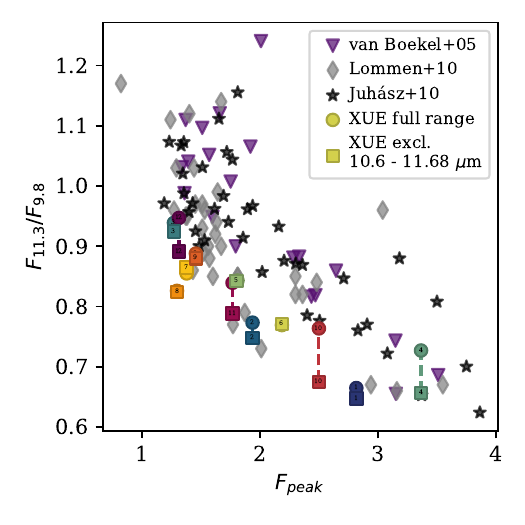}
    \caption{Silicate 10~\micron\ feature strength vs. shape. The purple triangles show the Herbig~AeBe sources from \citet{2005A&A...437..189V}, the gray diamonds show the sample of T-Tauri and Herbig-Ae/Be stars from \citet{2010A&A...515A..77L}, the black stars show the the Herbig~AeBe sources from \citet{2010ApJ...721..431J}, and the colored squares show the XUE sources.}
    \label{fig:shape_vs_strength}
\end{figure}

In Figure~\ref{fig:shape_vs_strength}, we compare the shape and strength of the silicate features in our sample with those observed in T Tauri and Herbig Ae/Be stars from \citet{2005A&A...437..189V}, \citet{2010A&A...515A..77L} and \citet{2010ApJ...721..431J}. Consistent with these previous studies, we find a negative correlation between the feature’s shape and its strength. Our sample spans a broad range of silicate-feature properties and generally exhibits lower F$_{11.3}$/F$_{9.8}$ ratios at a given F$_{\mathrm{peak}}$ compared to nearby T Tauri and Herbig stars.
However, \citet{2011A&A...526A.152V} demonstrated that the relationship between the 9.7~\micron\ optical depth ($\tau_{9.7}$) and the near-infrared color excess, E($J - K$), differs between molecular and diffuse sightlines \citep[see also][]{2007ApJ...666L..73C}. In our analysis, we adopt the extinction law from \citet{2023ApJ...950...86G}, which is derived from diffuse sightlines. 
This choice may introduce uncertainties in the dereddening process, making it difficult to assess whether the lower F$_{11.3}$/F$_{9.8}$ ratios observed in the XUE sample are truly due to variations in grain size. 
At fixed R$_V$, the adopted A$_V$ affects the shape and strength of dust features: stronger and broader features arise for higher A$_V$ values. For example, if A$_V$ would be 20\% higher, the XUE sample’s F$_{11.3}$/F$_{9.8}$ distribution diverges further from other datasets, while a 20\% lower A$_V$ shifts the XUE points upward but still leaves them at the lower edge of the other distributions. 
In the case of the comparison samples, the authors assume that extinction from nearby sources is negligible, so the spectra are presented without any extinction correction.
If the observed differences between the XUE and nearby samples persist, they may indicate that the disk surfaces in the XUE sample are dominated by smaller dust grains. 
Further investigation is required to determine whether the observed trend reflects an intrinsic dust property or is an artifact of the dereddening approach. 

\subsection{Spectral indices}
\label{sec:spectral_index}
\begin{figure}[htbp]
    \centering
    \includegraphics[width=0.48\textwidth]{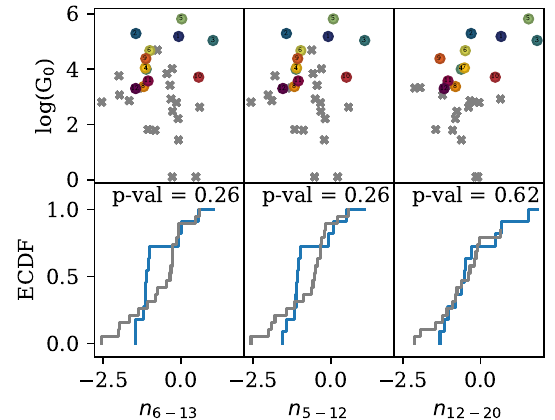}
    \caption{Comparison with a sample of disks with masses $>2$~\msun\ in Orion \citep{2016ApJS..226....8K}. The top panels show the external FUV flux as a function of each spectral index, the XUE sources are shown with the colored dots and \citet{2016ApJS..226....8K} data with the gray crosses. The bottom panels show the empirical cumulative distribution functions for each index, the p-values listed are the result of a Kolmogorov-Smirnov test.}
    \label{fig:sp_index_comparison}
\end{figure}

Figure~\ref{fig:sp_index_comparison} shows the spectral indices $n_{6-13}$, $n_{5-12}$ and $n_{12-20}$ defined as:

\begin{equation}
    n_{\lambda_1 - \lambda_2} = \frac{\log_{10}{(\lambda_1 F_1)} - \log_{10}{(\lambda_2 F_2)}}{\log_{10}{(\lambda_1)} - \log_{10}{(\lambda_2)}},
\end{equation}

\noindent compared to with those from the \textit{Spitzer} survey of PPDs in Orion derived by \citet{2016ApJS..226....8K}. The top panels show the spectral indices vs. the FUV irradiation experienced by the PPDs.
We do not detect any significant correlation between the spectral indices and the external FUV flux. 

The bottom panels show histograms of the spectral indices of our samples compared to those of the Orion disks. The XUE continuum slopes generally lie within the loci of the Orion slopes. XUE~3 is the only outlier on the $n_{5-12}$ and $n_{6-13}$ related plots, and XUE~3 and XUE~5 are possible outliers on the $n_{12-20}$ related plots. We performed Kolmogorow-Smirnow-Tests to verify the hypothesis that both samples are drawn from the same parent distribution. The obtained p-values do not allow us to reject the null hypothesis,
suggesting that, for the XUE sources, external irradiation does not significantly alter the disk geometry to the extent that it affects the MIR spectral index.

\subsection{Water line luminosities}\label{sec:H2O_lum}

\begin{figure*}
    \centering
    \includegraphics[width=\textwidth]{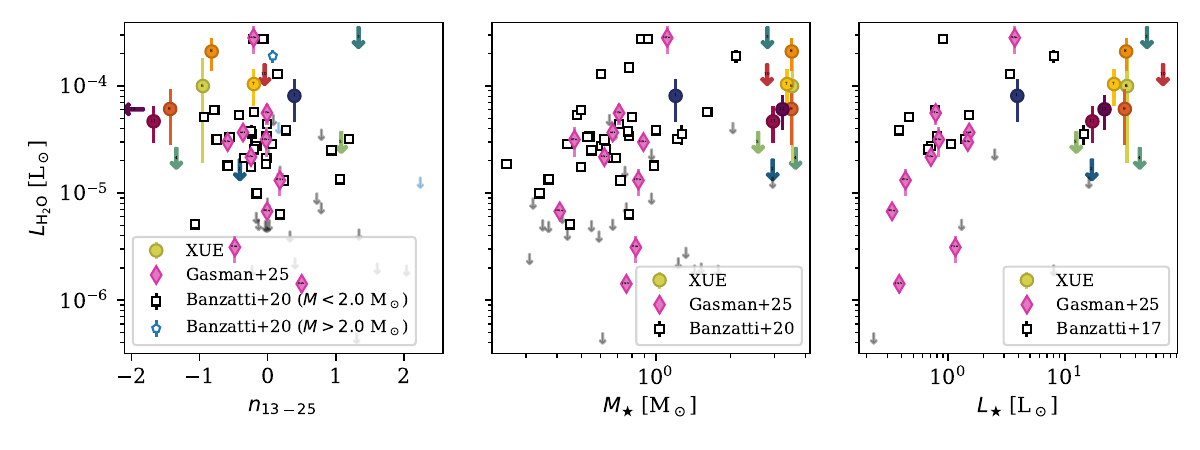}
    \caption{Comparison between the 17\,\micron\ \H2O\ luminosity and the spectral index n$_{13-25}$ (\textit{left}), the stellar mass (\textit{middle}) and the stellar luminosity (\textit{right})  of the XUE disks with the \textit{Spitzer} sample studied by \citet{2020ApJ...903..124B} shown with black squares ($M<2$~\msun) and blue pentagons ($M>2$~\msun) and the MIRI observations of nearby disks from the MINDS collaboration \citep{2025A&A...694A.147G}. The measurements for the XUE spectra are shown with the colored circles, following the same color scheme as the rest of the paper with arrows showing the upper limits. The upper limits from \citet{2020ApJ...903..124B} are indicated with gray arrows. The MINDS sources are shown as magenta diamonds.}
    \label{fig:H2Olum_vs_n13-25}
\end{figure*}

We measured the water line luminosity ($L_{H_2O}$) by integrating the spectra in the $[17.075 - 17.385]$~$\mu$m wavelength range and calculated the errors by measuring the noise between 17.333 and 17.350~\micron\ (see Appendix~\ref{sec:H2O_spectra}). For some sources (XUE 2, 3, 4, 5, 9 and 10), the \H2O\ lines are not clearly detected, and therefore we report only upper limits.
The assessment of line detection was made by visual inspection rather than by applying a specific sigma threshold. 
Figure~\ref{fig:H2Olum_vs_n13-25} shows $L_{H_2O}$, against the spectral index $n_{13-25}$, the stellar mass ($M_{\bigstar}$) and the stellar luminosity ($L_{\bigstar}$) of our XUE sources in comparison with a sample of T\,Tauri stars located in  nearby ($<200$\,pc), young (1-3\,Myr) star-forming regions from  \citet{2020ApJ...903..124B} and the MINDS collaboration \citep[][]{2025A&A...694A.147G}. 
The stellar luminosities for the \textit{Spitzer} sample were taken from \citet{2017ApJ...834..152B}. 
We find that in general the XUE spectral indices are lower than those found in the T~Tauri sample, and that the H$_2$O luminosity is at the high end of the distribution. Nearby sources exhibit a trend between H$_2$O luminosity and both $M_{\bigstar}$ and $L_{\bigstar}$, although this relationship is marked by substantial scatter \citep[e.g.][]{2020ApJ...903..124B, 2025A&A...694A.147G}. In contrast, the XUE sources do not follow this trend; instead, the relation appears to flatten at higher stellar masses and luminosities. 

\section{Discussion}\label{sec:discussion}

In this paper we present the first sample of IMTTs observed with MIRI-MRS on board JWST. 
To investigate the effect of external UV radiation on their inner disk properties, we observed 12 highly irradiated disks in the NGC~6357 region. We classify our sources based on previous studies of stars with similar masses observed with ISO and Spitzer. Our analysis reveals a rich and diverse molecular inventory, including \H2O, CO, HCN, \C2H2, OH and \CO2, along with some \CO2\ isotopologues detected in one of our sources. In Section~\ref{sec:comparisons}, we compare our sample to nearby disk populations to explore the influence of UV radiation. In this section, we discuss our findings and provide an interpretation.

\subsection{Nature of the XUE disks}

The disks in our sample are generally more massive ($\sim3$~\msun) than those observed around IMTT and Herbig stars \citep[$1.5–2$~\msun][]{}, and as a result, they are significantly younger than those samples, which can be as old as 10 Myr \citep[see][]{2023SSRv..219....7B} Despite their youth, the SEDs of the XUE sources suggest that some of these disks are already undergoing advanced stages of clearing. 
This could be due to our sample selection, as we used the $K$-band spectral type to identify the disk bearing sources. Thus, each target has a visible photosphere in the NIR. 
Therefore, it is possible that our sample is biased towards objects which have already (partially) lost their disks.  
Most of our sources do not exhibit clear signatures of irradiation effects, such as strongly enhanced gas emission lines or prominent PAH features \citep[e.g.][see Sect.~\ref{sec:lack_of_distinct_chem}]{2023Natur.621...56B, 2024A&A...692A.137A, 2025A&A...696A..99Z}. Additionally, their spectral energy distributions (SEDs) are consistent with being absent of large disk gaps, which are typically associated with group I sources.
Instead, we observe the presence of inner disk holes. This observation supports the hypothesis, in line with observations of the XUE~1 disk, that the disks are relatively small in size \citep[][]{2023ApJ...958L..30R, 2025ApJ...985...72P}.

Large, gapped disks are often associated with the formation of giant planets that block the inward drift of pebbles. In contrast, the potentially small sizes of the XUE disks could be explained by the rapid inward drift of solid material \citep[][]{2022A&A...658A.112S, 2025A&A...693A.286S}, or by the effect of external irradiation due to the FUV photons from the nearby massive stars (see Sect.~\ref{sec:truncation}). Therefore, it is possible that the XUE disks were never large, that they did not form gas giant planets capable of halting the inward drift, or that they dissipated in a short time scale due to the external irradiation. In all three scenarios, the presence of gas giant planets in wide orbits is unlikely.

Most notably, the XUE spectra suggest that terrestrial planet formation can proceed even in strongly irradiated environments, as the observed disks retain the chemical complexity necessary for forming rocky planets. This is consistent with studies like \citet{2015ApJ...811...10R, 2011ApJ...733..113R, 2012A&A...539A.119F}, which show that inner disks can survive in regions exposed to intense FUV radiation. Together, these observations support the idea that external irradiation does not necessarily inhibit the early stages of terrestrial planet formation.

\subsection{Comparison to nearby disks}

In Sections~\ref{sec:MIR_SED_shape} and \ref{sec:spectral_index}, we compare the spectral shape of the XUE sample with the disks found around stars of similar spectral type in Orion. In terms of spectral shape, the XUE disks appear to closely resemble their nearby counterparts. 
The 10~\micron\ silicate features of our sample show an offset toward lower values of F$_{11.3}$/F$_{9.8}$ compared to nearby T~Tauri and Herbig disks.
If the difference is robust, this would imply that the disk surfaces may be dominated by smaller grains compared to nearby disks. but if this is due to the extinction correction method then this would imply that external irradiation does not significantly affect the dust composition of the hot, inner disk in IMTTs. 

In terms of the molecular content, our sample is the only one in this mass range in which we are able to study the molecular emission with the unprecedented sensitivity of JWST. 
The XUE sources are cooler, with spectral types G and F, compared to the generally older Herbig Ae/Be stars (spectral types A and B) studied by \textit{Spitzer}. 
Disks for which similar observations with JWST exist have masses $\leq1$~\msun. Nevertheless, the XUE spectra have significant similarities to T~Tauri disks in nearby regions in terms of their molecular content, therefore in the following we will make a parallel between both samples, even if the central star masses are different. 
We detect hot ($\sim900$~K) \H2O in most of our Group~II disks, and in half of the Group~I disks. Only two disks of the XUE sample show signatures of warm ($\sim400$~K) \H2O\ and we do not detect any cold ($\sim200$~K) \H2O\ in our sample. This is in contrast with observations of nearby T~Tauri disks, in which most disks have a cold water component \citep[][]{2025AJ....169..165B, 2025A&A...694A.147G}. 
The lack of a cold water component in the XUE sample could either be due to the external irradiation impinging these disk, or to the strong radiation from the central stars in comparison to T~Tauri disks. In order to test this hypothesis, similar observations of IMTTs in non-irradiated environments are needed.

\subsection{Disk truncation of the XUE sources}\label{sec:truncation}

Radiation thermo-chemical models \citep[][Hernandez Arboleda, \textit{in prep.}]{2015A&A...582A.105A} predict that flared disks with large outer radii and FUV-heated surface layers can produce strong molecular line emission (e.g., H$_2$O, CO$_2$, C$_2$H$_2$) because the warm extended layers increase the emitting area. In such disks, PAH emission, which is frequently observed in Herbig Ae/Be stars, typically originates from the outer disk surface and traces the spatial distribution of gas \citep[e.g.][]{2006Sci...314..621L, 2010ApJ...718..558A, 2023A&A...674A..57Y}. Given the high stellar luminosities and strong external radiation fields in XUE sources, comparable to or exceeding those of Herbig stars, strong line and PAH emission across the full disk surface would be expected.

However, this is not observed. Only one XUE source (XUE~10) shows detectable PAH emission, and none display particularly high molecular line luminosities compared to nearby disk samples. This lack of enhanced emission features and the absence if PAH features suggests that the XUE disks are truncated, likely as a result of external photoevaporation \citep[see][]{2023ApJ...958L..30R}. 
Supporting this, \citet{2025ApJ...985...72P} modeled XUE~1 and found that it's spectrum can be explained by a truncated disk at approximately 10~au and significantly depleted in gas at larger radii. 
This result is consistent with observations of the DF Tau system, where the outer disk is truncated by binary interactions, yet the inner disk properties appear unaffected by either the outer disk evolution or the close binary nature \citep[][see also \citet{2024A&A...691A..32N}]{2024A&A...689A..85G}. 
Direct confirmation of disk truncation in the XUE disks requires ALMA observations to more robustly constrain their spatial extent and mass.

The rest of the XUE sample exhibits similar spectral features and levels of external irradiation, indicating that disk truncation may be common in this environment. In these compact configurations, the inner disk chemistry appears to be governed primarily by the central star rather than by the external radiation field. This interpretation is consistent with ALMA observations of a T~Tauri and a Herbig disk in the outskirts of the Orion Nebula Cluster, where external fields of around 100 G$_{0}$ do not dominate the observed chemical composition \citep{2024ApJ...969..165D}.

Radiation thermo-chemical models of truncated disks provide further support for this scenario (Hernandez Arboleda, \textit{in prep.}). As disks lose their outer regions, whether through photoevaporation or dynamical processes, the line emission becomes increasingly sensitive to stellar properties such as luminosity and less dependent on the external UV field. Compact disks with radii less than about 30 au can still maintain rich molecular inventories in their inner regions if the midplane remains warm due to stellar irradiation or accretion. Similarly, PAH emission is predicted to decline sharply when the UV-irradiated surface is reduced in size. In disks where the emitting surface is confined within 20 to 30 au, PAH features weaken significantly or disappear. The absence of PAH emission in 11 of the 12 XUE disks aligns with this prediction. XUE~10, the only exception, shows PAH emission which may indicate a somewhat less truncated or more flared geometry.

Although the XUE sample is small, it suggests that strong external irradiation environments such as NGC~6357 tend to host compact disks. This has implications for planet formation, as truncated disks may naturally lead to the formation of the compact planetary systems observed by Kepler \citep[e.g.][]{2011Natur.470...53L, 2018haex.bookE.114D}. However, disk truncation is not unique to irradiated regions. In Lupus, a region with relatively mild external radiation, up to 60 percent of the disk population is also compact \citep{2021A&A...651A..48M, 2025A&A...696A.232G}. Therefore, while external irradiation plays an important role, it is not the sole factor driving disk truncation.

\subsection{\H2O\ line luminosities}

The XUE sample displays a wide range of \H2O\ line luminosities and most of the disks show luminosities that are high but comparable to those found in non-irradiated disks in nearby star-forming regions \citep[see Figure~\ref{fig:H2Olum_vs_n13-25},][]{2020ApJ...903..124B, 2022AJ....163..174B, 2025A&A...694A.147G} even if the stellar masses and luminosities are higher than those of the central stars in nearby regions. 
This suggest that the relation between \H2O\ luminosity and $L_{\bigstar}$ (or $M_{\bigstar}$) flattens beyond $\sim2$~\msun. 
As discussed in Section~\ref{sec:truncation}, disk truncation reduces the warm disk surface layers and could explain why line luminosities are not strongly enhanced \citep[][Hernandez Arboleda, \textit{in prep}]{2023ApJ...958L..30R, 2025ApJ...985...72P}. 

Another possibility to explain the flattening of the \H2O\ line luminosity vs. $L_{\bigstar}$ (or  $M_{\bigstar}$) could be that in our sources, which span a higher mass range, the dust continuum emission is higher \citep[][]{2023SSRv..219....7B}, and therefore the \H2O\ lines do not seem as enhanced as expected.
However, other important parameter that determine the line luminosities is the accretion rate \citep[e.g.][]{2020ApJ...903..124B}, which is poorly constrained by current observations. 
JWST NIRSpec observations are needed to get an estimate of the accretion luminosity of these sources \citep[e.g.][]{2025A&A...698A.226R}.

\subsection{Lack of distinct chemistry due to external irradiation}
\label{sec:lack_of_distinct_chem}

JWST observations of irradiated disks around low-mass stars in the Orion Nebula Cluster (ONC) show clear signs of external irradiation, such as OH produced by photo-dissociation of water, CH$^+$ and \CH3+\ produced by photochemistry, as well as C~${\rm I}$ fluorescence and molecular H$_2$ \citep[][]{2023Natur.621...56B, 2023A&A...671A..41Z, 2024A&A...692A.137A, 2024A&A...689L...4G, 2025A&A...696A..99Z}. 
\CH3+\ has also been detected in the nearby disk TW Hya. By means of thermochemical models, \citet{2024PASP..136e4302H} conclude that the \CH3+\ emission must originate from the transition regime from optically thin to thick at the gap edge of a radially extended disk. 

Surprisingly, none of the XUE sources, even if exposed to similar levels of external irradiation as the proplyds in the ONC, show any of these external irradiation indicators. 
As discussed in Section~\ref{sec:truncation}, the XUE disks could be truncated. 
Additionally, the fact that we do not see signs of photo-evaporation in the spectra could indicate that these disks have already mostly evaporated;  
Given the mass-loss rates measured for proplyds in the ONC \citep[e.g.][]{1993ApJ...410..696O, 1998ApJ...499..758J,1999AJ....118.2350H,2016MNRAS.457.3593F,2018MNRAS.481..452H, 2023MNRAS.526.4315H}, it is expected that the evaporation phase is short lived \citep[see e.g.][]{2019MNRAS.490.5478W}, and therefore it is possible that the XUE sources are already at a post-proplyd state. 
The lack of an outer disk would imply that there is no volume of gas which is dense enough to produce \CH3+\ emission even in the presence of gaps like in the case of TW~Hya, explaining the lack of this molecule in our observations \citep[see Fig.~5 in][]{2025ApJ...985...72P}. 

If the XUE disks were at a post-evaporation stage, it is possible that by putting together observations of proplyds in the ONC and the XUE sample we are looking at an evolutionary sequence for externally irradiated disks. However, it is important to note that the central stars of the XUE sample have higher masses than those of the ONC disks observed with JWST. The latter could influence the physical conditions of the inner disks, which could also explain the differences in the observed signatures.

\section{Conclusions}\label{sec:conclusion}

We present the Extreme UV Environments (XUE) sample of 12 IMTTs located in the three sub-clusters of the giant $\sim1$~Myr old H~${\rm{II}}$ region NGC~6357. This is the first sample spanning this mass range ($1-4$~\msun) that has been observed with MIRI-MRS on board JWST and provides an opportunity to study the inner disk physical and chemical properties with the unprecedented sensitivity of JWST.
In the following, we summarize the main findings from this paper. 

\begin{itemize}
    
    \item Our observations of externally irradiated IMTTs show an inner disk chemical inventory similar to that of non-irradiated T~Tauri disks. To better understand the effect of the environment on the XUE disks, observations of non-irradiated IMTTs with similar sensitivity and spectral resolution are needed for comparison.

    \item The spectral indices of the XUE sample are similar to those of their nearby counterparts. All XUE sources show silicate 10~\micron\ emission, with generally lower F${11.3}$/F${9.8}$ ratios at a given F$_{peak}$ compared to nearby T~Tauri and Herbig stars. Further investigation is needed to determine whether this difference is due to the dust properties in the inner disk.

    \item None of the XUE sources show cold ($\sim200$~K) water emission and only two of them show warm ($\sim400$~K) water. $60-90$\% of our Group~II disks and both of the Group~I flaring disks show emission from hot ($\sim900$~K) water in their inner disks, showing that rocky planets can form in the presence of water, even in these extreme environments. 
    
    \item Unlike the proplyds in Orion, the XUE disks do not exhibit distinct chemistry compared to non-irradiated disks, suggesting that these disks might be truncated. This would reduce the emission area of molecules such as \H2O, HCN, and \C2H2, leading to line fluxes similar to those observed in nearby disks. This could imply that the XUE disks are in a later stage of evolution, where the outer disk has been fully photoevaporated.
    
\end{itemize}

Our observations show the potential of JWST to study PPDs in distant regions ($>1.5$~kpc). This opens a new window to study disks around intermediate and high-mass stars, most of whose nearby counterparts will saturate with JWST. 
They also reiterate the need to observe non-irradiated disks around stars of similar stellar masses as the XUE sample in order to have a comparison sample and determine the effect of external irradiation in the inner disk of PPDs.  

\begin{acknowledgements}

We thank the anonymous referee for their constructive comments that helped improve this manuscript. 
M.C.R-T. acknowledges support by the German Aerospace Center (DLR) and the Federal Ministry for Economic Affairs and Energy (BMWi) through program 50OR2314 ‘Physics and Chemistry of Planet-forming disks in extreme environments’.
A.B. and J.F. acknowledge support from the Swedish National Space Agency (2022-00154). 
Support for B.P-R. within program \#1759 was provided by NASA through a grant from the Space Telescope Science Institute, which is operated by the Association of Universities for Research in Astronomy, Inc., under NASA contract NAS 5–26555.
T.J.H. acknowledges UKRI guaranteed funding for a Horizon Europe ERC consolidator grant (EP/Y024710/1) and a Royal Society Dorothy Hodgkin Fellowship. 
The work of T.P. was partly supported by the  Excellence Cluster ORIGINS which is funded by the Deutsche Forschungsgemeinschaft (DFG, German Research Foundation) under Germany's Excellence Strategy - EXC-2094 - 390783311
V.R. acknowledges the support of the European Union’s Horizon 2020 research and innovation program and the European Research Council via the ERC Synergy Grant “ECOGAL” (project ID 855130).
E.S. is supported by the international Gemini Observatory, a program of NSF NOIRLab, which is managed by the Association of Universities for Research in Astronomy (AURA) under a cooperative agreement with the U.S. National Science Foundation, on behalf of the Gemini partnership of Argentina, Brazil, Canada, Chile, the Republic of Korea, and the United States of America. 

\noindent This work is based [in part] on observations made with the NASA/ESA/CSA James Webb Space Telescope. The data were obtained from the Mikulski Archive for Space Telescopes at the Space Telescope Science Institute, which is operated by the Association of Universities for Research in Astronomy, Inc., under NASA contract NAS 5-03127 for JWST. These observations are associated with program \#1759. 
This publication makes use of VOSA, developed under the Spanish Virtual Observatory (\href{https://svo.cab.inta-csic.es}{svo.cab.inta-csic.es}) project funded by MCIN/AEI/10.13039/501100011033/ through grant PID2020-112949GB-I00. 
This research has made use of the SIMBAD database,
operated at CDS, Strasbourg, France \citep{2000A&AS..143....9W}.
This research used the SpExoDisks Database at \href{https://spexodisks.com/}{spexodisks.com} \citep[][]{2024PASP..136k3002W}.

\noindent \textit{Software:} Matplotlib \citep{2007CSE.....9...90H}, NumPy \citep{2011CSE....13b..22V}, SciPy \citep{2020zndo...4406806V}, Astropy \citep{Astropy:2013, astropy:2018, astropy:2022}, dust\_extinction \citep[][]{2024JOSS....9.7023G}.

\end{acknowledgements}

\bibliographystyle{aa}
\bibliography{references}

\begin{thebibliography}{153}
\expandafter\ifx\csname natexlab\endcsname\relax\def\natexlab#1{#1}\fi

\bibitem[{{Acke} {et~al.}(2010){Acke}, {Bouwman}, {Juh{\'a}sz}, {Henning}, {van
  den Ancker}, {Meeus}, {Tielens}, \& {Waters}}]{2010ApJ...718..558A}
{Acke}, B., {Bouwman}, J., {Juh{\'a}sz}, A., {et~al.} 2010, \apj, 718, 558

\bibitem[{{Allen} {et~al.}(2025){Allen}, {Anania}, {Andersen}, {Aru},
  {Ballabio}, {Ballering}, {Beccari}, {Bern{\'e}}, {Bik}, {Boyden}, {Coleman},
  {D{\'\i}az-Berrios}, {Eatson}, {Frediani}, {Forbrich}, {Gkimisi},
  {Goicoechea}, {Gupta}, {Guarcello}, {Haworth}, {Henney}, {Isella}, {Itrich},
  {Keyte}, {Kim}, {Kuhn}, {Le Petit}, {Luo}, {Manara}, {Mauco}, {Meshaka},
  {Millstone}, {Owen}, {Paine}, {Parker}, {Peake}, {Peatt}, {Pinilla}, {Qiao},
  {Ram{\'\i}rez-Tannus}, {Ramsay}, {Reiter}, {Rogers}, {Rosotti}, {Schroetter},
  {Sellek}, {Testi}, {van Terwisga}, {Vicente}, {Walsh}, {Winter}, {Wright}, \&
  {Zeidler}}]{2025OJAp....8E..54A}
{Allen}, M., {Anania}, R., {Andersen}, M., {et~al.} 2025, The Open Journal of
  Astrophysics, 8, 54

\bibitem[{{Ansdell} {et~al.}(2017){Ansdell}, {Williams}, {Manara}, {Miotello},
  {Facchini}, {van der Marel}, {Testi}, \& {van
  Dishoeck}}]{2017AJ....153..240A}
{Ansdell}, M., {Williams}, J.~P., {Manara}, C.~F., {et~al.} 2017, \aj, 153, 240

\bibitem[{{Antonellini} {et~al.}(2016){Antonellini}, {Kamp}, {Lahuis},
  {Woitke}, {Thi}, {Meijerink}, {Aresu}, {Spaans}, {G{\"u}del}, \&
  {Liebhart}}]{2016A&A...585A..61A}
{Antonellini}, S., {Kamp}, I., {Lahuis}, F., {et~al.} 2016, \aap, 585, A61

\bibitem[{{Antonellini} {et~al.}(2015){Antonellini}, {Kamp},
  {Riviere-Marichalar}, {Meijerink}, {Woitke}, {Thi}, {Spaans}, {Aresu}, \&
  {Lee}}]{2015A&A...582A.105A}
{Antonellini}, S., {Kamp}, I., {Riviere-Marichalar}, P., {et~al.} 2015, \aap,
  582, A105

\bibitem[{{Appenzeller} \& {Mundt}(1989)}]{1989A&ARv...1..291A}
{Appenzeller}, I. \& {Mundt}, R. 1989, \aapr, 1, 291

\bibitem[{{Arabhavi} {et~al.}(2024){Arabhavi}, {Kamp}, {Henning}, {van
  Dishoeck}, {Christiaens}, {Gasman}, {Perrin}, {G{\"u}del}, {Tabone},
  {Kanwar}, {Waters}, {Pascucci}, {Samland}, {Perotti}, {Bettoni}, {Grant},
  {Lagage}, {Ray}, {Vandenbussche}, {Absil}, {Argyriou}, {Barrado},
  {Boccaletti}, {Bouwman}, {Caratti o Garatti}, {Glauser}, {Lahuis}, {Mueller},
  {Olofsson}, {Pantin}, {Scheithauer}, {Morales-Calder{\'o}n}, {Franceschi},
  {Jang}, {Pawellek}, {Rodgers-Lee}, {Schreiber}, {Schwarz}, {Temmink},
  {Vlasblom}, {Wright}, {Colina}, \& {{\"O}stlin}}]{2024Sci...384.1086A}
{Arabhavi}, A.~M., {Kamp}, I., {Henning}, T., {et~al.} 2024, Science, 384, 1086

\bibitem[{{Aru} {et~al.}(2024){Aru}, {Mauc{\'o}}, {Manara}, {Haworth},
  {Ballering}, {Boyden}, {Campbell-White}, {Facchini}, {Rosotti}, {Winter},
  {Miotello}, {McLeod}, {Robberto}, {Petr-Gotzens}, {Ballabio}, {Vicente},
  {Ansdell}, \& {Cleeves}}]{2024A&A...692A.137A}
{Aru}, M.~L., {Mauc{\'o}}, K., {Manara}, C.~F., {et~al.} 2024, \aap, 692, A137

\bibitem[{{Astropy Collaboration} {et~al.}(2022){Astropy Collaboration},
  {Price-Whelan}, {Lim}, {Earl}, {Starkman}, {Bradley}, {Shupe}, {Patil},
  {Corrales}, {Brasseur}, {N{"o}the}, {Donath}, {Tollerud}, {Morris},
  {Ginsburg}, {Vaher}, {Weaver}, {Tocknell}, {Jamieson}, {van Kerkwijk},
  {Robitaille}, {Merry}, {Bachetti}, {G{"u}nther}, {Aldcroft},
  {Alvarado-Montes}, {Archibald}, {B{'o}di}, {Bapat}, {Barentsen}, {Baz{'a}n},
  {Biswas}, {Boquien}, {Burke}, {Cara}, {Cara}, {Conroy}, {Conseil}, {Craig},
  {Cross}, {Cruz}, {D'Eugenio}, {Dencheva}, {Devillepoix}, {Dietrich},
  {Eigenbrot}, {Erben}, {Ferreira}, {Foreman-Mackey}, {Fox}, {Freij}, {Garg},
  {Geda}, {Glattly}, {Gondhalekar}, {Gordon}, {Grant}, {Greenfield}, {Groener},
  {Guest}, {Gurovich}, {Handberg}, {Hart}, {Hatfield-Dodds}, {Homeier},
  {Hosseinzadeh}, {Jenness}, {Jones}, {Joseph}, {Kalmbach}, {Karamehmetoglu},
  {Ka{l}uszy{'n}ski}, {Kelley}, {Kern}, {Kerzendorf}, {Koch}, {Kulumani},
  {Lee}, {Ly}, {Ma}, {MacBride}, {Maljaars}, {Muna}, {Murphy}, {Norman},
  {O'Steen}, {Oman}, {Pacifici}, {Pascual}, {Pascual-Granado}, {Patil},
  {Perren}, {Pickering}, {Rastogi}, {Roulston}, {Ryan}, {Rykoff}, {Sabater},
  {Sakurikar}, {Salgado}, {Sanghi}, {Saunders}, {Savchenko}, {Schwardt},
  {Seifert-Eckert}, {Shih}, {Jain}, {Shukla}, {Sick}, {Simpson},
  {Singanamalla}, {Singer}, {Singhal}, {Sinha}, {Sip{H{o}}cz}, {Spitler},
  {Stansby}, {Streicher}, {{{S}}umak}, {Swinbank}, {Taranu}, {Tewary},
  {Tremblay}, {Val-Borro}, {Van Kooten}, {Vasovi{'c}}, {Verma}, {de Miranda
  Cardoso}, {Williams}, {Wilson}, {Winkel}, {Wood-Vasey}, {Xue}, {Yoachim},
  {Zhang}, {Zonca}, \& {Astropy Project Contributors}}]{astropy:2022}
{Astropy Collaboration}, {Price-Whelan}, A.~M., {Lim}, P.~L., {et~al.} 2022,
  apj, 935, 167

\bibitem[{{Astropy Collaboration} {et~al.}(2018){Astropy Collaboration},
  {Price-Whelan}, {Sip{\"o}cz}, {G{\"u}nther}, {Lim}, {Crawford}, {Conseil},
  {Shupe}, {Craig}, {Dencheva}, {Ginsburg}, {Vand erPlas}, {Bradley},
  {P{\'e}rez-Su{\'a}rez}, {de Val-Borro}, {Aldcroft}, {Cruz}, {Robitaille},
  {Tollerud}, {Ardelean}, {Babej}, {Bach}, {Bachetti}, {Bakanov}, {Bamford},
  {Barentsen}, {Barmby}, {Baumbach}, {Berry}, {Biscani}, {Boquien}, {Bostroem},
  {Bouma}, {Brammer}, {Bray}, {Breytenbach}, {Buddelmeijer}, {Burke},
  {Calderone}, {Cano Rodr{\'\i}guez}, {Cara}, {Cardoso}, {Cheedella}, {Copin},
  {Corrales}, {Crichton}, {D'Avella}, {Deil}, {Depagne}, {Dietrich}, {Donath},
  {Droettboom}, {Earl}, {Erben}, {Fabbro}, {Ferreira}, {Finethy}, {Fox},
  {Garrison}, {Gibbons}, {Goldstein}, {Gommers}, {Greco}, {Greenfield},
  {Groener}, {Grollier}, {Hagen}, {Hirst}, {Homeier}, {Horton}, {Hosseinzadeh},
  {Hu}, {Hunkeler}, {Ivezi{\'c}}, {Jain}, {Jenness}, {Kanarek}, {Kendrew},
  {Kern}, {Kerzendorf}, {Khvalko}, {King}, {Kirkby}, {Kulkarni}, {Kumar},
  {Lee}, {Lenz}, {Littlefair}, {Ma}, {Macleod}, {Mastropietro}, {McCully},
  {Montagnac}, {Morris}, {Mueller}, {Mumford}, {Muna}, {Murphy}, {Nelson},
  {Nguyen}, {Ninan}, {N{\"o}the}, {Ogaz}, {Oh}, {Parejko}, {Parley}, {Pascual},
  {Patil}, {Patil}, {Plunkett}, {Prochaska}, {Rastogi}, {Reddy Janga},
  {Sabater}, {Sakurikar}, {Seifert}, {Sherbert}, {Sherwood-Taylor}, {Shih},
  {Sick}, {Silbiger}, {Singanamalla}, {Singer}, {Sladen}, {Sooley},
  {Sornarajah}, {Streicher}, {Teuben}, {Thomas}, {Tremblay}, {Turner},
  {Terr{\'o}n}, {van Kerkwijk}, {de la Vega}, {Watkins}, {Weaver}, {Whitmore},
  {Woillez}, {Zabalza}, \& {Astropy Contributors}}]{astropy:2018}
{Astropy Collaboration}, {Price-Whelan}, A.~M., {Sip{\"o}cz}, B.~M., {et~al.}
  2018, \aj, 156, 123

\bibitem[{{Astropy Collaboration} {et~al.}(2013){Astropy Collaboration},
  {Robitaille}, {Tollerud}, {Greenfield}, {Droettboom}, {Bray}, {Aldcroft},
  {Davis}, {Ginsburg}, {Price-Whelan}, {Kerzendorf}, {Conley}, {Crighton},
  {Barbary}, {Muna}, {Ferguson}, {Grollier}, {Parikh}, {Nair}, {Unther},
  {Deil}, {Woillez}, {Conseil}, {Kramer}, {Turner}, {Singer}, {Fox}, {Weaver},
  {Zabalza}, {Edwards}, {Azalee Bostroem}, {Burke}, {Casey}, {Crawford},
  {Dencheva}, {Ely}, {Jenness}, {Labrie}, {Lim}, {Pierfederici}, {Pontzen},
  {Ptak}, {Refsdal}, {Servillat}, \& {Streicher}}]{Astropy:2013}
{Astropy Collaboration}, {Robitaille}, T.~P., {Tollerud}, E.~J., {et~al.} 2013,
  \aap, 558, A33

\bibitem[{{Banzatti} {et~al.}(2022){Banzatti}, {Abernathy}, {Brittain},
  {Bosman}, {Pontoppidan}, {Boogert}, {Jensen}, {Carr}, {Najita}, {Grant},
  {Sigler}, {Sanchez}, {Kern}, \& {Rayner}}]{2022AJ....163..174B}
{Banzatti}, A., {Abernathy}, K.~M., {Brittain}, S., {et~al.} 2022, \aj, 163,
  174

\bibitem[{{Banzatti} {et~al.}(2020){Banzatti}, {Pascucci}, {Bosman}, {Pinilla},
  {Salyk}, {Herczeg}, {Pontoppidan}, {Vazquez}, {Watkins}, {Krijt}, {Hendler},
  \& {Long}}]{2020ApJ...903..124B}
{Banzatti}, A., {Pascucci}, I., {Bosman}, A.~D., {et~al.} 2020, \apj, 903, 124

\bibitem[{{Banzatti} {et~al.}(2023){Banzatti}, {Pontoppidan}, {Carr},
  {Jellison}, {Pascucci}, {Najita}, {Mu{\~n}oz-Romero}, {{\"O}berg}, {Kalyaan},
  {Pinilla}, {Krijt}, {Long}, {Lambrechts}, {Rosotti}, {Herczeg}, {Salyk},
  {Zhang}, {Bergin}, {Ballering}, {Meyer}, {Bruderer}, \& {Jdiscs
  Collaboration}}]{2023ApJ...957L..22B}
{Banzatti}, A., {Pontoppidan}, K.~M., {Carr}, J.~S., {et~al.} 2023, \apjl, 957,
  L22

\bibitem[{{Banzatti} {et~al.}(2017){Banzatti}, {Pontoppidan}, {Salyk},
  {Herczeg}, {van Dishoeck}, \& {Blake}}]{2017ApJ...834..152B}
{Banzatti}, A., {Pontoppidan}, K.~M., {Salyk}, C., {et~al.} 2017, \apj, 834,
  152

\bibitem[{{Banzatti} {et~al.}(2025){Banzatti}, {Salyk}, {Pontoppidan}, {Carr},
  {Zhang}, {Arulanantham}, {Krijt}, {{\"O}berg}, {Cleeves}, {Najita},
  {Pascucci}, {Blake}, {Romero-Mirza}, {Bergin}, {Cieza}, {Pinilla}, {Long},
  {Mallaney}, {Xie}, {Waggoner}, {Kaeufer}, \& {Jdiscs
  Collaboration}}]{2025AJ....169..165B}
{Banzatti}, A., {Salyk}, C., {Pontoppidan}, K.~M., {et~al.} 2025, \aj, 169, 165

\bibitem[{{Benjamin} {et~al.}(2003){Benjamin}, {Churchwell}, {Babler}, {Bania},
  {Clemens}, {Cohen}, {Dickey}, {Indebetouw}, {Jackson}, {Kobulnicky},
  {Lazarian}, {Marston}, {Mathis}, {Meade}, {Seager}, {Stolovy}, {Watson},
  {Whitney}, {Wolff}, \& {Wolfire}}]{2003PASP..115..953B}
{Benjamin}, R.~A., {Churchwell}, E., {Babler}, B.~L., {et~al.} 2003, \pasp,
  115, 953

\bibitem[{{Bern{\'e}} {et~al.}(2023){Bern{\'e}}, {Martin-Drumel}, {Schroetter},
  {Goicoechea}, {Jacovella}, {Gans}, {Dartois}, {Coudert}, {Bergin}, {Alarcon},
  {Cami}, {Roueff}, {Black}, {Asvany}, {Habart}, {Peeters}, {Canin}, {Trahin},
  {Joblin}, {Schlemmer}, {Thorwirth}, {Cernicharo}, {Gerin}, {Tielens},
  {Zannese}, {Abergel}, {Bernard-Salas}, {Boersma}, {Bron}, {Chown},
  {Cuadrado}, {Dicken}, {Elyajouri}, {Fuente}, {Gordon}, {Issa}, {Kannavou},
  {Khan}, {Lacinbala}, {Languignon}, {Le Gal}, {Maragkoudakis}, {Meshaka},
  {Okada}, {Onaka}, {Pasquini}, {Pound}, {Robberto}, {R{\"o}llig}, {Schefter},
  {Schirmer}, {Sidhu}, {Tabone}, {Van De Putte}, {Vicente}, \&
  {Wolfire}}]{2023Natur.621...56B}
{Bern{\'e}}, O., {Martin-Drumel}, M.-A., {Schroetter}, I., {et~al.} 2023, \nat,
  621, 56

\bibitem[{{Bohigas} {et~al.}(2004){Bohigas}, {Tapia}, {Roth}, \&
  {Ruiz}}]{2004AJ....127.2826B}
{Bohigas}, J., {Tapia}, M., {Roth}, M., \& {Ruiz}, M.~T. 2004, \aj, 127, 2826

\bibitem[{{Bressan} {et~al.}(2012){Bressan}, {Marigo}, {Girardi}, {Salasnich},
  {Dal Cero}, {Rubele}, \& {Nanni}}]{2012MNRAS.427..127B}
{Bressan}, A., {Marigo}, P., {Girardi}, L., {et~al.} 2012, \mnras, 427, 127

\bibitem[{{Brittain} {et~al.}(2023){Brittain}, {Kamp}, {Meeus}, {Oudmaijer}, \&
  {Waters}}]{2023SSRv..219....7B}
{Brittain}, S.~D., {Kamp}, I., {Meeus}, G., {Oudmaijer}, R.~D., \& {Waters},
  L.~B.~F.~M. 2023, \ssr, 219, 7

\bibitem[{{Broos} {et~al.}(2013){Broos}, {Getman}, {Povich}, {Feigelson},
  {Townsley}, {Naylor}, {Kuhn}, {King}, \& {Busk}}]{2013ApJS..209...32B}
{Broos}, P.~S., {Getman}, K.~V., {Povich}, M.~S., {et~al.} 2013, \apjs, 209, 32

\bibitem[{{Brott} \& {Hauschildt}(2005)}]{2005ESASP.576..565B}
{Brott}, I. \& {Hauschildt}, P.~H. 2005, in ESA Special Publication, Vol. 576,
  The Three-Dimensional Universe with Gaia, ed. C.~{Turon}, K.~S. {O'Flaherty},
  \& M.~A.~C. {Perryman}, 565

\bibitem[{{Bushouse} {et~al.}(2024){Bushouse}, {Eisenhamer}, {Dencheva},
  {Davies}, {Greenfield}, {Morrison}, {Hodge}, {Simon}, {Grumm}, {Droettboom},
  {Slavich}, {Sosey}, {Pauly}, {Miller}, {Jedrzejewski}, {Hack}, {Davis},
  {Crawford}, {Law}, {Gordon}, {Regan}, {Cara}, {MacDonald}, {Bradley},
  {Shanahan}, {Jamieson}, {Teodoro}, {Williams}, \&
  {Pena-Guerrero}}]{2024zndo..10870758B}
{Bushouse}, H., {Eisenhamer}, J., {Dencheva}, N., {et~al.} 2024, {JWST
  Calibration Pipeline}

\bibitem[{{Calvet} {et~al.}(2004){Calvet}, {Muzerolle}, {Brice{\~n}o},
  {Hern{\'a}ndez}, {Hartmann}, {Saucedo}, \& {Gordon}}]{2004AJ....128.1294C}
{Calvet}, N., {Muzerolle}, J., {Brice{\~n}o}, C., {et~al.} 2004, \aj, 128, 1294

\bibitem[{{Carr} \& {Najita}(2008)}]{2008Sci...319.1504C}
{Carr}, J.~S. \& {Najita}, J.~R. 2008, Science, 319, 1504

\bibitem[{{Chiar} {et~al.}(2007){Chiar}, {Ennico}, {Pendleton}, {Boogert},
  {Greene}, {Knez}, {Lada}, {Roellig}, {Tielens}, {Werner}, \&
  {Whittet}}]{2007ApJ...666L..73C}
{Chiar}, J.~E., {Ennico}, K., {Pendleton}, Y.~J., {et~al.} 2007, \apjl, 666,
  L73

\bibitem[{{Churchwell} {et~al.}(2007){Churchwell}, {Watson}, {Povich},
  {Taylor}, {Babler}, {Meade}, {Benjamin}, {Indebetouw}, \&
  {Whitney}}]{2007ApJ...670..428C}
{Churchwell}, E., {Watson}, D.~F., {Povich}, M.~S., {et~al.} 2007, \apj, 670,
  428

\bibitem[{{Dawson}(2018)}]{2018haex.bookE.114D}
{Dawson}, R.~I. 2018, in Handbook of Exoplanets, ed. H.~J. {Deeg} \& J.~A.
  {Belmonte}, 114

\bibitem[{{Decleir} {et~al.}(2022){Decleir}, {Gordon}, {Andrews}, {Clayton},
  {Cushing}, {Misselt}, {Pendleton}, {Rayner}, {Vacca}, \&
  {Whittet}}]{2022ApJ...930...15D}
{Decleir}, M., {Gordon}, K.~D., {Andrews}, J.~E., {et~al.} 2022, \apj, 930, 15

\bibitem[{{D{\'\i}az-Berr{\'\i}os} {et~al.}(2024){D{\'\i}az-Berr{\'\i}os},
  {Guzm{\'a}n}, {Walsh}, {{\"O}berg}, {Cleeves}, {Artur de la Villarmois}, \&
  {Carpenter}}]{2024ApJ...969..165D}
{D{\'\i}az-Berr{\'\i}os}, J.~K., {Guzm{\'a}n}, V.~V., {Walsh}, C., {et~al.}
  2024, \apj, 969, 165

\bibitem[{{Dotter}(2016)}]{2016ApJS..222....8D}
{Dotter}, A. 2016, \apjs, 222, 8

\bibitem[{{Facchini} {et~al.}(2016){Facchini}, {Clarke}, \&
  {Bisbas}}]{2016MNRAS.457.3593F}
{Facchini}, S., {Clarke}, C.~J., \& {Bisbas}, T.~G. 2016, \mnras, 457, 3593

\bibitem[{{Fang} {et~al.}(2012){Fang}, {van Boekel}, {King}, {Henning},
  {Bouwman}, {Doi}, {Okamoto}, {Roccatagliata}, \&
  {Sicilia-Aguilar}}]{2012A&A...539A.119F}
{Fang}, M., {van Boekel}, R., {King}, R.~R., {et~al.} 2012, \aap, 539, A119

\bibitem[{{Fatuzzo} \& {Adams}(2008)}]{2008ApJ...675.1361F}
{Fatuzzo}, M. \& {Adams}, F.~C. 2008, \apj, 675, 1361

\bibitem[{{Feigelson} {et~al.}(2013){Feigelson}, {Townsley}, {Broos}, {Busk},
  {Getman}, {King}, {Kuhn}, {Naylor}, {Povich}, {Baddeley}, {Bate},
  {Indebetouw}, {Luhman}, {McCaughrean}, {Pittard}, {Pudritz}, {Sills}, {Song},
  \& {Wadsley}}]{2013ApJS..209...26F}
{Feigelson}, E.~D., {Townsley}, L.~K., {Broos}, P.~S., {et~al.} 2013, \apjs,
  209, 26

\bibitem[{{Fitzpatrick} {et~al.}(2019){Fitzpatrick}, {Massa}, {Gordon},
  {Bohlin}, \& {Clayton}}]{2019ApJ...886..108F}
{Fitzpatrick}, E.~L., {Massa}, D., {Gordon}, K.~D., {Bohlin}, R., \& {Clayton},
  G.~C. 2019, \apj, 886, 108

\bibitem[{{Fouesneau} {et~al.}(2022){Fouesneau}, {Andrae}, {Dharmawardena},
  {Rybizki}, {Bailer-Jones}, \& {Demleitner}}]{2022A&A...662A.125F}
{Fouesneau}, M., {Andrae}, R., {Dharmawardena}, T., {et~al.} 2022, \aap, 662,
  A125

\bibitem[{{Frediani} {et~al.}(2025){Frediani}, {Bik}, {Ram{\'\i}rez-Tannus},
  {Waters}, {Getman}, {Feigelson}, {Portilla-Revelo}, {Tabone}, {Haworth},
  {Winter}, {Henning}, {Perotti}, {Brandeker}, {Chaparro}, {Cuartas-Restrepo},
  {Hern{\'a}ndez}, {Kuhn}, {Preibisch}, {Roccatagliata}, {van Terwisga}, \&
  {Zeidler}}]{2025arXiv250713921F}
{Frediani}, J., {Bik}, A., {Ram{\'\i}rez-Tannus}, M.~C., {et~al.} 2025, arXiv
  e-prints, arXiv:2507.13921

\bibitem[{{Garufi} {et~al.}(2017){Garufi}, {Meeus}, {Benisty}, {Quanz},
  {Banzatti}, {Kama}, {Canovas}, {Eiroa}, {Schmid}, {Stolker}, {Pohl},
  {Rigliaco}, {M{\'e}nard}, {Meyer}, {van Boekel}, \&
  {Dominik}}]{2017A&A...603A..21G}
{Garufi}, A., {Meeus}, G., {Benisty}, M., {et~al.} 2017, \aap, 603, A21

\bibitem[{{Gasman} {et~al.}(2025){Gasman}, {Temmink}, {van Dishoeck},
  {Kurtovic}, {Grant}, {Sellek}, {Tabone}, {Henning}, {Kamp}, {G{\"u}del},
  {Barrado}, {Caratti o Garatti}, {Glauser}, {Waters}, {Arabhavi}, {Jang},
  {Kanwar}, {Lienert}, {Perotti}, {Schwarz}, \&
  {Vlasblom}}]{2025A&A...694A.147G}
{Gasman}, D., {Temmink}, M., {van Dishoeck}, E.~F., {et~al.} 2025, \aap, 694,
  A147

\bibitem[{{Gasman} {et~al.}(2023){Gasman}, {van Dishoeck}, {Grant}, {Temmink},
  {Tabone}, {Henning}, {Kamp}, {G{\"u}del}, {Lagage}, {Perotti}, {Christiaens},
  {Samland}, {Arabhavi}, {Argyriou}, {Abergel}, {Absil}, {Barrado},
  {Boccaletti}, {Bouwman}, {Caratti o Garatti}, {Geers}, {Glauser},
  {Guadarrama}, {Jang}, {Kanwar}, {Lahuis}, {Morales-Calder{\'o}n}, {Mueller},
  {Nehm{\'e}}, {Olofsson}, {Pantin}, {Pawellek}, {Ray}, {Rodgers-Lee},
  {Scheithauer}, {Schreiber}, {Schwarz}, {Vandenbussche}, {Vlasblom}, {Waters},
  {Wright}, {Colina}, {Greve}, \& {{\"O}stlin}}]{2023A&A...679A.117G}
{Gasman}, D., {van Dishoeck}, E.~F., {Grant}, S.~L., {et~al.} 2023, \aap, 679,
  A117

\bibitem[{{Getman} {et~al.}(2022){Getman}, {Feigelson}, {Garmire}, {Broos},
  {Kuhn}, {Preibisch}, \& {Airapetian}}]{2022ApJ...935...43G}
{Getman}, K.~V., {Feigelson}, E.~D., {Garmire}, G.~P., {et~al.} 2022, \apj,
  935, 43

\bibitem[{{Getman} {et~al.}(2014{\natexlab{a}}){Getman}, {Feigelson}, \&
  {Kuhn}}]{2014ApJ...787..109G}
{Getman}, K.~V., {Feigelson}, E.~D., \& {Kuhn}, M.~A. 2014{\natexlab{a}}, \apj,
  787, 109

\bibitem[{{Getman} {et~al.}(2014{\natexlab{b}}){Getman}, {Feigelson}, {Kuhn},
  {Broos}, {Townsley}, {Naylor}, {Povich}, {Luhman}, \&
  {Garmire}}]{2014ApJ...787..108G}
{Getman}, K.~V., {Feigelson}, E.~D., {Kuhn}, M.~A., {et~al.}
  2014{\natexlab{b}}, \apj, 787, 108

\bibitem[{{Getman} {et~al.}(2005){Getman}, {Flaccomio}, {Broos}, {Grosso},
  {Tsujimoto}, {Townsley}, {Garmire}, {Kastner}, {Li}, {Harnden}, {Wolk},
  {Murray}, {Lada}, {Muench}, {McCaughrean}, {Meeus}, {Damiani}, {Micela},
  {Sciortino}, {Bally}, {Hillenbrand}, {Herbst}, {Preibisch}, \&
  {Feigelson}}]{2005ApJS..160..319G}
{Getman}, K.~V., {Flaccomio}, E., {Broos}, P.~S., {et~al.} 2005, \apjs, 160,
  319

\bibitem[{{Goicoechea} {et~al.}(2024){Goicoechea}, {Le Bourlot}, {Black},
  {Alarc{\'o}n}, {Bergin}, {Bern{\'e}}, {Bron}, {Canin}, {Chapillon}, {Chown},
  {Dartois}, {Gerin}, {Habart}, {Haworth}, {Joblin}, {Kannavou}, {Le Petit},
  {Onaka}, {Peeters}, {Pety}, {Roueff}, {Sidhu}, {Schroetter}, {Tabone},
  {Tielens}, {Trahin}, {Van De Putte}, {Vicente}, \&
  {Zannese}}]{2024A&A...689L...4G}
{Goicoechea}, J.~R., {Le Bourlot}, J., {Black}, J.~H., {et~al.} 2024, \aap,
  689, L4

\bibitem[{{Gordon}(2024)}]{2024JOSS....9.7023G}
{Gordon}, K. 2024, The Journal of Open Source Software, 9, 7023

\bibitem[{{Gordon} {et~al.}(2009){Gordon}, {Cartledge}, \&
  {Clayton}}]{2009ApJ...705.1320G}
{Gordon}, K.~D., {Cartledge}, S., \& {Clayton}, G.~C. 2009, \apj, 705, 1320

\bibitem[{{Gordon} {et~al.}(2023){Gordon}, {Clayton}, {Decleir}, {Fitzpatrick},
  {Massa}, {Misselt}, \& {Tollerud}}]{2023ApJ...950...86G}
{Gordon}, K.~D., {Clayton}, G.~C., {Decleir}, M., {et~al.} 2023, \apj, 950, 86

\bibitem[{{Gordon} {et~al.}(2021){Gordon}, {Misselt}, {Bouwman}, {Clayton},
  {Decleir}, {Hines}, {Pendleton}, {Rieke}, {Smith}, \&
  {Whittet}}]{2021ApJ...916...33G}
{Gordon}, K.~D., {Misselt}, K.~A., {Bouwman}, J., {et~al.} 2021, \apj, 916, 33

\bibitem[{{Grant} {et~al.}(2024){Grant}, {Kurtovic}, {van Dishoeck}, {Henning},
  {Kamp}, {Nowacki}, {Perraut}, {Banzatti}, {Temmink}, {Christiaens},
  {Samland}, {Gasman}, {Tabone}, {G{\"u}del}, {Lagage}, {Arabhavi}, {Barrado},
  {Caratti o Garatti}, {Glauser}, {Jang}, {Kanwar}, {Lahuis},
  {Morales-Calder{\'o}n}, {Olofsson}, {Perotti}, {Schwarz}, {Vlasblom}, {Garcia
  Lopez}, \& {Long}}]{2024A&A...689A..85G}
{Grant}, S.~L., {Kurtovic}, N.~T., {van Dishoeck}, E.~F., {et~al.} 2024, \aap,
  689, A85

\bibitem[{{Grant} {et~al.}(2023){Grant}, {van Dishoeck}, {Tabone}, {Gasman},
  {Henning}, {Kamp}, {G{\"u}del}, {Lagage}, {Bettoni}, {Perotti},
  {Christiaens}, {Samland}, {Arabhavi}, {Argyriou}, {Abergel}, {Absil},
  {Barrado}, {Boccaletti}, {Bouwman}, {o Garatti}, {Geers}, {Glauser},
  {Guadarrama}, {Jang}, {Kanwar}, {Lahuis}, {Morales-Calder{\'o}n}, {Mueller},
  {Nehm{\'e}}, {Olofsson}, {Pantin}, {Pawellek}, {Ray}, {Rodgers-Lee},
  {Scheithauer}, {Schreiber}, {Schwarz}, {Temmink}, {Vandenbussche},
  {Vlasblom}, {Waters}, {Wright}, {Colina}, {Greve}, {Justannont}, \&
  {{\"O}stlin}}]{2023ApJ...947L...6G}
{Grant}, S.~L., {van Dishoeck}, E.~F., {Tabone}, B., {et~al.} 2023, \apjl, 947,
  L6

\bibitem[{{Guarcello} {et~al.}(2023){Guarcello}, {Drake}, {Wright},
  {Albacete-Colombo}, {Clarke}, {Ercolano}, {Flaccomio}, {Kashyap}, {Micela},
  {Naylor}, {Schneider}, {Sciortino}, \& {Vink}}]{2023ApJS..269...13G}
{Guarcello}, M.~G., {Drake}, J.~J., {Wright}, N.~J., {et~al.} 2023, \apjs, 269,
  13

\bibitem[{{G{\"u}del} {et~al.}(2008){G{\"u}del}, {Briggs}, {Montmerle},
  {Audard}, {Rebull}, \& {Skinner}}]{2008Sci...319..309G}
{G{\"u}del}, M., {Briggs}, K.~R., {Montmerle}, T., {et~al.} 2008, Science, 319,
  309

\bibitem[{{Guerra-Alvarado} {et~al.}(2025){Guerra-Alvarado}, {van der Marel},
  {Williams}, {Pinilla}, {Mulders}, {Lambrechts}, \&
  {Sanchez}}]{2025A&A...696A.232G}
{Guerra-Alvarado}, O.~M., {van der Marel}, N., {Williams}, J.~P., {et~al.}
  2025, \aap, 696, A232

\bibitem[{{Gvaramadze} {et~al.}(2011){Gvaramadze}, {Kniazev}, {Kroupa}, \&
  {Oh}}]{2011A&A...535A..29G}
{Gvaramadze}, V.~V., {Kniazev}, A.~Y., {Kroupa}, P., \& {Oh}, S. 2011, \aap,
  535, A29

\bibitem[{{Haworth} {et~al.}(2018){Haworth}, {Clarke}, {Rahman}, {Winter}, \&
  {Facchini}}]{2018MNRAS.481..452H}
{Haworth}, T.~J., {Clarke}, C.~J., {Rahman}, W., {Winter}, A.~J., \&
  {Facchini}, S. 2018, \mnras, 481, 452

\bibitem[{{Haworth} {et~al.}(2023){Haworth}, {Coleman}, {Qiao}, {Sellek}, \&
  {Askari}}]{2023MNRAS.526.4315H}
{Haworth}, T.~J., {Coleman}, G. A.~L., {Qiao}, L., {Sellek}, A.~D., \&
  {Askari}, K. 2023, \mnras, 526, 4315

\bibitem[{{Henney} \& {O'Dell}(1999)}]{1999AJ....118.2350H}
{Henney}, W.~J. \& {O'Dell}, C.~R. 1999, \aj, 118, 2350

\bibitem[{{Henning} {et~al.}(2024){Henning}, {Kamp}, {Samland}, {Arabhavi},
  {Kanwar}, {van Dishoeck}, {G{\"u}del}, {Lagage}, {Waelkens}, {Abergel},
  {Absil}, {Barrado}, {Boccaletti}, {Bouwman}, {Caratti o Garatti}, {Geers},
  {Glauser}, {Lahuis}, {Mueller}, {Nehm{\'e}}, {Olofsson}, {Pantin}, {Ray},
  {Scheithauer}, {Vandenbussche}, {Waters}, {Wright}, {Argyriou},
  {Christiaens}, {Franceschi}, {Gasman}, {Grant}, {Guadarrama}, {Jang},
  {Morales-Calder{\'o}n}, {Pawellek}, {Perotti}, {Rodgers-Lee}, {Schreiber},
  {Schwarz}, {Tabone}, {Temmink}, {Vlasblom}, {Colina}, {Greve}, \&
  {{\"O}stlin}}]{2024PASP..136e4302H}
{Henning}, T., {Kamp}, I., {Samland}, M., {et~al.} 2024, \pasp, 136, 054302

\bibitem[{{Herbig}(1960)}]{1960ApJS....4..337H}
{Herbig}, G.~H. 1960, \apjs, 4, 337

\bibitem[{{Hunter}(2007)}]{2007CSE.....9...90H}
{Hunter}, J.~D. 2007, Computing in Science and Engineering, 9, 90

\bibitem[{{Husser} {et~al.}(2013){Husser}, {Wende-von Berg}, {Dreizler},
  {Homeier}, {Reiners}, {Barman}, \& {Hauschildt}}]{2013A&A...553A...6H}
{Husser}, T.~O., {Wende-von Berg}, S., {Dreizler}, S., {et~al.} 2013, \aap,
  553, A6

\bibitem[{{Johnson} {et~al.}(2010){Johnson}, {Aller}, {Howard}, \&
  {Crepp}}]{2010PASP..122..905J}
{Johnson}, J.~A., {Aller}, K.~M., {Howard}, A.~W., \& {Crepp}, J.~R. 2010,
  \pasp, 122, 905

\bibitem[{{Johnstone} {et~al.}(1998){Johnstone}, {Hollenbach}, \&
  {Bally}}]{1998ApJ...499..758J}
{Johnstone}, D., {Hollenbach}, D., \& {Bally}, J. 1998, \apj, 499, 758

\bibitem[{{Joy}(1945)}]{1945ApJ...102..168J}
{Joy}, A.~H. 1945, \apj, 102, 168

\bibitem[{{Juh{\'a}sz} {et~al.}(2010){Juh{\'a}sz}, {Bouwman}, {Henning},
  {Acke}, {van den Ancker}, {Meeus}, {Dominik}, {Min}, {Tielens}, \&
  {Waters}}]{2010ApJ...721..431J}
{Juh{\'a}sz}, A., {Bouwman}, J., {Henning}, T., {et~al.} 2010, \apj, 721, 431

\bibitem[{{Kessler-Silacci} {et~al.}(2006){Kessler-Silacci}, {Augereau},
  {Dullemond}, {Geers}, {Lahuis}, {Evans}, {van Dishoeck}, {Blake}, {Boogert},
  {Brown}, {J{\o}rgensen}, {Knez}, \& {Pontoppidan}}]{2006ApJ...639..275K}
{Kessler-Silacci}, J., {Augereau}, J.-C., {Dullemond}, C.~P., {et~al.} 2006,
  \apj, 639, 275

\bibitem[{{Kim} {et~al.}(2016){Kim}, {Watson}, {Manoj}, {Forrest}, {Furlan},
  {Najita}, {Sargent}, {Hern{\'a}ndez}, {Calvet}, {Adame}, {Espaillat},
  {Megeath}, {Muzerolle}, \& {McClure}}]{2016ApJS..226....8K}
{Kim}, K.~H., {Watson}, D.~M., {Manoj}, P., {et~al.} 2016, \apjs, 226, 8

\bibitem[{{King} {et~al.}(2013){King}, {Naylor}, {Broos}, {Getman}, \&
  {Feigelson}}]{2013ApJS..209...28K}
{King}, R.~R., {Naylor}, T., {Broos}, P.~S., {Getman}, K.~V., \& {Feigelson},
  E.~D. 2013, \apjs, 209, 28

\bibitem[{{K{\'o}sp{\'a}l} {et~al.}(2023){K{\'o}sp{\'a}l}, {{\'A}brah{\'a}m},
  {Diehl}, {Banzatti}, {Bouwman}, {Chen}, {Cruz-S{\'a}enz de Miera}, {Green},
  {Henning}, \& {Rab}}]{2023ApJ...945L...7K}
{K{\'o}sp{\'a}l}, {\'A}., {{\'A}brah{\'a}m}, P., {Diehl}, L., {et~al.} 2023,
  \apjl, 945, L7

\bibitem[{{Kuhn} {et~al.}(2013){Kuhn}, {Povich}, {Luhman}, {Getman}, {Busk}, \&
  {Feigelson}}]{2013ApJS..209...29K}
{Kuhn}, M.~A., {Povich}, M.~S., {Luhman}, K.~L., {et~al.} 2013, \apjs, 209, 29

\bibitem[{{Kurucz}(1993)}]{1993KurCD..13.....K}
{Kurucz}, R. 1993, Robert Kurucz CD-ROM, 13

\bibitem[{{Lagage} {et~al.}(2006){Lagage}, {Doucet}, {Pantin}, {Habart},
  {Duch{\^e}ne}, {M{\'e}nard}, {Pinte}, {Charnoz}, \&
  {Pel}}]{2006Sci...314..621L}
{Lagage}, P.-O., {Doucet}, C., {Pantin}, E., {et~al.} 2006, Science, 314, 621

\bibitem[{{Lawrence} {et~al.}(2007){Lawrence}, {Warren}, {Almaini}, {Edge},
  {Hambly}, {Jameson}, {Lucas}, {Casali}, {Adamson}, {Dye}, {Emerson},
  {Foucaud}, {Hewett}, {Hirst}, {Hodgkin}, {Irwin}, {Lodieu}, {McMahon},
  {Simpson}, {Smail}, {Mortlock}, \& {Folger}}]{2007MNRAS.379.1599L}
{Lawrence}, A., {Warren}, S.~J., {Almaini}, O., {et~al.} 2007, \mnras, 379,
  1599

\bibitem[{{Lindegren} {et~al.}(2021){Lindegren}, {Bastian}, {Biermann},
  {Bombrun}, {de Torres}, {Gerlach}, {Geyer}, {Hern{\'a}ndez}, {Hilger},
  {Hobbs}, {Klioner}, {Lammers}, {McMillan}, {Ramos-Lerate},
  {Steidelm{\"u}ller}, {Stephenson}, \& {van Leeuwen}}]{2021A&A...649A...4L}
{Lindegren}, L., {Bastian}, U., {Biermann}, M., {et~al.} 2021, \aap, 649, A4

\bibitem[{{Lindegren} {et~al.}(2018){Lindegren}, {Hern{\'a}ndez}, {Bombrun},
  {Klioner}, {Bastian}, {Ramos-Lerate}, {de Torres}, {Steidelm{\"u}ller},
  {Stephenson}, {Hobbs}, {Lammers}, {Biermann}, {Geyer}, {Hilger}, {Michalik},
  {Stampa}, {McMillan}, {Casta{\~n}eda}, {Clotet}, {Comoretto}, {Davidson},
  {Fabricius}, {Gracia}, {Hambly}, {Hutton}, {Mora}, {Portell}, {van Leeuwen},
  {Abbas}, {Abreu}, {Altmann}, {Andrei}, {Anglada}, {Balaguer-N{\'u}{\~n}ez},
  {Barache}, {Becciani}, {Bertone}, {Bianchi}, {Bouquillon}, {Bourda},
  {Br{\"u}semeister}, {Bucciarelli}, {Busonero}, {Buzzi}, {Cancelliere},
  {Carlucci}, {Charlot}, {Cheek}, {Crosta}, {Crowley}, {de Bruijne}, {de
  Felice}, {Drimmel}, {Esquej}, {Fienga}, {Fraile}, {Gai}, {Garralda},
  {Gonz{\'a}lez-Vidal}, {Guerra}, {Hauser}, {Hofmann}, {Holl}, {Jordan},
  {Lattanzi}, {Lenhardt}, {Liao}, {Licata}, {Lister}, {L{\"o}ffler},
  {Marchant}, {Martin-Fleitas}, {Messineo}, {Mignard}, {Morbidelli}, {Poggio},
  {Riva}, {Rowell}, {Salguero}, {Sarasso}, {Sciacca}, {Siddiqui}, {Smart},
  {Spagna}, {Steele}, {Taris}, {Torra}, {van Elteren}, {van Reeven}, \&
  {Vecchiato}}]{2018A&A...616A...2L}
{Lindegren}, L., {Hern{\'a}ndez}, J., {Bombrun}, A., {et~al.} 2018, \aap, 616,
  A2

\bibitem[{{Lissauer} {et~al.}(2011){Lissauer}, {Fabrycky}, {Ford}, {Borucki},
  {Fressin}, {Marcy}, {Orosz}, {Rowe}, {Torres}, {Welsh}, {Batalha}, {Bryson},
  {Buchhave}, {Caldwell}, {Carter}, {Charbonneau}, {Christiansen}, {Cochran},
  {Desert}, {Dunham}, {Fanelli}, {Fortney}, {Gautier}, {Geary}, {Gilliland},
  {Haas}, {Hall}, {Holman}, {Koch}, {Latham}, {Lopez}, {McCauliff}, {Miller},
  {Morehead}, {Quintana}, {Ragozzine}, {Sasselov}, {Short}, \&
  {Steffen}}]{2011Natur.470...53L}
{Lissauer}, J.~J., {Fabrycky}, D.~C., {Ford}, E.~B., {et~al.} 2011, \nat, 470,
  53

\bibitem[{{Lommen} {et~al.}(2010){Lommen}, {van Dishoeck}, {Wright},
  {Maddison}, {Min}, {Wilner}, {Salter}, {van Langevelde}, {Bourke}, {van der
  Burg}, \& {Blake}}]{2010A&A...515A..77L}
{Lommen}, D.~J.~P., {van Dishoeck}, E.~F., {Wright}, C.~M., {et~al.} 2010,
  \aap, 515, A77

\bibitem[{{Lucas} {et~al.}(2008){Lucas}, {Hoare}, {Longmore}, {Schr{\"o}der},
  {Davis}, {Adamson}, {Bandyopadhyay}, {de Grijs}, {Smith}, {Gosling},
  {Mitchison}, {G{\'a}sp{\'a}r}, {Coe}, {Tamura}, {Parker}, {Irwin}, {Hambly},
  {Bryant}, {Collins}, {Cross}, {Evans}, {Gonzalez-Solares}, {Hodgkin},
  {Lewis}, {Read}, {Riello}, {Sutorius}, {Lawrence}, {Drew}, {Dye}, \&
  {Thompson}}]{2008MNRAS.391..136L}
{Lucas}, P.~W., {Hoare}, M.~G., {Longmore}, A., {et~al.} 2008, \mnras, 391, 136

\bibitem[{{Maaskant} {et~al.}(2014){Maaskant}, {Min}, {Waters}, \&
  {Tielens}}]{2014A&A...563A..78M}
{Maaskant}, K.~M., {Min}, M., {Waters}, L.~B.~F.~M., \& {Tielens}, A.~G.~G.~M.
  2014, \aap, 563, A78

\bibitem[{{Ma{\'{\i}}z Apell{\'a}niz} {et~al.}(2007){Ma{\'{\i}}z
  Apell{\'a}niz}, {Walborn}, {Morrell}, {Niemela}, \&
  {Nelan}}]{2007ApJ...660.1480M}
{Ma{\'{\i}}z Apell{\'a}niz}, J., {Walborn}, N.~R., {Morrell}, N.~I., {Niemela},
  V.~S., \& {Nelan}, E.~P. 2007, \apj, 660, 1480

\bibitem[{{Mandell} {et~al.}(2012){Mandell}, {Bast}, {van Dishoeck}, {Blake},
  {Salyk}, {Mumma}, \& {Villanueva}}]{2012ApJ...747...92M}
{Mandell}, A.~M., {Bast}, J., {van Dishoeck}, E.~F., {et~al.} 2012, \apj, 747,
  92

\bibitem[{{Massi} {et~al.}(2015){Massi}, {Giannetti}, {Di Carlo}, {Brand},
  {Beltr{\'a}n}, \& {Marconi}}]{2015A&A...573A..95M}
{Massi}, F., {Giannetti}, A., {Di Carlo}, E., {et~al.} 2015, \aap, 573, A95

\bibitem[{{Meeus} {et~al.}(2001){Meeus}, {Waters}, {Bouwman}, {van den Ancker},
  {Waelkens}, \& {Malfait}}]{2001A&A...365..476M}
{Meeus}, G., {Waters}, L.~B.~F.~M., {Bouwman}, J., {et~al.} 2001, \aap, 365,
  476

\bibitem[{{Miotello} {et~al.}(2021){Miotello}, {Rosotti}, {Ansdell},
  {Facchini}, {Manara}, {Williams}, \& {Bruderer}}]{2021A&A...651A..48M}
{Miotello}, A., {Rosotti}, G., {Ansdell}, M., {et~al.} 2021, \aap, 651, A48

\bibitem[{{Najita} {et~al.}(2013){Najita}, {Carr}, {Pontoppidan}, {Salyk}, {van
  Dishoeck}, \& {Blake}}]{2013ApJ...766..134N}
{Najita}, J.~R., {Carr}, J.~S., {Pontoppidan}, K.~M., {et~al.} 2013, \apj, 766,
  134

\bibitem[{{Ndugu} {et~al.}(2024){Ndugu}, {Bitsch}, \&
  {Lienert}}]{2024A&A...691A..32N}
{Ndugu}, N., {Bitsch}, B., \& {Lienert}, J.~L. 2024, \aap, 691, A32

\bibitem[{{Nu{\~n}ez} {et~al.}(2021){Nu{\~n}ez}, {Povich}, {Binder},
  {Townsley}, \& {Broos}}]{2021AJ....162..153N}
{Nu{\~n}ez}, E.~H., {Povich}, M.~S., {Binder}, B.~A., {Townsley}, L.~K., \&
  {Broos}, P.~S. 2021, \aj, 162, 153

\bibitem[{{O'dell} {et~al.}(1993){O'dell}, {Wen}, \&
  {Hu}}]{1993ApJ...410..696O}
{O'dell}, C.~R., {Wen}, Z., \& {Hu}, X. 1993, \apj, 410, 696

\bibitem[{{Paine} {et~al.}(2025){Paine}, {Haworth}, \&
  {Nelson}}]{2025MNRAS.539.1414P}
{Paine}, S., {Haworth}, T.~J., \& {Nelson}, R.~P. 2025, \mnras, 539, 1414

\bibitem[{{Parravano} {et~al.}(2003){Parravano}, {Hollenbach}, \&
  {McKee}}]{2003ApJ...584..797P}
{Parravano}, A., {Hollenbach}, D.~J., \& {McKee}, C.~F. 2003, \apj, 584, 797

\bibitem[{{Pascucci} {et~al.}(2009){Pascucci}, {Apai}, {Luhman}, {Henning},
  {Bouwman}, {Meyer}, {Lahuis}, \& {Natta}}]{2009ApJ...696..143P}
{Pascucci}, I., {Apai}, D., {Luhman}, K., {et~al.} 2009, \apj, 696, 143

\bibitem[{{Pascucci} {et~al.}(2013){Pascucci}, {Herczeg}, {Carr}, \&
  {Bruderer}}]{2013ApJ...779..178P}
{Pascucci}, I., {Herczeg}, G., {Carr}, J.~S., \& {Bruderer}, S. 2013, \apj,
  779, 178

\bibitem[{{Perotti} {et~al.}(2023){Perotti}, {Christiaens}, {Henning},
  {Tabone}, {Waters}, {Kamp}, {Olofsson}, {Grant}, {Gasman}, {Bouwman},
  {Samland}, {Franceschi}, {van Dishoeck}, {Schwarz}, {G{\"u}del}, {Lagage},
  {Ray}, {Vandenbussche}, {Abergel}, {Absil}, {Arabhavi}, {Argyriou},
  {Barrado}, {Boccaletti}, {Caratti o Garatti}, {Geers}, {Glauser},
  {Justannont}, {Lahuis}, {Mueller}, {Nehm{\'e}}, {Pantin}, {Scheithauer},
  {Waelkens}, {Guadarrama}, {Jang}, {Kanwar}, {Morales-Calder{\'o}n},
  {Pawellek}, {Rodgers-Lee}, {Schreiber}, {Colina}, {Greve}, {{\"O}stlin}, \&
  {Wright}}]{2023Natur.620..516P}
{Perotti}, G., {Christiaens}, V., {Henning}, T., {et~al.} 2023, \nat, 620, 516

\bibitem[{{Pontoppidan} {et~al.}(2024){Pontoppidan}, {Salyk}, {Banzatti},
  {Zhang}, {Pascucci}, {{\"O}berg}, {Long}, {Romero-Mirza}, {Carr}, {Najita},
  {Blake}, {Arulanantham}, {Andrews}, {Ballering}, {Bergin}, {Calahan}, {Cobb},
  {Colmenares}, {Dickson-Vandervelde}, {Dignan}, {Green}, {Heretz}, {Herczeg},
  {Kalyaan}, {Krijt}, {Pauly}, {Pinilla}, {Trapman}, \&
  {Xie}}]{2024ApJ...963..158P}
{Pontoppidan}, K.~M., {Salyk}, C., {Banzatti}, A., {et~al.} 2024, \apj, 963,
  158

\bibitem[{{Pontoppidan} {et~al.}(2010){Pontoppidan}, {Salyk}, {Blake},
  {Meijerink}, {Carr}, \& {Najita}}]{2010ApJ...720..887P}
{Pontoppidan}, K.~M., {Salyk}, C., {Blake}, G.~A., {et~al.} 2010, \apj, 720,
  887

\bibitem[{{Portilla-Revelo} {et~al.}(2025){Portilla-Revelo}, {Getman},
  {Ram{\'\i}rez-Tannus}, {Haworth}, {Waters}, {Bik}, {Feigelson}, {Kamp}, {van
  Terwisga}, {Frediani}, {Henning}, {Winter}, {Roccatagliata}, {Preibisch},
  {Sabbi}, {Zeidler}, \& {Kuhn}}]{2025ApJ...985...72P}
{Portilla-Revelo}, B., {Getman}, K.~V., {Ram{\'\i}rez-Tannus}, M.~C., {et~al.}
  2025, \apj, 985, 72

\bibitem[{{Povich} {et~al.}(2017){Povich}, {Busk}, {Feigelson}, {Townsley}, \&
  {Kuhn}}]{2017ApJ...838...61P}
{Povich}, M.~S., {Busk}, H.~A., {Feigelson}, E.~D., {Townsley}, L.~K., \&
  {Kuhn}, M.~A. 2017, \apj, 838, 61

\bibitem[{{Povich} {et~al.}(2013){Povich}, {Kuhn}, {Getman}, {Busk},
  {Feigelson}, {Broos}, {Townsley}, {King}, \& {Naylor}}]{2013ApJS..209...31P}
{Povich}, M.~S., {Kuhn}, M.~A., {Getman}, K.~V., {et~al.} 2013, \apjs, 209, 31

\bibitem[{{Qiao} {et~al.}(2023){Qiao}, {Coleman}, \&
  {Haworth}}]{2023MNRAS.522.1939Q}
{Qiao}, L., {Coleman}, G. A.~L., \& {Haworth}, T.~J. 2023, \mnras, 522, 1939

\bibitem[{{Ramirez-Tannus} {et~al.}(2021){Ramirez-Tannus}, {Backs}, {Bik},
  {Bouwman}, {Brandner}, {Chevance}, {De Koter}, {Derkink}, {Feigelson},
  {Geen}, {Getman}, {Henning}, {Kamp}, {Kaper}, {Kruijssen}, {Kuhn},
  {Longmore}, {McLeod}, {Poorta}, {Povich}, {Preibisch}, {Roccatagliata},
  {Sabbi}, {Sana}, {Waters}, {Winter}, {Zari}, \& {van
  Terwisga}}]{2021jwst.prop.1759R}
{Ramirez-Tannus}, M.~C., {Backs}, F., {Bik}, A., {et~al.} 2021, {Physics and
  Chemistry of Planet-Forming Disks in Extreme Radiation Environments}, JWST
  Proposal. Cycle 1, ID. \#1759

\bibitem[{{Ram{\'\i}rez-Tannus} {et~al.}(2023){Ram{\'\i}rez-Tannus}, {Bik},
  {Cuijpers}, {Waters}, {G{\"o}ppl}, {Henning}, {Kamp}, {Preibisch}, {Getman},
  {Chaparro}, {Cuartas-Restrepo}, {de Koter}, {Feigelson}, {Grant}, {Haworth},
  {Hern{\'a}ndez}, {Kuhn}, {Perotti}, {Povich}, {Reiter}, {Roccatagliata},
  {Sabbi}, {Tabone}, {Winter}, {McLeod}, {van Boekel}, \& {van
  Terwisga}}]{2023ApJ...958L..30R}
{Ram{\'\i}rez-Tannus}, M.~C., {Bik}, A., {Cuijpers}, L., {et~al.} 2023, \apjl,
  958, L30

\bibitem[{{Ram{\'\i}rez-Tannus} {et~al.}(2020){Ram{\'\i}rez-Tannus}, {Poorta},
  {Bik}, {Kaper}, {de Koter}, {De Ridder}, {Beuther}, {Brandner}, {Davies},
  {Gennaro}, {Guo}, {Henning}, {Linz}, {Naylor}, {Pasquali},
  {Ram{\'\i}rez-Agudelo}, \& {Sana}}]{2020A&A...633A.155R}
{Ram{\'\i}rez-Tannus}, M.~C., {Poorta}, J., {Bik}, A., {et~al.} 2020, \aap,
  633, A155

\bibitem[{{Richert} {et~al.}(2015){Richert}, {Feigelson}, {Getman}, \&
  {Kuhn}}]{2015ApJ...811...10R}
{Richert}, A. J.~W., {Feigelson}, E.~D., {Getman}, K.~V., \& {Kuhn}, M.~A.
  2015, \apj, 811, 10

\bibitem[{{Richert} {et~al.}(2018){Richert}, {Getman}, {Feigelson}, {Kuhn},
  {Broos}, {Povich}, {Bate}, \& {Garmire}}]{2018MNRAS.477.5191R}
{Richert}, A.~J.~W., {Getman}, K.~V., {Feigelson}, E.~D., {et~al.} 2018,
  \mnras, 477, 5191

\bibitem[{{Rieke} {et~al.}(2015){Rieke}, {Wright}, {B{\"o}ker}, {Bouwman},
  {Colina}, {Glasse}, {Gordon}, {Greene}, {G{\"u}del}, {Henning}, {Justtanont},
  {Lagage}, {Meixner}, {N{\o}rgaard-Nielsen}, {Ray}, {Ressler}, {van Dishoeck},
  \& {Waelkens}}]{2015PASP..127..584R}
{Rieke}, G.~H., {Wright}, G.~S., {B{\"o}ker}, T., {et~al.} 2015, \pasp, 127,
  584

\bibitem[{{Rigby} {et~al.}(2023){Rigby}, {Perrin}, {McElwain}, {Kimble},
  {Friedman}, {Lallo}, {Doyon}, {Feinberg}, {Ferruit}, {Glasse}, {Rieke},
  {Rieke}, {Wright}, {Willott}, {Colon}, {Milam}, {Neff}, {Stark}, {Valenti},
  {Abell}, {Abney}, {Abul-Huda}, {Acton}, {Adams}, {Adler}, {Aguilar}, {Ahmed},
  {Albert}, {Alberts}, {Aldridge}, {Allen}, {Altenburg},
  {{\'A}lvarez-M{\'a}rquez}, {Alves de Oliveira}, {Andersen}, {Anderson},
  {Anderson}, {Argyriou}, {Armstrong}, {Arribas}, {Artigau}, {Arvai},
  {Atkinson}, {Bacon}, {Bair}, {Banks}, {Barrientes}, {Barringer}, {Bartosik},
  {Bast}, {Baudoz}, {Beatty}, {Bechtold}, {Beck}, {Bergeron}, {Bergkoetter},
  {Bhatawdekar}, {Birkmann}, {Blazek}, {Blome}, {Boccaletti}, {B{\"o}ker},
  {Boia}, {Bonaventura}, {Bond}, {Bosley}, {Boucarut}, {Bourque}, {Bouwman},
  {Bower}, {Bowers}, {Boyer}, {Bradley}, {Brady}, {Braun}, {Breda},
  {Bresnahan}, {Bright}, {Britt}, {Bromenschenkel}, {Brooks}, {Brooks},
  {Brown}, {Brown}, {Brown}, {Bunker}, {Burger}, {Bushouse}, {Cale}, {Cameron},
  {Cameron}, {Canipe}, {Caplinger}, {Caputo}, {Cara}, {Carey}, {Carniani},
  {Carrasquilla}, {Carruthers}, {Case}, {Catherine}, {Chance}, {Chapman},
  {Charlot}, {Charlow}, {Chayer}, {Chen}, {Cherinka}, {Chichester}, {Chilton},
  {Chonis}, {Clampin}, {Clark}, {Clark}, {Coe}, {Coleman}, {Comber}, {Comeau},
  {Connolly}, {Cooper}, {Cooper}, {Coppock}, {Correnti}, {Cossou}, {Coulais},
  {Coyle}, {Cracraft}, {Curti}, {Cuturic}, {Davis}, {Davis}, {Dean}, {DeLisa},
  {deMeester}, {Dencheva}, {Dencheva}, {DePasquale}, {Deschenes}, {Hunor
  Detre}, {Diaz}, {Dicken}, {DiFelice}, {Dillman}, {Dixon}, {Doggett},
  {Donaldson}, {Douglas}, {DuPrie}, {Dupuis}, {Durning}, {Easmin}, {Eck},
  {Edeani}, {Egami}, {Ehrenwinkler}, {Eisenhamer}, {Eisenhower}, {Elie},
  {Elliott}, {Elliott}, {Ellis}, {Engesser}, {Espinoza}, {Etienne}, {Etxaluze},
  {Falini}, {Feeney}, {Ferry}, {Filippazzo}, {Fincham}, {Fix}, {Flagey},
  {Florian}, {Flynn}, {Fontanella}, {Ford}, {Forshay}, {Fox}, {Franz}, {Fu},
  {Fullerton}, {Galkin}, {Galyer}, {Garc{\'\i}a Mar{\'\i}n}, {Gardner},
  {Gardner}, {Garland}, {Garrett}, {Gasman}, {Gaspar}, {Gaudreau}, {Gauthier},
  {Geers}, {Geithner}, {Gennaro}, {Giardino}, {Girard}, {Giuliano},
  {Glassmire}, \& {Glauser}}]{2023PASP..135d8001R}
{Rigby}, J., {Perrin}, M., {McElwain}, M., {et~al.} 2023, \pasp, 135, 048001

\bibitem[{{Roccatagliata} {et~al.}(2011){Roccatagliata}, {Bouwman}, {Henning},
  {Gennaro}, {Feigelson}, {Kim}, {Sicilia-Aguilar}, \&
  {Lawson}}]{2011ApJ...733..113R}
{Roccatagliata}, V., {Bouwman}, J., {Henning}, T., {et~al.} 2011, \apj, 733,
  113

\bibitem[{{Rogers} {et~al.}(2025){Rogers}, {Brandl}, \& {de
  Marchi}}]{2025A&A...698A.226R}
{Rogers}, C., {Brandl}, B., \& {de Marchi}, G. 2025, \aap, 698, A226

\bibitem[{{Romero-Mirza} {et~al.}(2024){Romero-Mirza}, {{\"O}berg}, {Banzatti},
  {Pontoppidan}, {Andrews}, {Wilner}, {Bergin}, {Czekala}, {Law}, {Salyk},
  {Teague}, {Qi}, {Bergner}, {Huang}, {Walsh}, {Guzm{\'a}n}, {Cleeves},
  {Aikawa}, {Bae}, {Booth}, {Cataldi}, {Ilee}, {Le Gal}, {Long}, {Loomis},
  {Menard}, \& {Liu}}]{2024ApJ...964...36R}
{Romero-Mirza}, C.~E., {{\"O}berg}, K.~I., {Banzatti}, A., {et~al.} 2024, \apj,
  964, 36

\bibitem[{{Russeil} {et~al.}(2017){Russeil}, {Adami}, {Bouret}, {Herv{\'e}},
  {Parker}, {Zavagno}, \& {Motte}}]{2017A&A...607A..86R}
{Russeil}, D., {Adami}, C., {Bouret}, J.~C., {et~al.} 2017, \aap, 607, A86

\bibitem[{{Russeil} {et~al.}(2012){Russeil}, {Zavagno}, {Adami}, {Anderson},
  {Bontemps}, {Motte}, {Rodon}, {Schneider}, {Ilmane}, \&
  {Murphy}}]{2012A&A...538A.142R}
{Russeil}, D., {Zavagno}, A., {Adami}, C., {et~al.} 2012, \aap, 538, A142

\bibitem[{{Russeil} {et~al.}(2010){Russeil}, {Zavagno}, {Motte}, {Schneider},
  {Bontemps}, \& {Walsh}}]{2010A&A...515A..55R}
{Russeil}, D., {Zavagno}, A., {Motte}, F., {et~al.} 2010, \aap, 515, A55

\bibitem[{{Salyk} {et~al.}(2025){Salyk}, {Pontoppidan}, {Banzatti}, {Bergin},
  {Arulanantham}, {Najita}, {Blake}, {Carr}, {Zhang}, \&
  {Xie}}]{2025AJ....169..184S}
{Salyk}, C., {Pontoppidan}, K.~M., {Banzatti}, A., {et~al.} 2025, \aj, 169, 184

\bibitem[{{Salyk} {et~al.}(2008){Salyk}, {Pontoppidan}, {Blake}, {Lahuis}, {van
  Dishoeck}, \& {Evans}}]{2008ApJ...676L..49S}
{Salyk}, C., {Pontoppidan}, K.~M., {Blake}, G.~A., {et~al.} 2008, \apjl, 676,
  L49

\bibitem[{{Salyk} {et~al.}(2011){Salyk}, {Pontoppidan}, {Blake}, {Najita}, \&
  {Carr}}]{2011ApJ...731..130S}
{Salyk}, C., {Pontoppidan}, K.~M., {Blake}, G.~A., {Najita}, J.~R., \& {Carr},
  J.~S. 2011, \apj, 731, 130

\bibitem[{{Sellek} {et~al.}(2020){Sellek}, {Booth}, \&
  {Clarke}}]{2020MNRAS.492.1279S}
{Sellek}, A.~D., {Booth}, R.~A., \& {Clarke}, C.~J. 2020, \mnras, 492, 1279

\bibitem[{{Sharples} {et~al.}(2013){Sharples}, {Bender}, {Agudo Berbel},
  {Bezawada}, {Castillo}, {Cirasuolo}, {Davidson}, {Davies}, {Dubbeldam},
  {Fairley}, {Finger}, {F{\"o}rster Schreiber}, {Gonte}, {Hess}, {Jung},
  {Lewis}, {Lizon}, {Muschielok}, {Pasquini}, {Pirard}, {Popovic}, {Ramsay},
  {Rees}, {Richter}, {Riquelme}, {Rodrigues}, {Saviane}, {Schlichter},
  {Schmidtobreick}, {Segovia}, {Smette}, {Szeifert}, {van Kesteren}, {Wegner},
  \& {Wiezorrek}}]{2013Msngr.151...21S}
{Sharples}, R., {Bender}, R., {Agudo Berbel}, A., {et~al.} 2013, The Messenger,
  151, 21

\bibitem[{{Sicilia-Aguilar} {et~al.}(2006){Sicilia-Aguilar}, {Hartmann},
  {Calvet}, {Megeath}, {Muzerolle}, {Allen}, {D'Alessio}, {Mer{\'\i}n},
  {Stauffer}, {Young}, \& {Lada}}]{2006ApJ...638..897S}
{Sicilia-Aguilar}, A., {Hartmann}, L., {Calvet}, N., {et~al.} 2006, \apj, 638,
  897

\bibitem[{{Sicilia-Aguilar} {et~al.}(2007){Sicilia-Aguilar}, {Hartmann},
  {Watson}, {Bohac}, {Henning}, \& {Bouwman}}]{2007ApJ...659.1637S}
{Sicilia-Aguilar}, A., {Hartmann}, L.~W., {Watson}, D., {et~al.} 2007, \apj,
  659, 1637

\bibitem[{{Sicilia-Aguilar} {et~al.}(2013){Sicilia-Aguilar}, {Kim}, {Sobolev},
  {Getman}, {Henning}, \& {Fang}}]{2013A&A...559A...3S}
{Sicilia-Aguilar}, A., {Kim}, J.~S., {Sobolev}, A., {et~al.} 2013, \aap, 559,
  A3

\bibitem[{{Siess} {et~al.}(2000){Siess}, {Dufour}, \&
  {Forestini}}]{2000A&A...358..593S}
{Siess}, L., {Dufour}, E., \& {Forestini}, M. 2000, \aap, 358, 593

\bibitem[{{Stapper} {et~al.}(2025{\natexlab{a}}){Stapper}, {Hogerheijde}, {van
  Dishoeck}, {Booth}, {Grant}, \& {van Terwisga}}]{2025A&A...693A..49S}
{Stapper}, L.~M., {Hogerheijde}, M.~R., {van Dishoeck}, E.~F., {et~al.}
  2025{\natexlab{a}}, \aap, 693, A49

\bibitem[{{Stapper} {et~al.}(2022){Stapper}, {Hogerheijde}, {van Dishoeck}, \&
  {Mentel}}]{2022A&A...658A.112S}
{Stapper}, L.~M., {Hogerheijde}, M.~R., {van Dishoeck}, E.~F., \& {Mentel}, R.
  2022, \aap, 658, A112

\bibitem[{{Stapper} {et~al.}(2025{\natexlab{b}}){Stapper}, {Hogerheijde}, {van
  Dishoeck}, {Vioque}, {Williams}, \& {Ginski}}]{2025A&A...693A.286S}
{Stapper}, L.~M., {Hogerheijde}, M.~R., {van Dishoeck}, E.~F., {et~al.}
  2025{\natexlab{b}}, \aap, 693, A286

\bibitem[{{Tabone} {et~al.}(2023){Tabone}, {Bettoni}, {van Dishoeck},
  {Arabhavi}, {Grant}, {Gasman}, {Henning}, {Kamp}, {G{\"u}del}, {Lagage},
  {Ray}, {Vandenbussche}, {Abergel}, {Absil}, {Argyriou}, {Barrado},
  {Boccaletti}, {Bouwman}, {Caratti o Garatti}, {Geers}, {Glauser},
  {Justannont}, {Lahuis}, {Mueller}, {Nehm{\'e}}, {Olofsson}, {Pantin},
  {Scheithauer}, {Waelkens}, {Waters}, {Black}, {Christiaens}, {Guadarrama},
  {Morales-Calder{\'o}n}, {Jang}, {Kanwar}, {Pawellek}, {Perotti}, {Perrin},
  {Rodgers-Lee}, {Samland}, {Schreiber}, {Schwarz}, {Colina}, {{\"O}stlin}, \&
  {Wright}}]{2023NatAs...7..805T}
{Tabone}, B., {Bettoni}, G., {van Dishoeck}, E.~F., {et~al.} 2023, Nature
  Astronomy, 7, 805

\bibitem[{{Temmink} {et~al.}(2024){Temmink}, {van Dishoeck}, {Gasman}, {Grant},
  {Tabone}, {G{\"u}del}, {Henning}, {Barrado}, {Caratti o Garatti}, {Glauser},
  {Kamp}, {Arabhavi}, {Jang}, {Kurtovic}, {Perotti}, {Schwarz}, \&
  {Vlasblom}}]{2024A&A...689A.330T}
{Temmink}, M., {van Dishoeck}, E.~F., {Gasman}, D., {et~al.} 2024, \aap, 689,
  A330

\bibitem[{{Townsley} {et~al.}(2018){Townsley}, {Broos}, {Garmire}, {Anderson},
  {Feigelson}, {Naylor}, \& {Povich}}]{2018ApJS..235...43T}
{Townsley}, L.~K., {Broos}, P.~S., {Garmire}, G.~P., {et~al.} 2018, \apjs, 235,
  43

\bibitem[{{Townsley} {et~al.}(2019){Townsley}, {Broos}, {Garmire}, \&
  {Povich}}]{2019ApJS..244...28T}
{Townsley}, L.~K., {Broos}, P.~S., {Garmire}, G.~P., \& {Povich}, M.~S. 2019,
  \apjs, 244, 28

\bibitem[{{Valeg{\r{a}}rd} {et~al.}(2021){Valeg{\r{a}}rd}, {Waters}, \&
  {Dominik}}]{2021A&A...652A.133V}
{Valeg{\r{a}}rd}, P.~G., {Waters}, L.~B.~F.~M., \& {Dominik}, C. 2021, \aap,
  652, A133

\bibitem[{{van Boekel} {et~al.}(2005){van Boekel}, {Min}, {Waters}, {de Koter},
  {Dominik}, {van den Ancker}, \& {Bouwman}}]{2005A&A...437..189V}
{van Boekel}, R., {Min}, M., {Waters}, L.~B.~F.~M., {et~al.} 2005, \aap, 437,
  189

\bibitem[{{van Breemen} {et~al.}(2011){van Breemen}, {Min}, {Chiar}, {Waters},
  {Kemper}, {Boogert}, {Cami}, {Decin}, {Knez}, {Sloan}, \&
  {Tielens}}]{2011A&A...526A.152V}
{van Breemen}, J.~M., {Min}, M., {Chiar}, J.~E., {et~al.} 2011, \aap, 526, A152

\bibitem[{{van der Walt} {et~al.}(2011){van der Walt}, {Colbert}, \&
  {Varoquaux}}]{2011CSE....13b..22V}
{van der Walt}, S., {Colbert}, S.~C., \& {Varoquaux}, G. 2011, Computing in
  Science and Engineering, 13, 22

\bibitem[{{van Terwisga} {et~al.}(2019){van Terwisga}, {Hacar}, \& {van
  Dishoeck}}]{2019A&A...628A..85V}
{van Terwisga}, S.~E., {Hacar}, A., \& {van Dishoeck}, E.~F. 2019, \aap, 628,
  A85

\bibitem[{{van Terwisga} {et~al.}(2020){van Terwisga}, {van Dishoeck}, {Mann},
  {Di Francesco}, {van der Marel}, {Meyer}, {Andrews}, {Carpenter}, {Eisner},
  {Manara}, \& {Williams}}]{2020A&A...640A..27V}
{van Terwisga}, S.~E., {van Dishoeck}, E.~F., {Mann}, R.~K., {et~al.} 2020,
  \aap, 640, A27

\bibitem[{{Virtanen} {et~al.}(2020){Virtanen}, {Gommers}, {Burovski},
  {Oliphant}, {Weckesser}, {Cournapeau}, {Alexbrc}, {Peterson}, {Reddy},
  {Haberland}, {Wilson}, {Nelson}, {Endolith}, {Mayorov}, {Van Der Walt},
  {Laxalde}, {Polat}, {Brett}, {Larson}, {Millman}, {Lars}, {Van Mulbregt},
  {Eric-Jones}, {Carey}, {Moore}, {Kern}, {Leslie}, {Perktold}, {Striega}, \&
  {Feng}}]{2020zndo...4406806V}
{Virtanen}, P., {Gommers}, R., {Burovski}, E., {et~al.} 2020, {scipy/scipy:
  SciPy 1.6.0}

\bibitem[{{Vlasblom} {et~al.}(2025){Vlasblom}, {Temmink}, {Grant}, {Kurtovic},
  {Sellek}, {van Dishoeck}, {G{\"u}del}, {Henning}, {Lagage}, {Barrado},
  {Caratti o Garatti}, {Glauser}, {Kamp}, {Lahuis}, {Olofsson}, {Arabhavi},
  {Christiaens}, {Gasman}, {Jang}, {Morales-Calder{\'o}n}, {Perotti},
  {Schwarz}, \& {Tabone}}]{2025A&A...693A.278V}
{Vlasblom}, M., {Temmink}, M., {Grant}, S.~L., {et~al.} 2025, \aap, 693, A278

\bibitem[{{Walborn}(2003)}]{2003IAUS..212...13W}
{Walborn}, N.~R. 2003, in IAU Symposium, Vol. 212, A Massive Star Odyssey: From
  Main Sequence to Supernova, ed. K.~{van der Hucht}, A.~{Herrero}, \&
  C.~{Esteban}, 13

\bibitem[{{Wang} {et~al.}(2007){Wang}, {Townsley}, {Feigelson}, {Getman},
  {Broos}, {Garmire}, \& {Tsujimoto}}]{2007ApJS..168..100W}
{Wang}, J., {Townsley}, L.~K., {Feigelson}, E.~D., {et~al.} 2007, \apjs, 168,
  100

\bibitem[{{Wells} {et~al.}(2015){Wells}, {Pel}, {Glasse}, {Wright},
  {Aitink-Kroes}, {Azzollini}, {Beard}, {Brandl}, {Gallie}, {Geers}, {Glauser},
  {Hastings}, {Henning}, {Jager}, {Justtanont}, {Kruizinga}, {Lahuis}, {Lee},
  {Martinez-Delgado}, {Mart{\'\i}nez-Galarza}, {Meijers}, {Morrison},
  {M{\"u}ller}, {Nakos}, {O'Sullivan}, {Oudenhuysen}, {Parr-Burman}, {Pauwels},
  {Rohloff}, {Schmalzl}, {Sykes}, {Thelen}, {van Dishoeck}, {Vandenbussche},
  {Venema}, {Visser}, {Waters}, \& {Wright}}]{2015PASP..127..646W}
{Wells}, M., {Pel}, J.~W., {Glasse}, A., {et~al.} 2015, \pasp, 127, 646

\bibitem[{{Wenger} {et~al.}(2000){Wenger}, {Ochsenbein}, {Egret}, {Dubois},
  {Bonnarel}, {Borde}, {Genova}, {Jasniewicz}, {Lalo{\"e}}, {Lesteven}, \&
  {Monier}}]{2000A&AS..143....9W}
{Wenger}, M., {Ochsenbein}, F., {Egret}, D., {et~al.} 2000, \aaps, 143, 9

\bibitem[{{Westmoquette} {et~al.}(2010){Westmoquette}, {Slavin}, {Smith}, \&
  {Gallagher}}]{2010MNRAS.402..152W}
{Westmoquette}, M.~S., {Slavin}, J.~D., {Smith}, L.~J., \& {Gallagher}, III,
  J.~S. 2010, \mnras, 402, 152

\bibitem[{{Wheeler} {et~al.}(2024){Wheeler}, {Hinkel}, \&
  {Banzatti}}]{2024PASP..136k3002W}
{Wheeler}, C.~H., {Hinkel}, N.~R., \& {Banzatti}, A. 2024, \pasp, 136, 113002

\bibitem[{{Winter} {et~al.}(2019){Winter}, {Clarke}, {Rosotti}, {Hacar}, \&
  {Alexander}}]{2019MNRAS.490.5478W}
{Winter}, A.~J., {Clarke}, C.~J., {Rosotti}, G.~P., {Hacar}, A., \&
  {Alexander}, R. 2019, \mnras, 490, 5478

\bibitem[{{Winter} \& {Haworth}(2022)}]{2022EPJP..137.1132W}
{Winter}, A.~J. \& {Haworth}, T.~J. 2022, European Physical Journal Plus, 137,
  1132

\bibitem[{{Winter} {et~al.}(2020){Winter}, {Kruijssen}, {Chevance}, {Keller},
  \& {Longmore}}]{2020MNRAS.491..903W}
{Winter}, A.~J., {Kruijssen}, J.~M.~D., {Chevance}, M., {Keller}, B.~W., \&
  {Longmore}, S.~N. 2020, \mnras, 491, 903

\bibitem[{{Wright} {et~al.}(2023){Wright}, {Rieke}, {Glasse}, {Ressler},
  {Garc{\'\i}a Mar{\'\i}n}, {Aguilar}, {Alberts}, {{\'A}lvarez-M{\'a}rquez},
  {Argyriou}, {Banks}, {Baudoz}, {Boccaletti}, {Bouchet}, {Bouwman}, {Brandl},
  {Breda}, {Bright}, {Cale}, {Colina}, {Cossou}, {Coulais}, {Cracraft}, {De
  Meester}, {Dicken}, {Engesser}, {Etxaluze}, {Fox}, {Friedman}, {Fu},
  {Gasman}, {G{\'a}sp{\'a}r}, {Gastaud}, {Geers}, {Glauser}, {Gordon},
  {Greene}, {Greve}, {Grundy}, {G{\"u}del}, {Guillard}, {Haderlein},
  {Hashimoto}, {Henning}, {Hines}, {Holler}, {Detre}, {Jahromi}, {James},
  {Jones}, {Justtanont}, {Kavanagh}, {Kendrew}, {Klaassen}, {Krause},
  {Labiano}, {Lagage}, {Lambros}, {Larson}, {Law}, {Lee}, {Libralato}, {Lorenzo
  Alverez}, {Meixner}, {Morrison}, {Mueller}, {Murray}, {Mycroft}, {Myers},
  {Nayak}, {Naylor}, {Nickson}, {Noriega-Crespo}, {{\"O}stlin}, {O'Sullivan},
  {Ottens}, {Patapis}, {Penanen}, {Pietraszkiewicz}, {Ray}, {Regan},
  {Roteliuk}, {Royer}, {Samara-Ratna}, {Samuelson}, {Sargent}, {Scheithauer},
  {Schneider}, {Schreiber}, {Shaughnessy}, {Sheehan}, {Shivaei}, {Sloan},
  {Tamas}, {Teague}, {Temim}, {Tikkanen}, {Tustain}, {van Dishoeck},
  {Vandenbussche}, {Weilert}, {Whitehouse}, \& {Wolff}}]{2023PASP..135d8003W}
{Wright}, G.~S., {Rieke}, G.~H., {Glasse}, A., {et~al.} 2023, \pasp, 135,
  048003

\bibitem[{{Wright} {et~al.}(2015){Wright}, {Wright}, {Goodson}, {Rieke},
  {Aitink-Kroes}, {Amiaux}, {Aricha-Yanguas}, {Azzollini}, {Banks},
  {Barrado-Navascues}, {Belenguer-Davila}, {Bloemmart}, {Bouchet}, {Brandl},
  {Colina}, {Detre}, {Diaz-Catala}, {Eccleston}, {Friedman},
  {Garc{\'\i}a-Mar{\'\i}n}, {G{\"u}del}, {Glasse}, {Glauser}, {Greene},
  {Groezinger}, {Grundy}, {Hastings}, {Henning}, {Hofferbert}, {Hunter},
  {Jessen}, {Justtanont}, {Karnik}, {Khorrami}, {Krause}, {Labiano}, {Lagage},
  {Langer}, {Lemke}, {Lim}, {Lorenzo-Alvarez}, {Mazy}, {McGowan}, {Meixner},
  {Morris}, {Morrison}, {M{\"u}ller}, {rgaard-Nielson}, {Olofsson},
  {O'Sullivan}, {Pel}, {Penanen}, {Petach}, {Pye}, {Ray}, {Renotte}, {Renouf},
  {Ressler}, {Samara-Ratna}, {Scheithauer}, {Schneider}, {Shaughnessy},
  {Stevenson}, {Sukhatme}, {Swinyard}, {Sykes}, {Thatcher}, {Tikkanen}, {van
  Dishoeck}, {Waelkens}, {Walker}, {Wells}, \& {Zhender}}]{2015PASP..127..595W}
{Wright}, G.~S., {Wright}, D., {Goodson}, G.~B., {et~al.} 2015, \pasp, 127, 595

\bibitem[{{Yoffe} {et~al.}(2023){Yoffe}, {van Boekel}, {Li}, {Waters},
  {Maaskant}, {Siebenmorgen}, {van den Ancker}, {dit de la Roche}, {Lopez},
  {Matter}, {Varga}, {Hogerheijde}, {Weigelt}, {Oudmaijer}, {Pantin}, {Meyer},
  {Augereau}, \& {Henning}}]{2023A&A...674A..57Y}
{Yoffe}, G., {van Boekel}, R., {Li}, A., {et~al.} 2023, \aap, 674, A57

\bibitem[{{Zannese} {et~al.}(2025){Zannese}, {Tabone}, {Habart}, {Dartois},
  {Goicoechea}, {Coudert}, {Gans}, {Martin-Drumel}, {Jacovella}, {Faure},
  {Godard}, {Tielens}, {Le Gal}, {Black}, {Vicente}, {Bern{\'e}}, {Peeters},
  {Van De Putte}, {Chown}, {Sidhu}, {Schroetter}, {Canin}, \&
  {Kannavou}}]{2025A&A...696A..99Z}
{Zannese}, M., {Tabone}, B., {Habart}, E., {et~al.} 2025, \aap, 696, A99

\bibitem[{{Zannese} {et~al.}(2023){Zannese}, {Tabone}, {Habart}, {Le Petit},
  {van Dishoeck}, \& {Bron}}]{2023A&A...671A..41Z}
{Zannese}, M., {Tabone}, B., {Habart}, E., {et~al.} 2023, \aap, 671, A41

\end{thebibliography}

\begin{appendix}
    
\section{Cluster membership and mass determination}
\label{sec:Gaia_mass_det}

\subsection{Gaia counterparts of the XUE sources}

We searched for Gaia counterparts within 1\arcsec of our remaining 12 targets using the coordinates from the \mys\ catalog. We found Gaia DR3 counterparts for all targets.
In order to test whether the Gaia matches are really the counterparts to the XUE targets, we inspected the available HST and 2MASS images. HST images were only available for some of the disks in Pis~24 (XUE\,1, 2, and 3), the remaining targets were inspected using 2MASS. This confirmed that the Gaia matches found within 1\arcsec are the counterparts of our targets.

\begin{figure}[ht]
    \centering
    \includegraphics[width=0.48\textwidth]{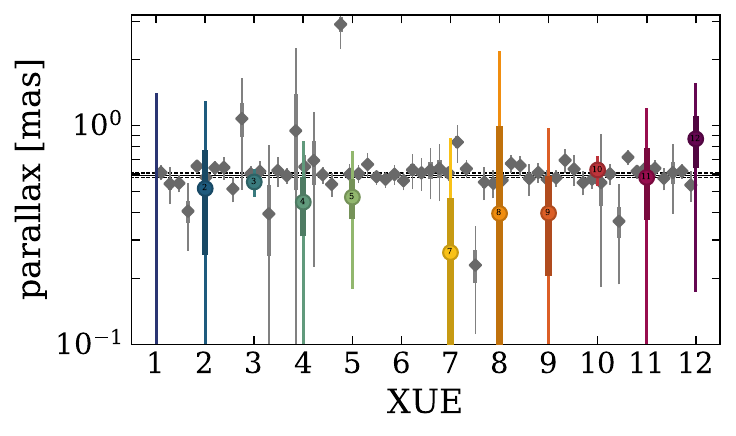}
    \caption{Gaia DR3 parallaxes of our sources in comparison to those of the O stars in the region (grey). The vertical thick and thin lines correspond to the 1$\sigma$ and 3$\sigma$ uncertainty in the parallax measurement. The horizontal black line shows the parallax at 1/1.663~mas corresponding to our determined mean distance and the dashed black lines show the parallax of the interval at [1/1.763, 1/1.563]~mas.}
    \label{fig:Gaia_par}
\end{figure}

The Gaia DR3 parallaxes of XUE~3, 4, 5, 9, and 10 were corrected using the algorithm by \citet{2021A&A...649A...4L} for the other targets, this was not possible. Most of
the parallaxes agree within their 3$\sigma$ uncertainty interval with the distance interval of $[1.635-1.705]$~kpc based on the known OB population of NGC~6357 \citep[][]{2023ApJ...958L..30R}. Only the parallax of XUE~6 does not agree with the interval, however this target has a Gaia DR3 re-normalized unit weight error (RUWE) larger than 1.4 which indicates a poor fit of the astrometric solution \citep[][]{2018A&A...616A...2L}. Figure~\ref{fig:Gaia_par} shows the parallaxes of the XUE sources in comparison to those of the O-type stars in the region. And the Gaia parallaxes of our sources are listed in Table~\ref{tab:full_properties}.

\subsection{Photometric data}\label{sec:photometry}

The photometric data were taken from the \mys\ catalogue \citep{2013ApJS..209...26F, 2013ApJS..209...28K, 2013ApJS..209...29K} which includes the $JHK$-band magnitudes from the {\it United Kingdom Infra-red Telescope} (UKIRT) wide field camera \citep{2007MNRAS.379.1599L, 2008MNRAS.391..136L}, plus the four {\it Spitzer} IRAC bands \citep[GLIMPSE;][]{2003PASP..115..953B}. In the specific case of XUE~10, we used VVV NIR photometry for the $J$, $H$ and $K$-bands. 
The near- and mid-infrared colour-magnitude diagrams (CMDs) and colour-colour diagrams (CCDs) are shown in Figure~\ref{fig:CMD_CCD} and are discussed in the next section.

\subsection{Extinction and mass determination} \label{sec:stellar_props_section}

Values for the total to selective extinction R$_V$ vary in the literature from 3.1 \citep[][]{2007ApJ...660.1480M} to 3.7 \citep{2004AJ....127.2826B, 2012A&A...538A.142R}. In this paper, we calculated an average $R_V = 3.30^{+0.13}_{-0.19}$ by selecting all 78 sources within a cone of 15\arcmin\ radius centered at (RA, Dec) = (261.269027, $-$34.315231) from the \citet{2022A&A...662A.125F} catalog, which offers a uniformly derived set of astrophysical parameters for over 123 million stars, enabling detailed studies of dust extinction and average grain size along the line of sight. The catalog provides stellar parameters, including $R_V$, based on data from \textit{Gaia} DR2, 2MASS, and ALLWISE. We used all available sources in this region without applying any additional quality cuts, and computed the median $R_V$ value and its uncertainties from the 16th and 84th percentiles to characterize the local extinction properties.

\begin{figure*}
    \centering
    \includegraphics[width=\textwidth]{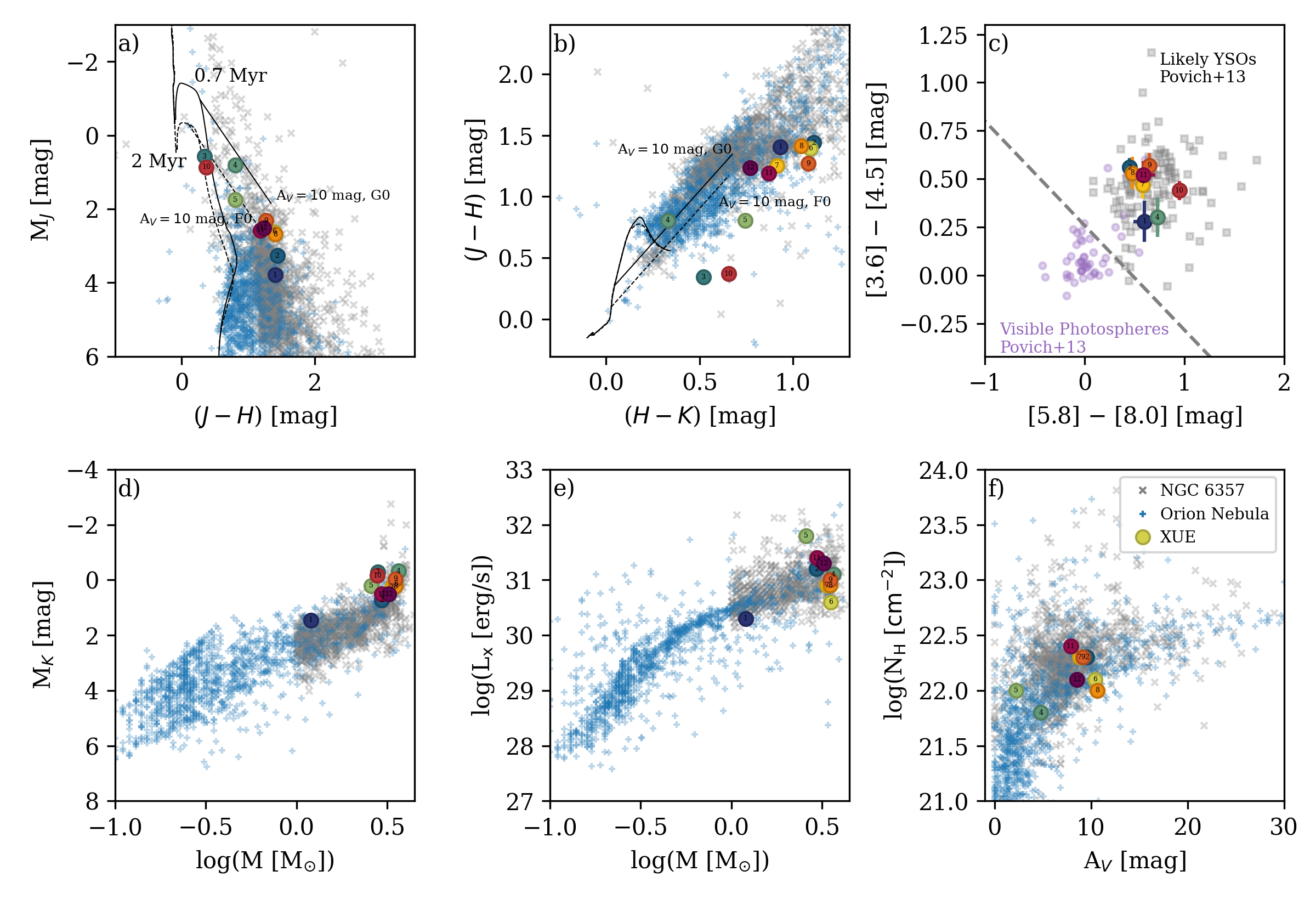}
    \caption{Characteristics of XUE stars. In all panels, XUE stars are color-coded and labeled for clarity. The Orion Nebula young stars are represented by blue plus symbols, and NGC~6357 young stars are shown as grey crosses \citep[][]{2005ApJS..160..319G, 2013ApJS..209...32B, 2013ApJS..209...28K, 2019ApJS..244...28T, 2022ApJ...935...43G}. (a, b) UKIRT/VVV NIR color-magnitude and color-color diagrams. PARSEC 1.2S PMS isochrones for 0.7 and 2~Myr are shown as solid and dashed curves, respectively. Reddening vectors corresponding to $A_V = 10$~mag are drawn as solid and dashed lines, originating from G0 stars at 0.7~Myr and F0 stars at 2~Myr. (c) Spitzer-IRAC MIR color-color diagram of NGC~6357 young stars with known IRAC photometry. Grey squares represent known disky stars, while purple circles indicate diskless stars \citep[][]{2013ApJS..209...31P}. (d) Absolute $K$-band magnitude (uncorrected for extinction) as a function of stellar mass. (e) Intrinsic X-ray luminosity as a function of stellar mass. (f) X-ray column density plotted against visual-band extinction.
    The names of the sources are displayed on the markers and are visible when zooming in.}
    \label{fig:CMD_CCD}
\end{figure*}

Using the spectral classification from \citet{2020A&A...633A.155R} and R$_V = 3.3$~mag for NGC~6357 \citep[][]{2015A&A...573A..95M, 2017A&A...607A..86R, 2022A&A...662A.125F}, we compared the NIR photometry (Section~\ref{sec:photometry}) of XUE stars with PARSEC 1.2S pre-main sequence (PMS) evolutionary models \citep[][]{2012MNRAS.427..127B} on the $J$ vs. $J-H$ color-magnitude diagram (Figure~\ref{fig:CMD_CCD}a). This analysis, which minimizes the influence of circumstellar disks, provided estimates for stellar ages, extinctions, masses, and bolometric luminosities.

The results indicate that the NIR photometry of XUE G/K-type stars corresponds to a stellar age of $\sim 0.7$~Myr, while the observed properties of A/F-type stars suggest slightly older ages of $(1.5-2)$~Myr. The $0.7$~Myr age for G/K-type stars agrees with the median age for hundreds of NGC~6357 members derived using PARSEC models by \citet{2022ApJ...935...43G}. However, the PARSEC-based ages for G/K-type stars are systematically lower than the Siess \citep[][]{2000A&A...358..593S} model-based ages of $(1-1.5)$~Myr previously calculated for NGC~6357 members \citep[][]{2014ApJ...787..108G}. Regardless of the PMS evolutionary model used, NGC~6357 stars have ages comparable to the Orion Nebula Cluster (ONC) and the stellar populations of its underlying Orion Molecular Cloud One \citep[OMC-1][]{2014ApJ...787..109G, 2022ApJ...935...43G}. These young ages align with the fact that neither NGC~6357 nor the ONC has experienced supernova explosions (SNe) yet. 
This is supported by the absence of known supernova remnants, non-thermal emission, or chemically enriched hot plasma in their diffuse X-ray and radio emission, as well as the continued presence of massive O-type stars and lack of known pulsars or neutron stars \citep[e.g.,][]{2008Sci...319..309G, 2007ApJS..168..100W, 2010MNRAS.402..152W, 2011A&A...535A..29G, 2017ApJ...838...61P}.

In contrast to the ONC, which is dominated by a single O6-type star, $\theta^1$OriC, NGC~6357 hosts at least a dozen O-type stars with spectral types earlier than O7. These massive stars have efficiently ablated central regions of their parental molecular cloud, carving out large cavities and leaving behind smaller, dense, bright-rimmed clouds. Several stellar clusters, including the prominent Pismis 24, are now exposed, yet remain surrounded by extensive reservoirs of dense molecular material. 

Despite this molecular erosion, many X-ray- and NIR-detected cluster members in NGC~6357 still exhibit significant extinction, attributed to both local dust and the interstellar medium. Figures~\ref{fig:CMD_CCD}a and \ref{fig:CMD_CCD}b show that typical extinction values in NGC~6357 range from several to over ten magnitudes --- higher than those in the ONC but comparable to OMC-1. Similarly, the XUE stars display high extinction values, aligning with the correlation between X-ray column densities and visual-band extinctions observed for stars in both NGC~6357 and the Orion Nebula (Figure~\ref{fig:CMD_CCD}f).

NIR and mid-infrared (MIR) color-color diagrams (Figures~\ref{fig:CMD_CCD}b, \ref{fig:CMD_CCD}c) indicate that most XUE stars host optically thick disks, as evidenced by strong $K_s$-band excesses. These stars, along with other NGC~6357 disky members, are clearly distinguishable from diskless X-ray-emitting young stars.

The X-ray luminosities of G/K-type XUE stars are consistent with their intermediate masses (Figure~\ref{fig:CMD_CCD}e). In contrast, the A/F-type stars XUE~3 and XUE~10 are undetected in X-rays, likely due to their developed radiative cores, which suppress magnetic dynamos and reduce X-ray emission \citep[][]{2021AJ....162..153N, 2022ApJ...935...43G}.

\subsection{Assumed OB stars properties}

Section~\ref{sec:selection} describes how we calculated the external FUV flux to which the XUE sources are exposed. Table~\ref{tab:OB_stars_props} lists the assumed OB stars' properties to obtain the Phoenix atmosphere models.

\begin{table*}[ht!]
    \centering
    \caption{Assumed properties for the massive stars in the NGC\,6357 region.}
\begin{tabular}{cccccccc}
    \hline
    Object & R.A. & DEC. & Sp. Type & T$_{\rm{eff}}$ & $\log{L}$ & Mass\tablefootmark{a} & FUV luminosity \\
    & J(2000) & J(2000) & & K & L$_{\odot}$ & M$_{\odot}$ & erg\,s$^{-1}$\\
    \hline
    
Pis24 15 & 17:24:28.95 & -34:14:50.6 & O7.5Vz & 35000 & 4.93 & 24 & $2.20 \times 10^{38}$ \\
Pis24 3 & 17:24:42.30 & -34:13:21.4 & O8V & 35000 & 4.96 & 24 & $2.20 \times 10^{38}$ \\
Pis24 2 & 17:24:43.28 & -34:12:43.9 & O5.5V((f)) & 40000 & 5.37 & 36 & $4.47 \times 10^{38}$ \\
Pis24 1NE & 17:24:43.497 & -34:11:56.89 & O3.5If* & 40000 & 5.89 & 65 & $1.53 \times 10^{39}$ \\
Pis24 1SW & 17:24:43.481 & -34:11:57.21 & O4III & 44000 & 5.81 & 60 & $1.06 \times 10^{39}$ \\
Pis24 16 & 17:24:44.48 & -34:11:58.8 & O7.5V & 36000 & 5.0 & 26 & $2.56 \times 10^{38}$ \\
Pis24 17 & 17:24:44.72 & -34:12:02.6 & O3.5III(f*) & 44500 & 5.92 & 70 & $1.39 \times 10^{39}$ \\
Pis24 13 & 17:24:45.77 & -34:09:39.8 & O6.5V((f)) & 38000 & 5.25 & 32 & $3.97 \times 10^{38}$ \\
16 & 17:24:36.039 & -34:14:00.48 & O9-B0 & 32000 & 4.76 & 20 & $1.57 \times 10^{38}$ \\
B0 & 17:24:34.800 & -34:13:17.91 & O9.5-B1V & 29000 & 4.25 & 14 & $5.00 \times 10^{37}$ \\
B1 & 17:24:40.491 & -34:12:06.47 & O9.5-B2V & 26300 & 4.35 & 14 & $6.69 \times 10^{37}$ \\
B2 & 17:24:42.878 & -34:09:11.90 & O9.5-B1V & 29000 & 4.62 & 17 & $1.18 \times 10^{38}$ \\
B4 & 17:24:55.086 & -34:11:11.61 & O9.5-B1V & 29000 & 4.71 & 18 & $1.46 \times 10^{38}$ \\
O1 & 17:24:43.289 & -34:11:41.89 & O9-B1 & 29250 & 4.62 & 17 & $1.16 \times 10^{38}$ \\
N78 51 & 17:25:29.16 & -34:25:15.6 & O6Vn((f)) & 41000 & 5.35 & 36 & $3.94 \times 10^{38}$ \\
N78 49 & 17:25:34.23 & -34:23:11.7 & O5.5IV(f) & 39000 & 5.87 & 60 & $1.38 \times 10^{39}$ \\
118 & 17:26:01.897 & -34:16:31.41 & B0-B3 & 24250 & 4.3 & 13 & $5.15 \times 10^{37}$ \\
B10 & 17:25:58.088 & -34:16:06.05 & O7-O9V & 34500 & 4.43 & 17 & $7.11 \times 10^{37}$ \\
B9 & 17:25:57.298 & -34:18:50.77 & O9-B3V & 24750 & 4.33 & 13 & $5.15 \times 10^{37}$ \\
        \hline
    \end{tabular}
    \tablefoot{
\tablefoottext{a}{Mass of the adopted MIST isochrones.}}
    \label{tab:OB_stars_props}
\end{table*}

\subsection{Properties of the XUE sample}

Table~\ref{tab:full_properties} contains all the data used and derived in this paper for the XUE sources. The full table is available at the CDS.

\begin{table}[]
    \centering
    \caption{Adopted and derived properties of the XUE sources.}
    \setlength{\tabcolsep}{5pt}
    \begin{tabular}{cccccccc}
    \hline
     XUE  &   RA  & ... &  $\log{G_{0}}$  &   A$_{V}$  &   ...  &   $\varpi$ & $\varpi_{error}$ \\
      & deg & ... & & mag & ... & mas & mas \\
      \hline
    1  &   261.1671  & ... &   5.186  &   9.2  &   ...  &   -0.6595  &   0.6870 \\
    2  &   261.1734  & ... &   5.292  &   9.6  &   ...  &   0.5154  &   0.2594\\

\hline
    \end{tabular}
    \tablefoot{The full table containing 40 columns per source is available at the CDS.}
    \label{tab:full_properties}
\end{table}

\section{Full Spectral energy distributions}\label{sec:full_SEDs}

\begin{figure*}[ht]
    \centering
    \includegraphics[width=\textwidth]{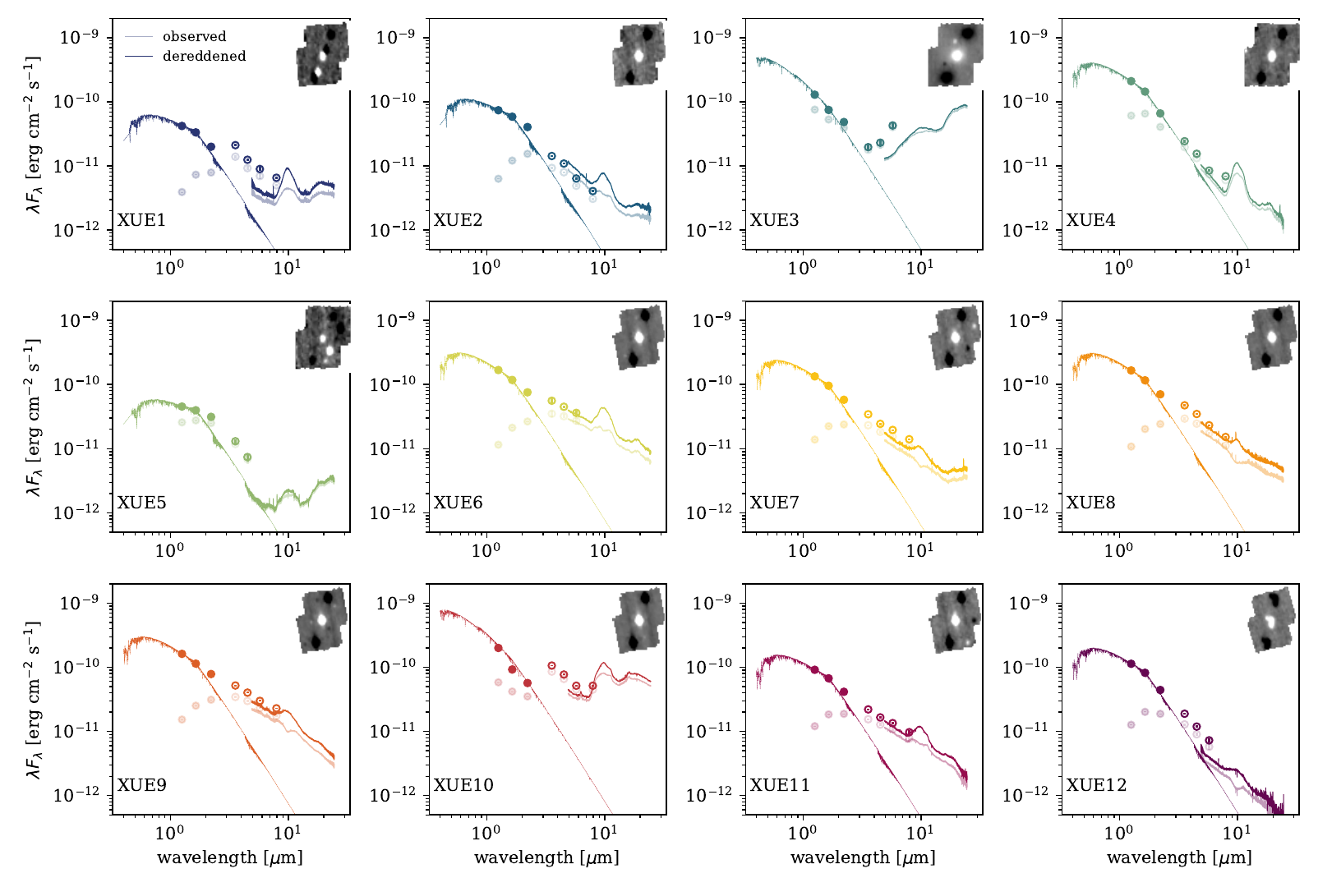}
    \caption{Spectral energy distributions of the XUE sources. Each column shows the sources in one region: Pis~24, G353.1+0.6 and G353.1+0.7 from left to right. The dashed lines show Phoenix \citep[][]{2005ESASP.576..565B}and Kurucz \citep[][]{1993KurCD..13.....K} stellar photospheres with the parameters derived in Section~\ref{sec:photometry}. The dots show UKIRT and IRAC photometry, with the exception of XUE~10 in which VISTA photometry is displayed. The solid line shows the MIRI MRS spectrum. The light colors show the observed data, the dark ones show the dereddened data. The panels in the top-left of each subfigure show the MRS data cube at 5.5~\micron.}
    \label{fig:SED_wStellarPhot_overview}
\end{figure*}

Figure~\ref{fig:SED_wStellarPhot_overview} shows the SEDs of our sources together with the existing photometry and stellar models corresponding to the luminosities and temperatures derived in Section~\ref{sec:stellar_props_section}.

\textit{Spitzer} observations of disks in nearby regions  \citep[e.g.][]{2007ApJ...659.1637S, 2006ApJ...638..897S} found a large diversity in spectral shapes. Our sample of highly irradiated disks also resembles the above mentioned diversity. 
Given that our selection criteria included having determined the spectral types of the central sources based on their $K-$band spectrum (Section~\ref{sec:selection}), it is expected that they lack a strong NIR excess indicating that the might have undergone substantial clearing. 
Based on Figure~\ref{fig:SED_overview} we can classify our sources into three categories: XUE~2, 4, and 5 are disks that show NIR and MIR emission similar to stellar photospheres. The SEDs XUE~1, 6, 7, 8, 9, 11, and 12 show moderate emission in the $K-$band and significant MIR emission, which XUE~1 and 7 showing increasing flux beyond 20~\micron. Finally, XUE~3 and 10 display SEDs with little to no NIR excess and steadily increasing emission in the MIR, especially in the case of XUE~3.

\section{Key regions of molecular emission }
\label{sec:other_molecules}

\begin{figure*}
    \centering
    \includegraphics[width=\textwidth]{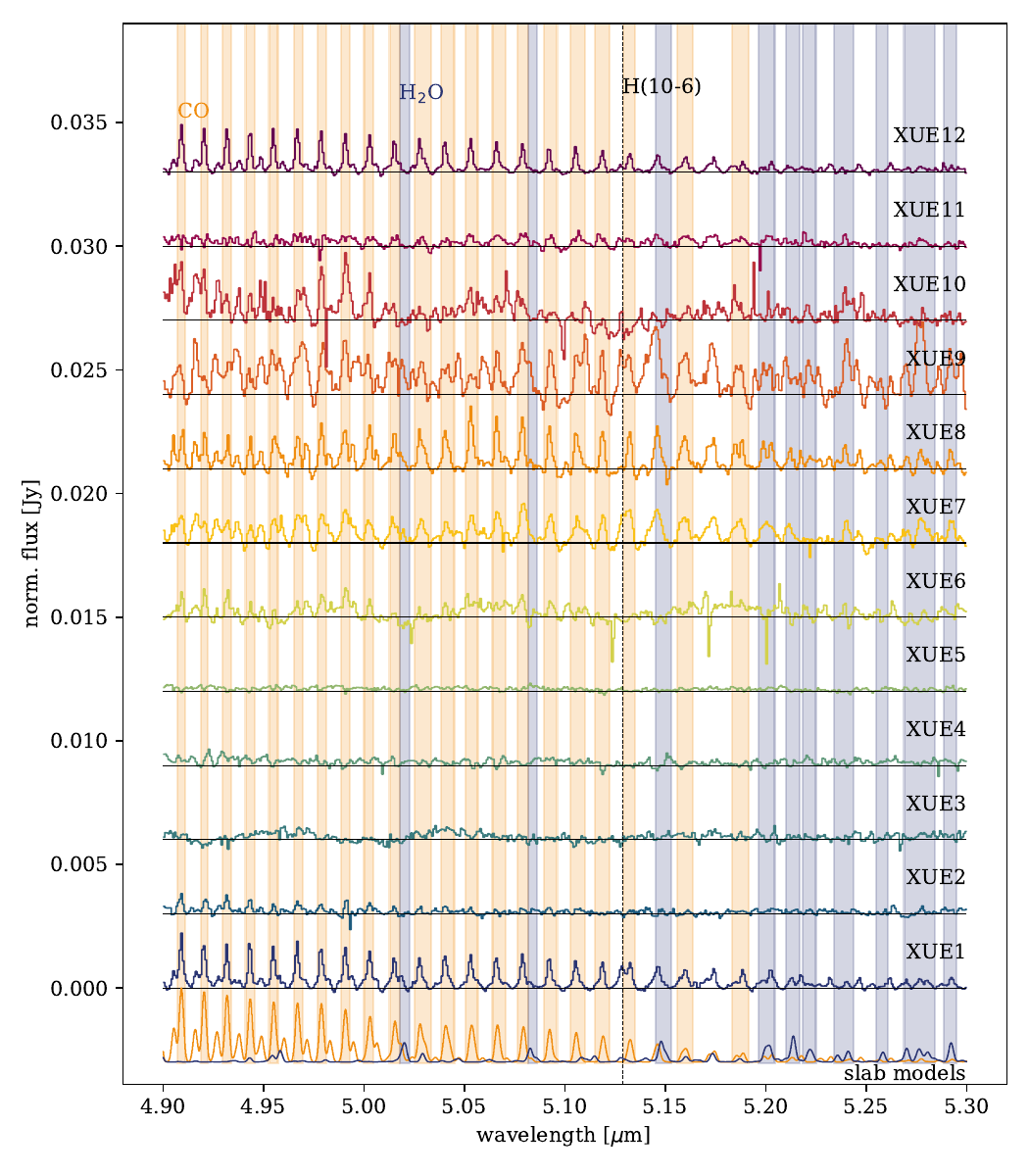}
    \caption{Overview of the spectral region between 4.9 and 5.3~\micron\ for all XUE sources. The lowermost spectrum shows the most prominent molecules in this region; CO at a temperature of 1500~K and a column density of 3$\times10^{17}$~cm$^{-2}$ (yellow) and H$_2$O at 850~K and $10^{18}$~cm$^{-2}$ (blue).}
    \label{fig:4.9-5.3_region}
\end{figure*}

\begin{figure}
    \centering
    \includegraphics[width=0.5\textwidth]{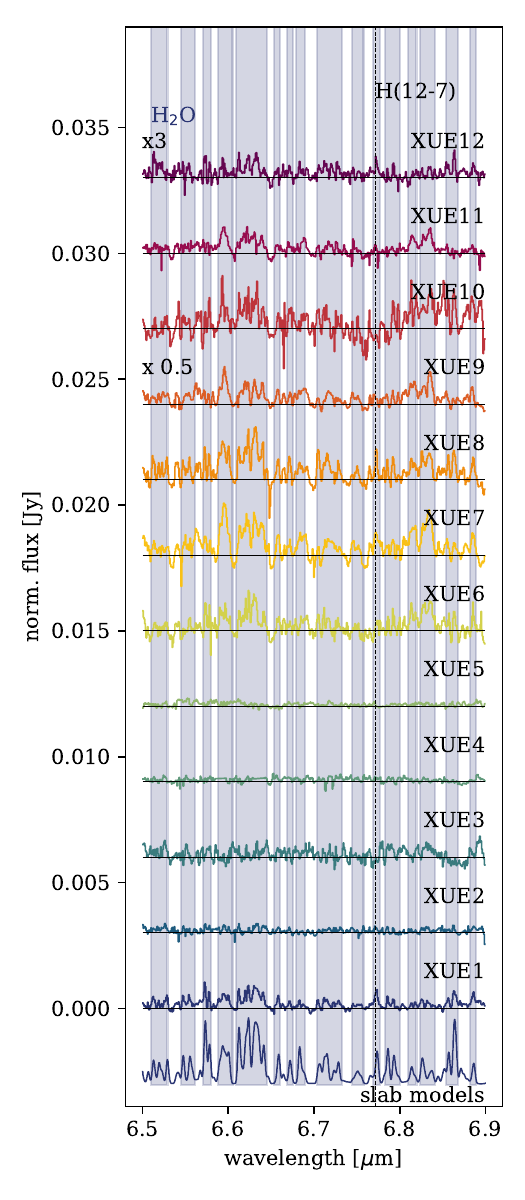}
    \caption{Same as Figure~\ref{fig:4.9-5.3_region} but for the region between 6.5 and 6.9~\micron. The lowermost spectrum shows H$_2$O at 850~K and $10^{18}$~cm$^{-2}$ (blue).}
    \label{fig:6.5-6.9_region}
\end{figure}

\begin{figure*}
    \centering
    \includegraphics[width=\textwidth]{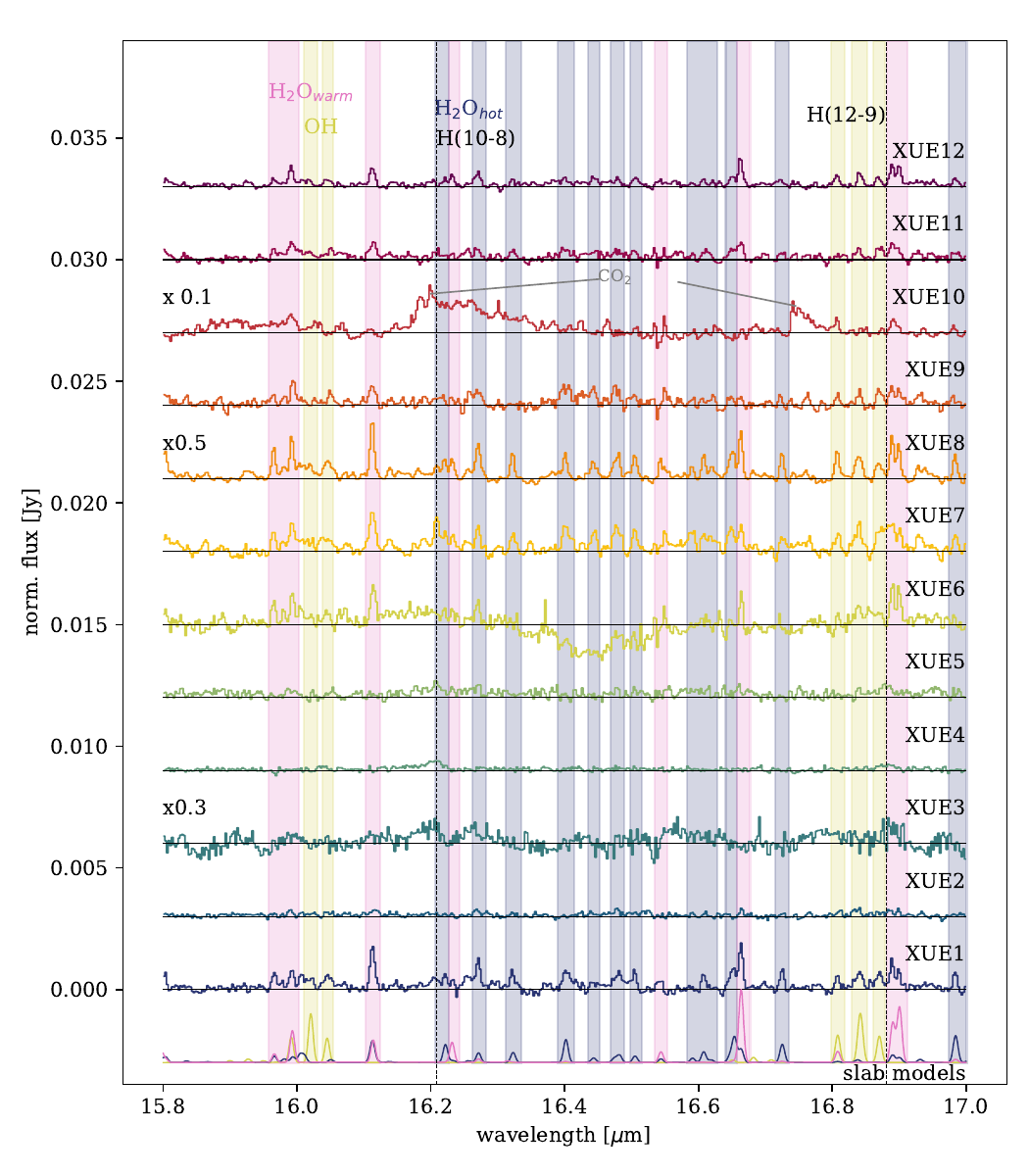}
    \caption{Same as Figure~\ref{fig:4.9-5.3_region} but for the region between 15.8 and 17~\micron. The lowermost spectrum shows OH at 600~K and $10^{16}$~cm$^{-2}$ (green).}
    \label{fig:15.8-17.0_region}
\end{figure*}

\begin{figure}
    \centering
    \includegraphics[width=0.5\textwidth]{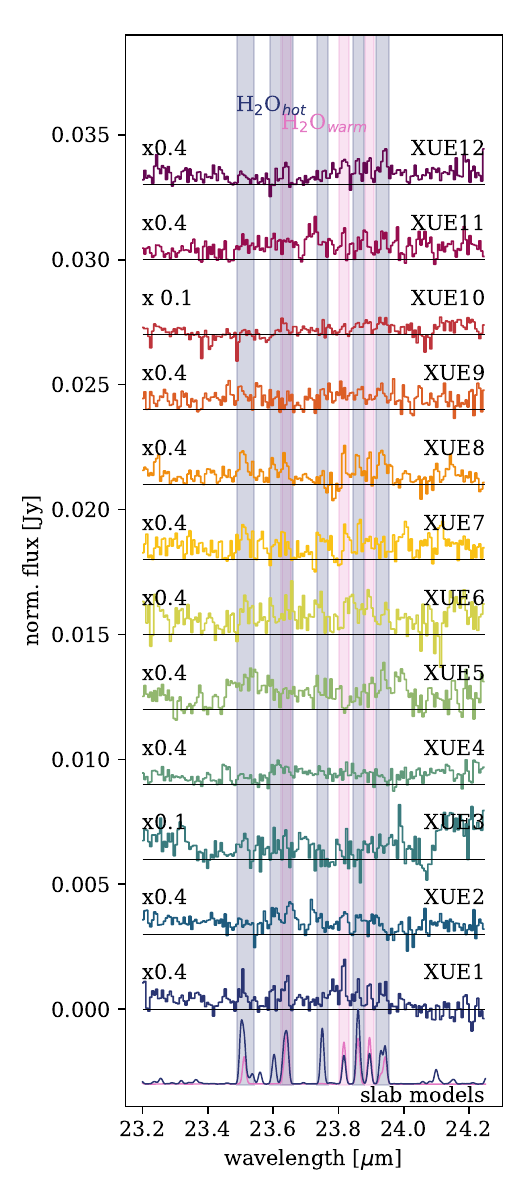}
    \caption{Same as Figure~\ref{fig:4.9-5.3_region} but for the region between 23.2 and 24.3~\micron. The lowermost spectrum shows warm and hot H$_2$O at 400~K and $2.2\times10^{16}$~cm$^{-2}$ (blue) and at 900~K and $2.2\times10^{16}$~cm$^{-2}$ (purple).}
    \label{fig:23.2-24.25_region}
\end{figure}

Figures \ref{fig:4.9-5.3_region}, \ref{fig:6.5-6.9_region}, \ref{fig:15.8-17.0_region} and \ref{fig:23.2-24.25_region} show overviews of regions of the spectra where key molecular lines are present.

\section{Description of individual sources}\label{sec:individual_sources}

In the following we describe the SED shape and molecular content of each source individually.\\

\textbf{XUE~1:} This source was analyzed in detail by \citet{2023ApJ...958L..30R} and \citet{2025ApJ...985...72P}. There are three sources in the MRS data cube, one at 1.5\arcsec\ from the disk bearing source and another one 0.2\arcsec\ away. The two closest sources are not resolved in the MIRI cubes, and source three is not detected redwards of $\sim9$~\micron \citep[see][for a detailed discussion]{2023ApJ...958L..30R}. The prominent 10~\micron\ silicate emission indicates the presence of small, partially crystalline silicate dust at the disk surface. They find abundance of water, CO, $^{12}$CO$_2$, HCN, C$_2$H$_2$. Here we confirm the presence of two water components at different temperatures around 24~\micron\ as indicated by \citet{2023ApJ...957L..22B} (Figure~\ref{fig:23.2-24.25_region}). In addition to the molecules reported in \citet{2023ApJ...958L..30R}, we detect weak OH lines at 16 and 16.80~\micron\ (Figure~\ref{fig:15.8-17.0_region}). The spectrum also contains hydrogen recombination lines. In particular we detect the Pfund line (H(6-5) at 7.4598~\micron) and some lines of the Humpreys series: H(10-6) at 5.1286~\micron, H(9-6) at 5.9082~\micron, H(8-6) at 7.5025~\micron, H(7-6) at 12.3719~\micron. Additionally we detect weak lines corresponding to $H_2$ (0-0) transitions, S(3) at 9.66~\micron\ is prominent, and S(1), S(4), S(5), S(7) are potentially detected.  \\

\textbf{XUE~2:} This is a single source in the MRS cube up to channel 4 where a few bright sources are detected around XUE~2. 
None of these additional sources overlap with XUE2, and thus they do not contaminate the extracted spectrum.
The MIRI flux is higher than the \textit{Spitzer} photometry, which might indicate that this source is variable. Examples of variability detected with \textit{Spitzer} and MIRI have been also observed in nearby disks \citep[e.g.][]{2023Natur.620..516P}. 
The SED shape has a change of slope between 8-12 and 24~\micron, which could indicate the presence of an inner hole. 
This source is relatively poor in its molecular content. We detect a few lines corresponding to the P-branch CO fundamental emission between 4.9 and 5.0~\micron\ for the $v = 1 - 0$ transitions, unlike the case of XUE~1 there is no significant detection of CO lines corresponding to the $v = 2 - 1$ transitions. Due to the low signal-to-noise of the MIRI spectrum, we cannot confirm or reject the presence of water emission between 6 and 7~\micron, HCN at 14~\micron, CO$_2$ at 15~\micron\ or OH at 16 and 16.80~\micron. We do not detect any presence of C$_2$H$_2$, water beyond 15~\micron\ nor $^{13}$CO$_2$. There is an excess emission in the spectrum at $\sim$7.2~\micron\ which is consistent with the location of CH$_3^+$ as detected by \citet{2023Natur.621...56B}, nevertheless due to the low S/N and to the potential presence of water at the same wavelengths, we are not able to accurately model this molecule and therefore cannot confirm its presence in the spectrum. The spectrum contains the Pfund H(6-5) line in emission.\\

\textbf{XUE~3}: In channels 1 and 2 XUE~3 is the only source in the detector, in channel 3 two extra sources are detected, the background in channel 4 is very bright, making the source almost undetectable. 
The extra sources detected in channel 3 do not contaminate the extraced spectrum. 
The \textit{Spitzer} photometric points are higher than the MIRI spectrum, these could be due to the \textit{Spitzer} photometry being contaminated by the strong background emission. The spectrum does not show any molecular emission and shows weak 10~\micron\ silicate emission. The strong rising of the MIRI SED at larger wavelengths together with the lack of strong NIR excess suggests the presence of a inner hole or a large disk gap.\\

\textbf{XUE~4:} This source is the only one in the detector and it is prominent in all channels. Its SED does not show any $K$-band excess and the MIRI MRS spectrum has a negative slope between 5 and 8~\micron. Making it a good candidate for a transition disk. It has a prominent 10~\micron\ silicate feature, with a red shoulder that indicates the presence of crystalline silicates. We also identify clear forsterite bands at 16 and 19~\micron\ in the MIRI spectrum. It is not possible to identify any molecular emission in the spectrum. \\

\textbf{XUE~5:} There are 3 non-overlapping sources clearly detected in all channels, one to the lower right part of the detector (SW), and the third source located to the lower-left of the second source. The third source is only detected at the shortest wavelengths. At the longest wavelength, there is very bright emission on the top left corner of the detector (NE) that dominates the emission in the data cube. 
The background spectrum of XUE~5 is very rich in PAHs and there are some remnants of the background subtraction left in the spectrum. The blue part of the MIRI spectrum has a negative slope and might present absorption lines coming from the stellar photosphere. With the exception of H$_2$ (0-0) S(4) and S(5), there are no molecular emission lines detected in the MIRI spectrum. The lack of any NIR excess, together with the lack of molecular emission that XUE~5 has a very gas and dust-depleted inner disk, consistent with a transitional disk.\\

\textbf{XUE~6:} This is a single source in detector in all channels, but there is a lot of nebulosity visible in the redder channels at the top-right of the detector. The PAH emission from the background is very strong, and some remnants of it are left in absorption in the spectrum after the background subtraction. The SED shape is consistent with a continuous disk, with the flux constantly decreasing toward longer wavelengths. XUE~6 has a relatively rich molecular inventory: we detect some transitions corresponding to the P-branch of the CO fundamental, H$_2$O emission between 6 and 7~\micron, as well as HCN, C$_2$H$_2$, and CO$_2$. The H$_2$ corresponding to the 0-0 S(5) transition at 6.90952~\micron\ is also visible in the spectrum.\\

\textbf{XUE~7:} There are several fainter point sources in the detector close to XUE~7, which could explain the fact that the \textit{Spitzer} fluxes are slightly higher than the MIRI spectrum. Towards the red end of channel 3 at around 15~\micron, there is a bright source overlapping with XUE~7 which could be the reason why the SED has a change of slope at around 16~\micron. XUE~7 has a relatively weak 10~\micron\ emission feature and shows prominent CO, CO$_2$, and H$_2$O lines. Weaker emission from HCN, C$_2$H$_2$ and OH is also detected in the MIRI spectrum.\\

\textbf{XUE~8:} In most channels XUE~8 appears as a single source in the detector. As well as for XUE~7, the SED shape suggests a full disk. 
XUE~8's silicate emission feature at 10~\micron\ has an uncommon shape, having a lower flux at 11.3~\micron\ than expected for its peak flux.
Its spectrum is extremely rich in molecular emission. It has very strong water emission across the spectrum as well as prominent CO, HCN, CO$_2$, H$_2$ and OH lines. Due to the abundance of water lines in the 13 to 16~\micron\ range, we cannot confirm or reject the presence of C$_2$H$_2$ nor $^{13}$CO$_2$ in the spectrum.\\

\textbf{XUE~9:} This source appears single at all wavelengths and there is very strong nebular emission in the redder channels. 
The SED slope decreases constantly towards longer wavelengths consistent with a full disk. 
This source does not show a lot of molecular richness; There is emission from hot water at around 6~\micron, but we cannot confirm the presence of other molecules.\\

\textbf{XUE~10:} This is the brightest source in our sample and it is single in the detector at all wavelengths. The \textit{Spitzer} fluxes are higher than the MIRI SED indicating possible variability. XUE~10 SED's flux increases steeply red wards of 12~\micron. 
XUE~10 has a strong silicate feature at 10~\micron\ with a prominent emission feature at 11.3~\micron\ that is probably a combination of forsterite and PAH emission. The spectrum also shows strong PAH emission at 6.2~\micron. 
XUE~10 has a very peculiar chemical content, it is very water poor but shows very prominent emission from \CO2, including \13CO2, and even the rarer isotopologues $^{16}$OC$^{18}$O and $^{16}$OC$^{17}$O \citep[see][for a full analysis of this spectrum]{2025arXiv250713921F}. \\

\textbf{XUE~11:} This source is single in the MRS field-of-view, its brightness with respect to the background decreases significantly in channel~4. 
XUE~11's SED is consistent with that of a full disk, with the flux decreasing towards longer wavelengths. 
We tentatively detect CO around 5~\micron, hot water at $\sim6$~\micron\ and HCN at 14~\micron. Given the quality of the spectrum, it is not possible to identify other molecules.\\

\textbf{XUE~12:} This source consists of three stars very close to each other, therefore the photometric points, as well as our mass estimates need to be taken as upper limits. 
Around $\sim15$~\micron, the three sources become indistinguishable in the MIRI detector. This is likely because the two additional sources are foreground contaminants that appear very faint at longer wavelengths. However, there remains a possibility that, beyond this wavelength, our 1D spectrum may be affected by emission lines from these nearby sources.  
The source becomes extremely faint towards channel~4, making it indistinguishable from the background. 
The MIRI SED decreases continuously with wavelength, but given that the stellar SED is uncertain, it is not possible to determine where the infrared excess starts to determine the inner disk structure. 
XUE~12 is rich in molecules: we detect prominent CO and \CO2, including the possible presence of \13CO2. We also detect hot and warm \H2O\ and HCN. We cannot confirm or reject the presence of \C2H2.  

\section{Water lines at 17~\micron}\label{sec:H2O_spectra}

In this section we show the spectra used to determine the water line luminosity (L$_{H_2O}$) for all the XUE sources. As in \citet{2013ApJ...766..134N} and \citet{2020ApJ...903..124B}, we integrated the spectrum over the full wavelength range shown in the plots. The measurement error was determined from the [17.333–17.350]~\micron\ region. Dashed lines in the figure indicate three distinct emission features at 17.101~\micron, 17.223~\micron, and 17.356~\micron. Each of those lines consists of multiple \H2O\ transitions, and is dominated by the rotational transitions $12_{5\,8} \rightarrow 11_{2\,9}$, $11_{3\,9} \rightarrow 10_{0\,10}$, and $11_{2\,9} \rightarrow 10_{1\,10}$, respectively \citep[e.g. Table~3 in][]{2017ApJ...834..152B}.

\begin{figure}[ht]
    \centering
    \includegraphics[width=0.45\textwidth]{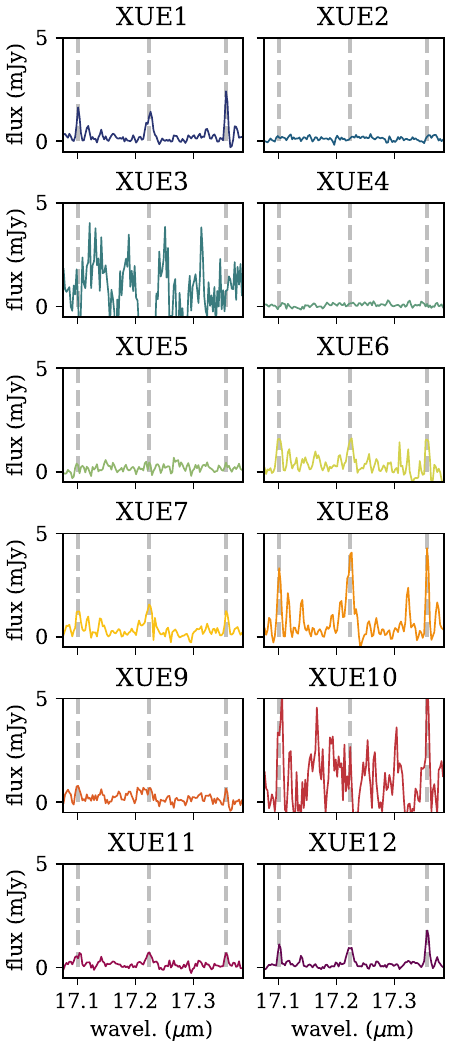}
    \caption{Water emission at 17\,\micron\ for the XUE sources. For XUE~2, 3, 4, 5, 9 and 10 it was not possible to significantly detect the water lines, we therefore report only upper limits. The vertical lines show the location of three well-separated emission features at 17.101~\micron, 17.223~\micron, and 17.356~\micron.}
    \label{fig:H2O_spectra}
\end{figure}
\end{appendix}

\end{document}